\documentclass[sigconf,9pt]{acmart}

\usepackage{blindtext}
\usepackage{graphicx}
\usepackage{textcomp}
\usepackage{xcolor}
\usepackage[ruled,linesnumbered,vlined]{algorithm2e}
\usepackage{algpseudocode}
 
\usepackage{hyperref}
\usepackage{breqn,xspace}
\usepackage{bm}
\usepackage{multirow}
\usepackage{threeparttable}
\usepackage{mathrsfs}
\usepackage{pifont}
\usepackage{subfig}
\usepackage{lipsum,multicol}
\usepackage{makecell}
\usepackage{booktabs}
\usepackage{duckuments}

\DeclareMathOperator*{\argmin}{arg\,min}

\newcommand{\name}{SigChord\xspace}

\ifodd 0
\definecolor{darkgreen}{RGB}{0,200,0}
\newcommand{\rev}[1]{{\color{blue}#1}}

\else
\newcommand{\rev}[1]{#1}

\fi

\AtBeginDocument{%
  }

\setcopyright{acmlicensed}
\copyrightyear{2025}
\acmYear{2025}
\acmDOI{XXXXXXX.XXXXXXX}

\acmConference[MobiSys '25]{The 23rd ACM International Conference on Mobile Systems, Applications, and Services}{June 23--27,
  2025}{Anaheim, California, US}
\acmISBN{xxx-x-xxxx-XXXX-X/25/06}

\acmSubmissionID{615}

\begin{document}

%%
%% The "title" command has an optional parameter,
%% allowing the author to define a "short title" to be used in page headers.
\title{\name: Sniffing Wide Non-sparse Multiband Signals for Terrestrial and Non-terrestrial Wireless Networks}

% \author{
% 	Conditional Accepted Paper \#615 to ACM MobiSys 2025
% }

%%
%% The "author" command and its associated commands are used to define
%% the authors and their affiliations.
%% Of note is the shared affiliation of the first two authors, and the
%% "authornote" and "authornotemark" commands
%% used to denote shared contribution to the research.
\author{Jinbo Peng}
\email{jbpeng22@m.fudan.edu.cn}
\affiliation{
  \institution{Fudan University}
  \city{Shanghai}
  \country{China}
}

\author{Junwen Duan}
\email{jwduan24@m.fudan.edu.cn}
\affiliation{
  \institution{Fudan University}
  \city{Shanghai}
  \country{China}
}

\author{Zheng Lin}
\email{zlin20@fudan.edu.cn}
\affiliation{
  \institution{Fudan University}
  \city{Shanghai}
  \country{China}
}

\author{Haoxuan Yuan}
\email{hxyuan22@m.fudan.edu.cn}
\affiliation{
  \institution{Fudan University}
  \city{Shanghai}
  \country{China}
}

\author{Yue Gao}
\email{gao.yue@fudan.edu.cn}
\affiliation{
  \institution{Fudan University}
  \city{Shanghai}
  \country{China}
}

\author{Zhe Chen}
\email{zhechen@fudan.edu.cn}
\affiliation{
  \institution{Fudan University}
  \city{Shanghai}
  \country{China}
}

% \author{G.K.M. Tobin}

%%
%% By default, the full list of authors will be used in the page
%% headers. Often, this list is too long, and will overlap
%% other information printed in the page headers. This command allows
%% the author to define a more concise list
%% of authors' names for this purpose.
% \renewcommand{\shortauthors}{Trovato et al.}

%%
%% The abstract is a short summary of the work to be presented in the
%% article.
% 
% https://en.wikipedia.org/wiki/LTE_frequency_bands 
% 4G (LTE) 400MHz - 6GHz
% https://en.wikipedia.org/wiki/5G_NR_frequency_bands
% 5G 450MHz - 7GHz
% sat 4-8GHz (C band), 12-18GHz (Ku band), 27-40GHz (Ka band)
\begin{abstract}
While unencrypted information inspection in physical layer (e.g., open headers) can provide deep insights for optimizing wireless networks, the state-of-the-art~(SOTA) methods heavily depend on full sampling rate (a.k.a Nyquist rate), and high-cost radios, 
due to terrestrial and non-terrestrial networks densely occupying multiple bands across large bandwidth~(e.g., from 4G/5G at 0.4-7~\!GHz to LEO satellite at 4-40~\!GHz). To this end, we present 
\name, an efficient physical layer inspection system built on low-cost and sub-Nyquist sampling radios. We first design 
a deep and rule-based interleaving algorithm based on Transformer network to perform spectrum sensing and signal recovery under sub-Nyquist sampling rate, and second, cascade protocol identifier and decoder based on Transformer neural networks to help physical layer packets analysis. We implement \name using software-defined radio platforms, and extensively evaluate it on over-the-air terrestrial and non-terrestrial wireless signals. The experiments demonstrate that \name delivers over 99\% accuracy in detecting and decoding, while still decreasing 34\% sampling rate, compared with the SOTA approaches.
\end{abstract}
\thanks{\rev{Our code is open-sourced at \url{https://anonymous.4open.science/r/Anonymous-A3F5}.}}

%%
%% The code below is generated by the tool at http://dl.acm.org/ccs.cfm.
%% Please copy and paste the code instead of the example below.
%%


%%
%% Keywords. The author(s) should pick words that accurately describe
%% the work being presented. Separate the keywords with commas.
%% A "teaser" image appears between the author and affiliation
%% information and the body of the document, and typically spans the
%% page.
% \begin{teaserfigure}
%   \includegraphics[width=\textwidth]{sampleteaser}
%   \caption{Seattle Mariners at Spring Training, 2010.}
%   \Description{Enjoying the baseball game from the third-base
%   seats. Ichiro Suzuki preparing to bat.}
%   \label{fig:teaser}
% \end{teaserfigure}

% \received{20 February 2007}
% \received[revised]{12 March 2009}
% \received[accepted]{5 June 2009}

%%
%% This command processes the author and affiliation and title
%% information and builds the first part of the formatted document.
\maketitle

\setlength{\abovedisplayskip}{1.5pt}
\setlength{\belowdisplayskip}{1.5pt}

% \vspace{-2ex}
\section{Introduction}

\begin{figure}[t]
    \centering
    \includegraphics[width=0.48\textwidth]{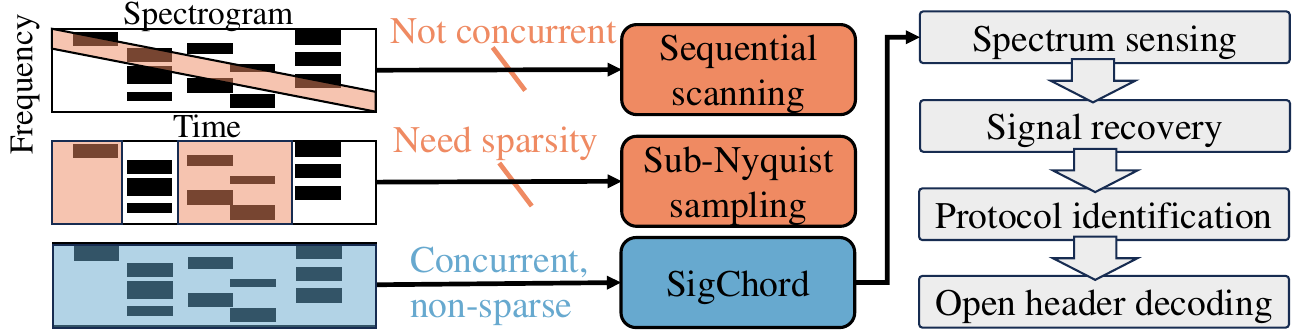}
    \caption{
        The \name wide multiband sniffing system. Unlike sequential scanning methods~\cite{shi2015beyond,guddeti2019sweepsense, subbaraman2023crescendo,wireshark} that capture only one signal at a time, or sub-Nyquist methods~\cite{mishali2009blind,hassanieh2014ghz,qin2018sparse,song2022approaching} limited to sparse spectra, \name detects and decodes concurrent, non-sparse wide multiband signals at low sampling rates.
    }
    \label{fig:teaser}
\end{figure}

Wireless communication plays a vital role in modern network infrastructure, underpinning countless applications and services~\cite{dvbs2,ieee802.11,lin2021spatial,ding2023hidden,yang2020magprint,yang2023slnet,lin2022tracking}. In this context, a lack of precise situational awareness will complicate management and undermine the ability to maintain network quality~\cite{huang2013depth,li2016mobileinsight}. Therefore, physical layer inspection tools are important in these environments. Such tools play a role similar to the upper layer sniffing and monitoring tools~\cite{wireshark,li2021experience,li2016mobileinsight,yaseen2021towards,huang2017sketchvisor} which help dissect and optimize network communication. Similarly, by enabling researchers and network operators to analyze wireless signals directly for deep packet inspection, these tools facilitate better understanding, optimization, network configurations, and protocol designs~\cite{gadre2020frequency, zhang2023dependent,guirguis2018primary}, ultimately improving the performance and reliability of wireless networks.

However, current packet inspection tools, such as Wireshark~\cite{wireshark}, provide narrow and isolated views of individual signals, lacking the holistic perspective needed to address the complexities of modern wireless environments, which are characterized by heterogeneous networks sharing spectrum resources to improve spectrum utilization~\cite{cano2017fair,zheng2005collaboration,pawelczak2005cognitive,brik2005dsap,guan2017smart,visser2008multinode,valls2016maximizing}.
On the one hand, terrestrial networks, including Wi-Fi, LTE, and 5G NR, share the unlicensed spectrum around the 5~\!GHz band for flexible network deployments~\cite{peng2024sums}. This leads to a complex and dynamic environment, creating significant management challenges and making it increasingly difficult for individual systems to assess and adapt to the coexistence of heterogeneous wireless networks~\cite{yuan2023graph,peng2024sums}.
On the other hand, the rise of non-terrestrial networks has exacerbated these challenges. Low-Earth Orbit (LEO) satellite networks, although typically assigned dedicated frequency bands, have been observed occupying unlicensed bands, leading to coexistence with terrestrial networks~\cite{al-jumaily2022evaluation,yuan2024satsense,yuan2025constructing}. Furthermore, the wide beam coverage of satellites often overlaps with multiple terrestrial networks. In bent-pipe communications~\cite{hauri2020internet}, satellites function as relay nodes to extend coverage, further complicating the dynamics of wireless coexistence and heterogeneity.
Therefore, there is an urgent demand for a tool that provides \textit{holistic} insights into the rapidly evolving and complex wireless environments.

% base stations cannot directly sense the network conditions within the coverage area, rendering it more challenging for network configuration.

% The rapid evolution of wireless networks has significantly intensified the demand for spectrum resources, leading to an ever-increasing pressing issue of spectrum scarcity. To combat this challenge, spectrum-sharing techniques have been introduced in terrestrial networks to enhance spectrum utilization efficiency~\cite{cano2017fair,zheng2005collaboration,pawelczak2005cognitive,brik2005dsap,guan2017smart,visser2008multinode}.

However, designing such a tool is non-trivial due to the prohibitively high cost of traditional sampling for wide bandwidths.
According to Nyquist sampling theory, capturing GHz spectra requires IQ sampling rates at least equal to the spectrum bandwidth, which confines sniffing to narrow and homogeneous signals~\cite{gao2023swirls,kochut2004sniffing,li2021experience,li2016mobileinsight,albazrqaoe2016practical,xie2022ng,hu2019nb}, or necessitates sacrificing detailed information~\cite{nika2014towards,zhang2022machine,guan2022efficient,zhang2014distributed,zhang2014vehicle,pu2021spectrum}. For the democratization of effective sniffing tools, existing studies for wideband signal sniffing reduces the sampling overhead through two strategies: {{rapid spectrum sweeping}}~\cite{guddeti2019sweepsense,subbaraman2023crescendo,shi2015beyond} and {{sub-Nyquist sampling}}~\cite{mishali2009blind,mishali2010from,hassanieh2014ghz,qin2018sparse,song2022approaching}, but both have significant limitations. Spectrum sweeping employs narrowband radios to scan the spectrum rapidly but cannot capture multiple signals simultaneously, restricting its ability to analyze concurrent behaviors~\cite{zhang2023dependent}. Moreover, while it expedites data collection by acquiring only limited information for each signal, this insufficiency restricts detailed physical-layer analysis such as packet header decoding. Sub-Nyquist sampling techniques, such as Sparse Fourier Transform~\cite{hassanieh2014ghz,hassanieh2012faster} and Compressed Sensing~(CS)~\cite{candes2006robust,mishali2010from}, can recover signals below the Nyquist rate by exploiting spectrum sparsity. However, the rapid expansion of wireless networks has introduced increasingly non-sparse conditions, where the occupied bandwidth exceeds the capabilities of sub-Nyquist sampling.
% \rev{Generally, \textit{non-sparse} scenarios refer to cases where the IQ sampling rate falls below twice the Landau rate~\cite{landau1967necessary,mishali2009blind}}.
In \textit{non-sparse} scenarios, the IQ sampling rate falls below twice the Landau rate~\cite{landau1967necessary,mishali2009blind}. \textit{Sub-Nyquist sampling techniques fail to detect and recover signals below twice the Landau rate}~\cite{mishali2009blind}.
% In such \textit{non-sparse} scenarios, the IQ sampling rate falls below twice the Landau rate~\cite{landau1967necessary,mishali2009blind}, causing \textit{Sub-Nyquist sampling techniques to fail in detecting and recovering signals}~\cite{mishali2009blind}.

% Although state-of-the-art~(SOTA) recovery algorithms~\cite{wu2019deep,bora2017compressed} use generative models for better performance, they involve time-consuming iterations and struggle to preserve complex signal structures (e.g., modulation and encoding) even with advanced generative models~\cite{zhou2024larger,lewkowycz2022solving,satpute2024can,chi2024rf}.

To overcome the aforementioned limitations, we propose \name, a Transformer-based wireless signal sniffer capable of \textit{real-time} and \textit{concurrent} sniffing of \textit{non-sparse} wideband signals, as shown in Figure~\ref{fig:teaser}. \name employs multi-coset sub-Nyquist sampling at the frontend. At the backend, \name first uses a Multi Layer Perceptron (MLP) network to embed the IQ samples into a latent space. Then, to enable \text{non-sparse} signal sniffing, unlike state-of-the-art~(SOTA) end-to-end and generative model based algorithms~\cite{wu2019deep,bora2017compressed} that involve time-consuming iterations and struggle to preserve complex signal structures (e.g., modulation and encoding)~\cite{zhou2024larger,lewkowycz2022solving,satpute2024can,chi2024rf}, we divide the signal recovery into two stages. First, a deep Transformer network predicts the critical information for recovery, i.e., the spectrum occupancy. Then, with this information, we make signal recovery below twice the Landau rate feasible through rule-based least squares estimation. After that, \name uses Transformer-based protocol identifiers and decoders for protocol classification and physical layer packet decoding. By decoding open headers, \name extracts sufficient unencrypted data for wireless network measurement.

\begin{itemize}
    \item We design a physical layer sniffing system \name, capable of spectrum sensing, signal recovery, protocol identification, and decoding (including terrestrial, OFDM signals such as Wi-Fi~\cite{ieee802.11} and non-terrestrial, single-carrier signals such as DVB-S2~\cite{dvbs2}) with low sampling overhead. \name enables detailed signal sniffing in wide and non-sparse spectra for the first time.

    % \item We devise a rule-based Transformer network for spectrum sensing and signal recovery, capable of operating below the sub-Nyquist sampling limit. Unlike end-to-end recovery approaches, our algorithm trains the neural network exclusively to predict spectrum occupancy, eliminating the need for labeled original Nyquist-rate signals. This rule-based design is both efficient and robust, that generalizes well to entirely unseen signals.
    \item We design a deep and rule-based algorithm for signal recovery that breaks the sub-Nyquist sampling limit, i.e., twice the Landau rate. The neural network does not require Nyquist-rate original signals as training labels, and the recovery algorithm generalizes well to entirely unseen signals.
    
    \item The cascaded signal analysis Transformer networks eliminate the need for complex protocol-specific preamble correlation and traditional signal processing algorithms. With minimal adjustments to the model architecture, \name seamlessly adapts to a wide range of protocols for decoding physical layer headers.
    \item We implement \name using software-defined radio platforms. Experiments show that \name is both highly effective and efficient, enabling accurate and real-time performance in physical layer inspection.
    % non-sparse wideband spectrum sensing, signal recovery, protocol identification, and decoding.

% \item \needrev{We design a multitask-learning framework, where the features of protocol identification model are fused with embeddings of protocol decoding model. The proposed framework significantly improves the decoding accuracy.}

% \needrev{
%     \item We show that using a Multi Layer Perceptron (MLP) to embed the IQ samples into a latent space can effectively exploit the redundancy in physical layer signals and enable separation and decoding of heavily aliased signals.
%     
%     \item Experiments show that \name not only achieves higher sniffing performance in commonly sparse spectrum scenarios, but also maintains accurate in non-sparse scenarios where CS fails.
% }
\end{itemize}

This paper is organized as follows. We give a brief introduction to sub-Nyquist sampling and physical layer protocol headers, and reveal our motivation in Section~\ref{sec:motiv}. In Section~\ref{sec:sys}, we demonstrate the design details of \name. We introduce the implementation and setup in Section~\ref{sec:impl}, show the experiment results in Section~\ref{sec:exp}. Related studies are reviewed in Section~\ref{sec:related} and we finally conclude this paper in Section~\ref{sec:conclusion}.

\begin{figure}[t]
    \centering
    \includegraphics[width=0.48\textwidth]{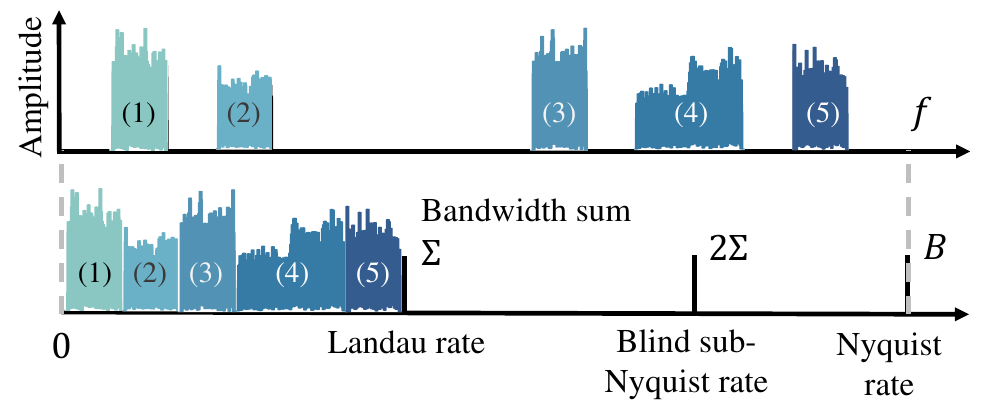}
    \caption{Nyquist rate, Landau rate~\cite{landau1967necessary} and blind sub-Nyquist rate~\cite{mishali2009blind} for multiband signals. The Nyquist rate equals to the bandwidth of the whole spectrum. The Landau rate equals to the sum of the bandwidth of each signal. The blind sub-Nyquist rate in \cite{mishali2009blind} is twice the Landau rate.}
    \label{fig:motiv:sampling_rate}
% \vspace{-3ex}
\end{figure}

\section{Motivation and Background} \label{sec:motiv}
\subsection{Sub-Nyquist Sampling}
% (简要介绍压缩感知，介绍Mishali的2倍支撑理论，展示CS在非稀疏下的糟糕性能)
% 由Candes等人提出的压缩感知是一种能在Nyquist速率下采集信号的方法~\cite{}。形式化地说，令$X \in \mathbb{C}^{L, N}$表示待测量的信号，$A \in \mathbb{C}^{P, L}$表示$P$次测量，其中$P << L$，压缩感知过程可表述为：
% \begin{equation}
%    Y = AX
% \end{equation}
% 其中$Y$表示测量结果。当$X$行稀疏，即部分行非零，且测量$A$的各列互相关性较弱时，可以通过$Y$和$A$唯一且准确恢复出$X$。一些工作基于压缩感知在认知无线电领域实现了低开销的频谱感知~\cite{}，但它们也因此对信号有稀疏性的要求。
% 对于压缩感知采样率和频谱稀疏度的关系，Mishali等人的理论性工作指出，系统总采样率至少要大于所有信号频宽总和的两倍才可准确恢复信号。我们在Figure~\ref{fig:motiv:cs}中展示了在信号非稀疏时压缩感知的性能恶化情况。
% Figure~\ref{fig:motiv:cs:ss}给出了频谱感知的检测率，可以看到在采样率接近理论极限时，检测准确率骤然下降。而在Figure~\ref{fig:motiv:cs:rec}中可以看到当信号不满足稀疏性的时候，即使在无噪声情况下CS也不能准确恢复信号。
Traditional signal sampling is limited by the Nyquist rate, requiring a sampling rate at least equal to the full spectrum bandwidth, necessitating costly high-speed ADCs. Landau~\cite{landau1967necessary} showed that when spectrum occupancy is known, the minimum sampling rate can be reduced to the sum of occupied bandwidths, known as the Landau rate.
Mishali et al.~\cite{mishali2009blind} extended this concept to blind sampling, where the occupied locations are unknown, showing that universal sub-Nyquist sampling is feasible. They proved that the overall sampling rate can be reduced to twice the Landau rate by CS techniques, enabling low-cost blind wideband signal sniffing. The relationship between Nyquist rate $B$, non-blind sub-Nyquist Landau rate $\Sigma$, and blind sub-Nyquist rate $2\Sigma$ is illustrated in Figure~\ref{fig:motiv:sampling_rate}.

CS~\cite{candes2006robust} practically facilitates sparse signal recovery at sub-Nyquist rates. Formally, let $X \in \mathbb{C}^{L\times N}$ denote the signal to be sampled, and $A \in \mathbb{C}^{P\times  L}$ be the $P$ measurement vectors, where $P \ll L$. The CS process can be expressed as
\begin{equation}
   Y = AX + n,
\label{equ:cs}
\end{equation}
where $Y$ represents the measurement results and $n$ is the noise. When $X$ is row-sparse, meaning that only a few rows are non-zero, and the measurement matrix $A$ satisfies the Restricted Isometric Property (RIP), $X$ can be uniquely and accurately recovered by Eq.~\ref{equ:cs_opt}, where $\|\ast\|_k$ represents $k$-norm and $\epsilon$ represents the noise threshold.
\begin{equation}
    \hat{X} = \argmin_{\tilde{X}} \|\tilde{X}\|_0 \quad \text{s.t.} \quad \|Y - A\tilde{X}\|_2 \leq \epsilon
    \label{equ:cs_opt}
\end{equation}

\begin{figure}[t]
    \centering
    \subfloat[]{
        \includegraphics[width=0.32\linewidth]{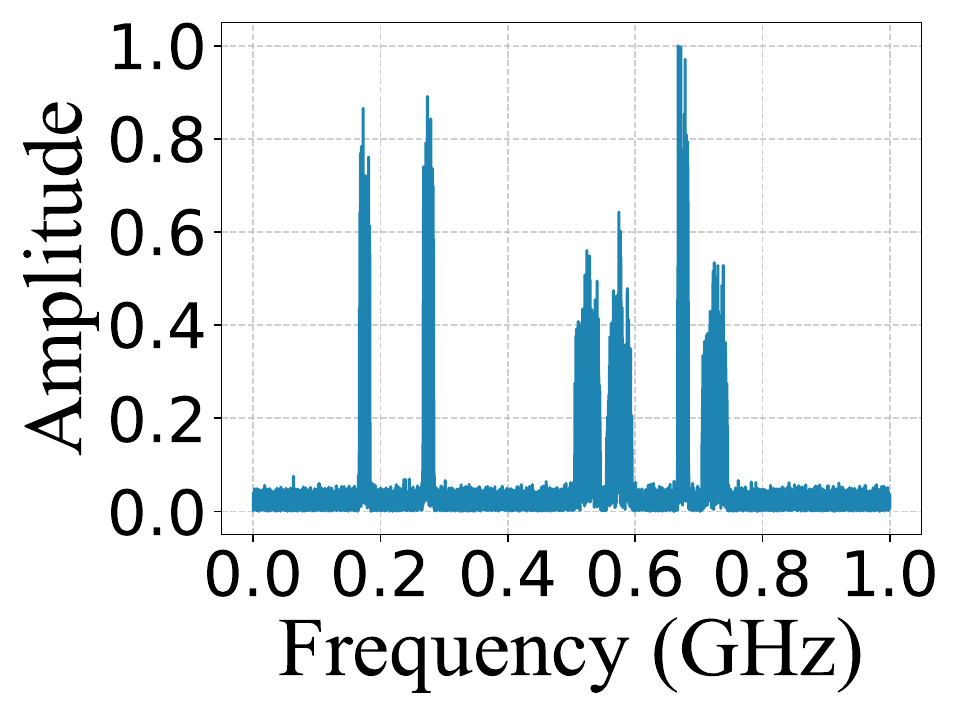}
        \label{fig:motiv:cs:a}
    }
    \subfloat[]{
        \includegraphics[width=0.32\linewidth]{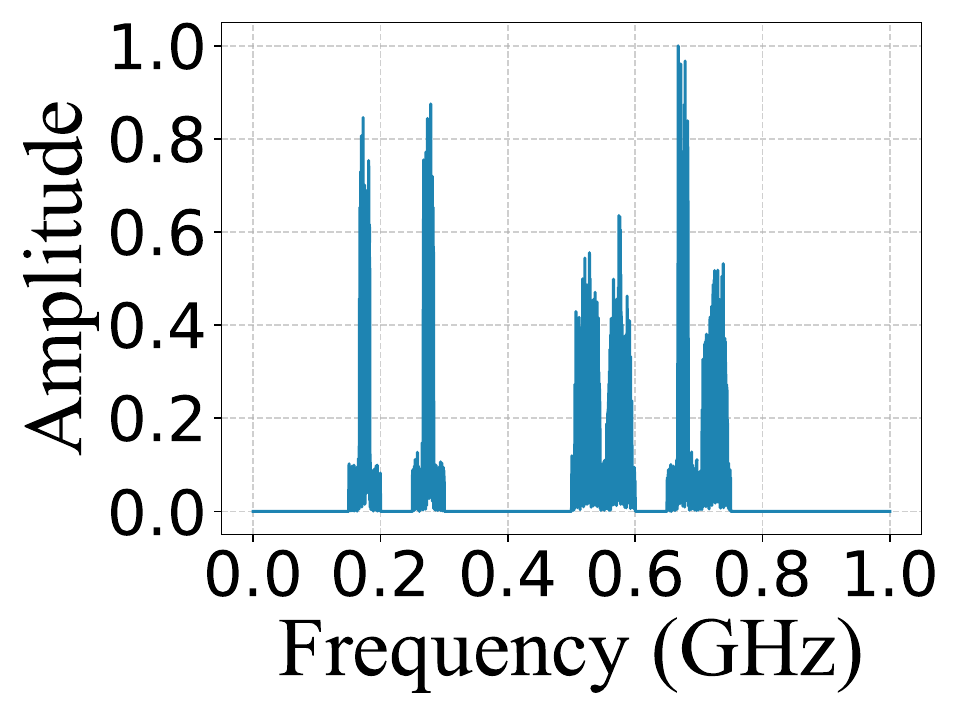}
        \label{fig:motiv:cs:b}
    }
    \subfloat[]{
        \includegraphics[width=0.32\linewidth]{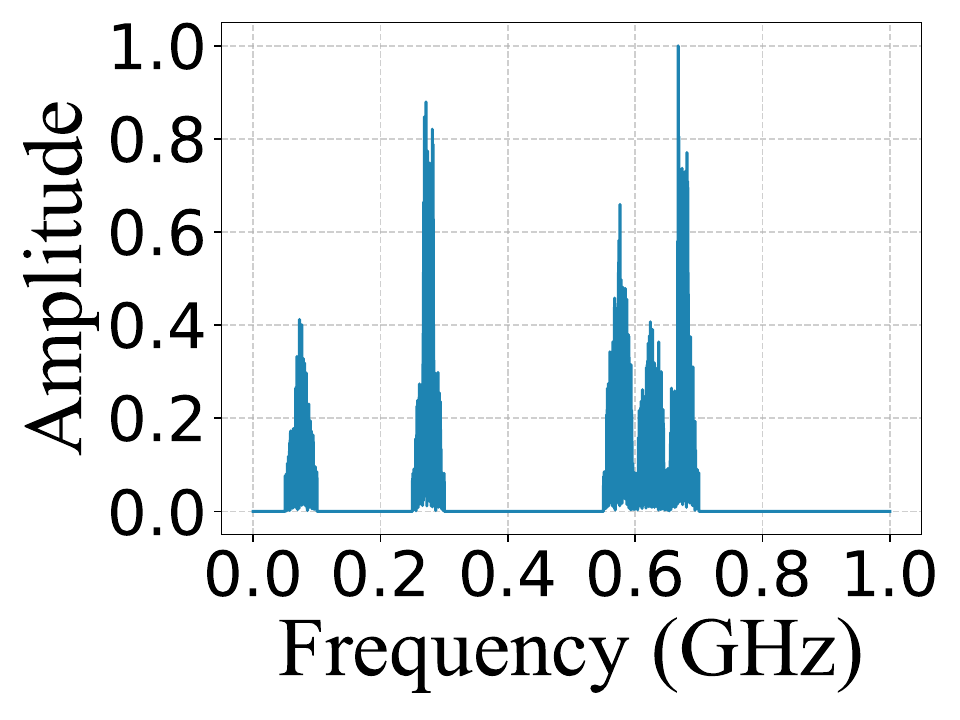}
        \label{fig:motiv:cs:c}
    }
    \caption{CS recovered~\cite{tropp2005simultaneous} spectrum under different sampling rates with SNR of 10~\!dB. (a) Original; (b) Recovered at 146\% of the blind sub-Nyquist limit; (c) Recovered at 83\% of the blind sub-Nyquist limit.}
    \label{fig:motiv:cs}
\end{figure}

Several studies~\cite{mishali2009blind,mishali2010from,qin2018sparse,song2022approaching} have proposed low cost sub-Nyquist sampling methods to formulate sampling and recovery into standard CS problems. The measurement matrix $A$ and measurement result $Y$ are determined by specific sampling schemes. The recovery target $X$ typically represents the signal spectrum, where each row of $X$ corresponds to the spectrum of a sub-band.
While these methods perform well under sparse spectrum conditions, their effectiveness diminishes significantly for non-sparse spectra that exceed the blind sub-Nyquist sampling capacity, leading to pronounced performance degradation.
Figure~\ref{fig:motiv:cs} demonstrates the CS recovery performance. We randomly generate DVB-S2 signals and Wi-Fi signals in a 1~\!GHz spectrum. We recover $X$ by a classic algorithm SOMP~\cite{tropp2005simultaneous}. With sufficient sampling rate, as shown in Figure~\ref{fig:motiv:cs:b}, CS algorithm basically recovers the signals. But under non-sparse scenario shown in Figure~\ref{fig:motiv:cs:c}, i.e., when the sampling rate is below the theoretical limit, the CS algorithm fails to recover the signal and cannot even correctly predict the spectrum occupancy.

% CS algorithm fails to recover that it cannot even correctly predict the spectrum occupancy.
% \needrev{
% The blind recovery performance will significantly drop when approaching the sub-Nyquist limit~\cite{mishali2009blind,song2024nonuniform,song2022approaching}.
% }

Despite degradation of existing methods near twice the Landau rate, we argue that this bound can be reduced. Recall that in non-blind recovery, i.e., where spectrum occupancy is known, the sampling limit is just one time the Landau rate.
Instead of abrupt direct recovery, there should be a smoother path to \textit{blind} recovery that passes through \textit{non-blind} recovery: spectrum sensing first, then non-blind recovery. The effective bound becomes the maximum of the Landau rate and the rate required for spectrum sensing.
If accurate spectrum sensing is feasible below twice the Landau rate, we can break the blind recovery bound.
To realize this approach, deep learning~\cite{lin2024efficient,hu2024accelerating,lin2022channel,zhang2024satfed,lin2024split} offers a promising avenue to for uncovering hidden structures in signals~\cite{zhang2022machine,zhang2024spectrum,yang2020magprint,yang2023slnet,gao2021fedswap,ding2023hidden}. 
For example, Zhang et al.~\cite{zhang2022machine} show that deep learning requires less sampling resource for spectrum sensing.
Follow the discussions above, \name uses deep learning for spectrum sensing to enable subsequent non-sparse recovery.

% 深度学习能够工作（最核心的频谱感知）的原因似乎不在于协议冗余，随机生成的根升余弦信号也能准确识别。
\subsection{Headers of Physical Layer Signals}
% 通常来说，为了增强信号的抗噪能力，物理层协议会有用于纠错的冗余设计，特别是对于承载了关键元信息的信号头会有更多的冗余。例如用于卫星通信的DVB-S2协议头将记载调制编码方式和帧类型的7个比特编码为64个比特。Wi-Fi信号头中的L-SIG字段会将24比特的信息经卷积码编码为48比特。在\needrev{IEEE 802.11/xx}中提出的20/40~\!MHz宽的High Throughput (HT) Wi-Fi在HT-SIG字段也采用了类似的编码。并且，40~\!MHz宽的HT Wi-Fi信号直接重复20~\!MHz信号的对应字段，提供更多冗余度

% 物理层协议携带的负载部分通常会用用户特定的伪随机序列打乱以防止窃听。而物理层协议头的编码通常是固定的，其携带了物理层通信参数的元信息。例如DVB-S2协议头用7个bit记录负载的调制方式，编码速率，包长和pilot的有无。These 7 bits are bi-orthogonally coded into 64 bits, which are then appended to the fixed SOF preamble.

In physical layer protocols, payloads are often scrambled with user-specific pseudo-random sequences for security, while headers remain accessible, carrying critical metadata.
For example, the DVB-S2 protocol used for satellite communications encodes the modulation and coding scheme~(MCS), packet length, and pilot state into 7 bits, which are bi-orthogonally expanded into 64 bits, scrambled by predefined sequence and appended to a fixed SOF preamble.
Similarly, IEEE 802.11 Wi-Fi signals encode MCS and packet length into 24 bits for non-HT packets, and 48 bits for HT packets which include additional parameters such as transmission duration and bandwidth. These bits are convolutionally encoded into 48 bits (non-HT) or 96 bits (HT) across the L-SIG and HT-SIG1 fields, scrambled and appended to fixed L-STF and L-LTF preambles.
% HT packets also encode parameters such as smoothing, sounding and CRC into HT-SIG2 fields following HT-SIG1.
The scrambling sequences and encoding schemes are open to the public, so it is feasible to recover the headers. The detailed fields are listed in Table~\ref{tab:header_fields}.

\begin{table}[t]
\centering
\caption{Fields of DVB-S2 and Wi-Fi Headers}
\label{tab:header_fields}
\begin{tabular}{|c|c|c|c|c|c|}
\hline
\multicolumn{6}{|c|}{\textbf{DVB-S2}} \\ \hline
\multicolumn{4}{|c|}{Bit 0-4} & \multicolumn{1}{c|}{Bit 5} & \multicolumn{1}{c|}{Bit 6} \\ \hline
\multicolumn{4}{|c|}{MCS} & \multicolumn{1}{c|}{Frame Size} & \multicolumn{1}{c|}{Pilot State} \\ \hline

\multicolumn{6}{|c|}{\textbf{Wi-Fi L-SIG}} \\ \hline
\multicolumn{2}{|c|}{Bit 0-3} & \multicolumn{1}{c|}{Bit 4} & \multicolumn{3}{c|}{Bit 5-16} \\ \hline
\multicolumn{2}{|c|}{\makecell{MCS (Non-HT)\\ Fixed (HT)}} & \multicolumn{1}{c|}{Reserved} & \multicolumn{3}{c|}{\makecell{Packet Length (Non-HT)\\Tx Duration (HT)}} \\ \hline
\multicolumn{2}{|c|}{Bit 17} & \multicolumn{4}{c|}{Bit 18-23}\\ \hline
\multicolumn{2}{|c|}{Parity} & \multicolumn{4}{c|}{Padding}\\ \hline

\multicolumn{6}{|c|}{\textbf{Wi-Fi HT-SIG1}} \\ \hline
\multicolumn{2}{|c|}{Bit 0-6} & \multicolumn{1}{c|}{Bit 7} & \multicolumn{3}{c|}{Bit 8-23} \\ \hline
\multicolumn{2}{|c|}{MCS} & \multicolumn{1}{c|}{Bandwidth} & \multicolumn{3}{c|}{Packet Length} \\ \hline
\end{tabular}
\end{table}

A blind sniffer operates non-cooperatively without prior protocol knowledge. To detect physical layer packets, traditional methods involve enumerating preambles to identify potential headers, yielding low detection accuracy~\cite{belgiovine2024tprime}. Additionally, subsequent signal processing requires various synchronization, compensation, equalization, demodulation, and decoding algorithms for each protocol, significantly increasing system complexity. To address these, after signal recovery, \name employs end-to-end neural networks for signal classification and processing, including synchronization, compensation, equalization, demodulation and decoding, enabling accurate sniffing and simplified system designs.

% 然而，盲嗅探器是以非合作的方式工作的，并不能提前知道要接收什么协议的信号。因此，传统的方法需要枚举协议preamble做关联去寻找可能的协议头，这样分辨信号的准确率并不高。并且，之后的信号解析需要针对不同的协议实现不同的同步，补偿，均衡，解调和解码算法，会带来较大的系统复杂度。\name使用神经网络来完成端到端的信号分类，以及信号的同步，补偿，均衡，解调和解码，以实现准确的嗅探和简化的系统设计。

% 实际的多频带信号比Figure3所示的场景更加复杂，多种异质的信号共存，存在噪声、信道衰减等干扰。近年来深度学习在无线网络领域展现出了强大的性能，在信号分类，信道估计，信号接收等方面能够提取出信号的深层特征，实现端到端的信号分析任务~\cite{}。因此，深度学习有望从复杂的采样中发掘信号的冗余结构，从而准确解析各个信号。

\section{System Design} \label{sec:sys}
% 我们的系统设计主要针对两方面。一是考虑到嗅探器需要能够方便地收集信号，\name的采样前端使用了Multicoset欠采样，减轻嗅探器前端负担；二是信号分析后端需要能够完成信号频谱感知、信号分离、协议识别、解调和解码等一系列复杂任务，因此在\name中我们基于近年来展现出强大语义分析能力的Transformer神经网络来设计各个模块。
\name consists of a low-cost sub-Nyquist sampling frontend and a signal analysis backend. Firstly, \name utilizes multi-coset sub-Nyquist sampling to alleviate the burden on the sniffing frontend. Secondly, the backend must handle complex sequential analysis tasks from limited sub-Nyquist samples. \rev{That is, we need to effectively predict the spectrum occupancy in order to enable smoother signal recovery path through non-blind recovery. And we need to effectively capture signal features from time-series data in order to classify, compensate, demodulate and finally decode signals. To this end, \name employs Transformer-based modules for their proven strengths in semantic analysis~\cite{vaswani2017attention,ling2022multi,zaheer2020big}. Besides, the well-established and optimized tool-chains for Transformers could enhance the availability and deployment efficiency of \name.}

% 我们将信号嗅探任务划分为三个子任务：频谱感知，协议识别和协议解析。频谱感知任务是从采样的IQ序列中确认哪些频谱被占用。协议识别和协议解析任务根据频谱占用信息分离各个信号。协议识别任务识别信号的协议类型，定位协议头在时域上的相对位置。协议解析任务完成对协议信号补偿、恢复、解码，最终解析出物理层公开包头的字段。我们的系统设计如图~\ref{fig:sys:overview}所示。

\begin{figure}[t]
    \centering
    \includegraphics[width=0.48\textwidth]{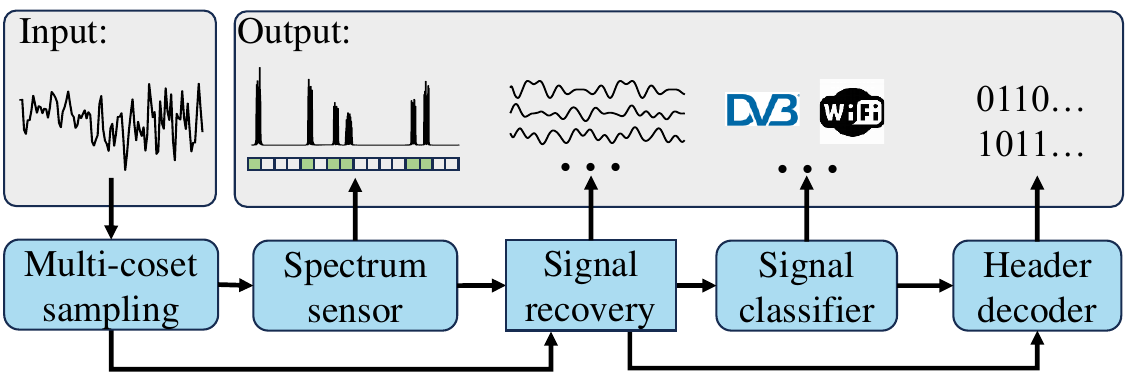}
    \caption{Signal processing pipeline of \name}
    \label{fig:sys:overview}
\end{figure}

The backend of \name comprises three submodules: spectrum sensing, protocol identification, and header decoding. The spectrum sensing module identifies occupied sub-bands from sub-Nyquist IQ-sampled data. With this information, \name recovers signals from limited sub-Nyquist samples and provides downstream models with separated signals. The protocol identification module classifies frames with intact headers, filtering out those without, while the header decoding module calibrates, demodulates, and decodes physical layer packet headers. The overall pipeline is illustrated in Figure~\ref{fig:sys:overview}.
\rev{With a modular and software-based design, the backend could run flexibly on PCs, servers, or even the cloud, reducing \name's reliance on resource-limited devices. Additionally, the modular architecture ensures flexible integration and extension for new protocols.}

% 协议解析部分用query decoder，其中query做两种embedding。一种是非固定字段，每个位置对应一个embedding（类似position embedding）；另一种是固定字段，按照本来的0，1做两类embedding（目前实验没跑，不知道效果）

\subsection{Sub-Nyquist Sampling and Preprocess} \label{sec:sys:pre}

The sub-Nyquist sampling frontend of \name utilizes multi-coset sampling, which employs a multiplexer and multiple low-speed ADCs with unique time delays to sample the signal in parallel. Let the multiband signal be $x(t) = \sum_j x_j(t)$, where $x(t)$ is band-limited within $[-B/2, B/2]$. The Nyquist rate for sampling $x(t)$ is $B$. The low-speed ADCs operate at $B/L$. multi-coset sampling employs $P$ ADCs ($P \ll L$), so the total sampling rate becomes $P/L$ of the Nyquist rate. The samples are captured as $y \in \mathbb{C}^{P \times N}$, with the $j$-th ADC's samples given by
\begin{equation}
y_{j,n} = x\left(\frac{nL + c_j}{B}\right), \quad n = 1, 2, \dots, N
\label{equ:multicoset}
\end{equation}
where $c_j$ is the unique offset for the $j$-th ADC, and $N$ is the number of samples per ADC. Applying Fourier transform to Eq.~\eqref{equ:multicoset}, we get the compressed sensing model $Y = AX$, where $Y \in \mathbb{C}^{P \times N}$ is the transformed samples, $A \in \mathbb{C}^{P \times L}$ is the measurement matrix formed by Fourier bases determined by the offsets $c_j$, and $X \in \mathbb{C}^{L \times N}$ is the spectrum matrix to be recovered. Detailed discussions can be found in \cite{mishali2009blind, song2024nonuniform}, and we omit them here for brevity.

% \needrev{The core ideal is that, in principle, the spectrum matrix $X$ is recoverable from sub-Nyquist multi-coset samples. Instead of seeking CS methods,} in \name we design Transformer-based algorithms to analyze signals, as will be introduced below.

% embedding
% Transformer layer使用自注意力机制处理表示成token的序列化的数据，令每个位置的token能够与其他位置的token做关联，得到新的序列数据，并呈递给下一层。通常的Transformer layer具有多个自注意力头，以从多个角度关联序列数据。Transformer网络根据不同的任务将输入的序列化数据映射到高维空间~\cite{}，然后经过多层的Transformer layer提取深层特征。通常的Transformer形成encoder和decoder两部分，encoder用于序列数据自我关联，decoder用于已有序列数据（例如encoder提取的特征）和新输入序列数据的关联。

\begin{figure}[b]
    \centering
    \includegraphics[width=0.45\textwidth]{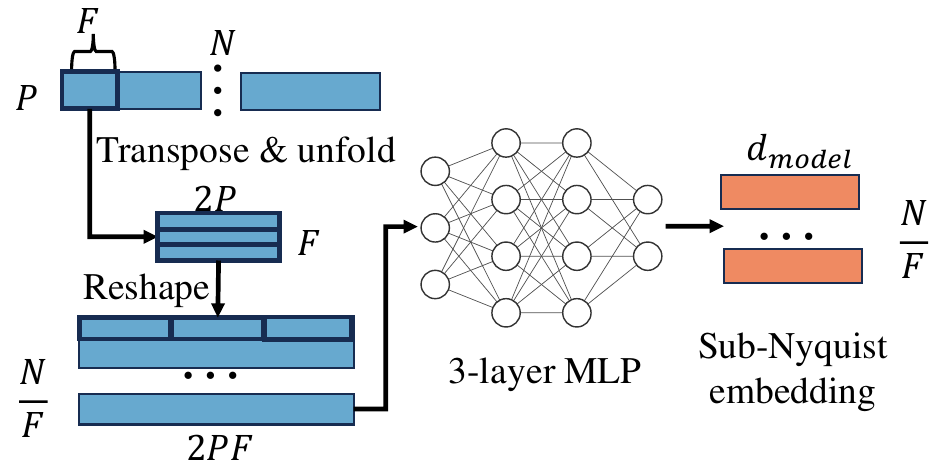}
    \caption{
        The embedding process of \name.
    }
    \label{fig:sys:embedding}
\end{figure}

Instead of relying on CS methods, we incorporate a data-driven Transformer with rule-based signal recovery. The Transformer layer processes a sequence of high-dimensional token vectors using an attention mechanism, where each token is correlated with others, producing a new token sequence. The standard Transformer architecture includes encoders and decoders: the encoder applies self-attention to the input sequence, producing an output sequence of the same length, while the decoder uses attention to correlate the existing features (e.g., encoder features) with new query tokens, generating an output sequence matching the query length. Each Transformer layer distributes the attention mechanism across multiple heads, allowing attention from diverse perspectives.

% 将采样序列输入到Transformer网络中并不straightforward。为了解析信号，模型需要处理包含完整协议头的较长的IQ序列。此前的信号分类工作提出了包括，reshaping, Linear embedding layer, convolutional neural network等方法来处理IQ序列。然而，面对高度混叠的multiband signal samples，此前简单的embedding方案不能很好地表示采样序列特征，并且不能灵活适应不同欠采样率的输入。为此，我们提出非线性的3层MLP embedding。记输入的multi-coset采样序列为$s \in \mathbb{C}^{P \times N}$，我们首先将其reshape为$s' \in \mathbb{R}^{N \times 2P}。
\rev{To process IQ samples with a Transformer, previous methods include reshaping~\cite{belgiovine2024tprime,cai2022signal}, linear embeddings~\cite{cai2022signal}, and convolutional neural networks (CNN)~\cite{hamidi-rad2021mcformer} for Nyquist sampled data. Given the severe aliasing in sub-Nyquist samples, we adopt reshaping and a 3-layer MLP for nonlinear embedding, in seek of both effectiveness and efficiency}, as shown in Figure~\ref{fig:sys:embedding}.
Given the multi-coset samples $y \in \mathbb{C}^{P \times N}$, we transpose and unfold $y$ into $z \in \mathbb{R}^{N \times 2P}$ such that $z_{j,2k +1} + iz_{j, 2k+2} = y_{k, j}$, where $i$ is the imaginary unit. Longer input sequences to Transformer significantly increase memory and processing demands, therefore, we further reshape $z$ into $z' \in \mathbb{R}^{\frac{N}{F} \times 2PF}$, where $F$ is the folding factor to group IQ samples into patches and reduce the input sequence length.

We then feed $z'$ into the MLP. The MLP consists of 3 linear layers with Gaussian Error Linear Units~(GELU) activation and dropout after each of the first 2 layers. \rev{Chosen for its smooth non-linearity, GELU improves gradient flow and enhances model expressiveness compared to ReLU.} The input feature size is $2PF$, and the output sizes are $2d_{model}$, $2d_{model}$, and $d_{model}$, respectively. Each layer includes layer normalization. Positional embeddings are added to the output. \rev{Each subsequent module in \name is equipped with its own reshaping and a 3-layer MLP for IQ sample embedding.}

\subsection{Spectrum Sensing and Signal Recovery} \label{sec:sys:ss}

\begin{figure}[t]
    \centering
    \includegraphics[width=0.45\textwidth]{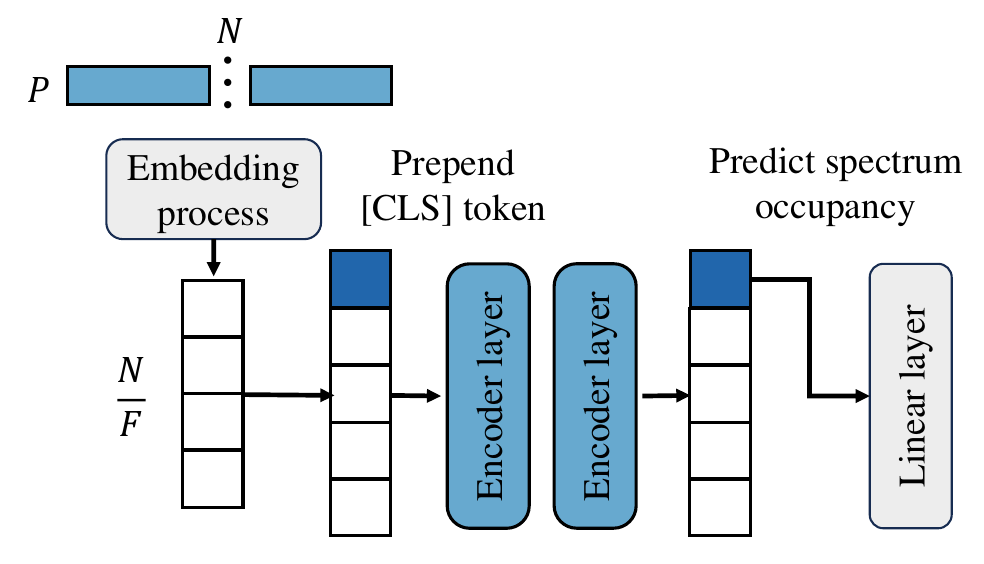}
    \caption{The architecture for spectrum sensing.}
    \label{fig:sys:ss}
\end{figure}

% 此前我们提到若能在低于2倍landau rate的采样率下准确感知信号频谱，就可通过non-blind recovery突破\cite{}中的bound。将待检测频谱均匀划分为$L$个子频带, 这与Section3.A中待恢复的频谱矩阵$X$的行数相同，即子频带划分与压缩感知方法的划分相同。\name将频谱感知视作多标签的二分类任务，预测每个子频带的占用情况。记真实频谱占用为$S \in \{0,1\}^{L}$，其中$S_{j} = 1$表示第$j$个子频带被占用。\name使用如图所示的模型。为简化模型规模，我们在preprocessing embedding基础上prepend一个与输入无关可学习的[CLS] token，经过两层Transformer encoder之后，仅该[CLS] token对应的hidden state输入到单层线性层，经sigmoid激活作为输出。模型输出为$\hat{S} \in \mathbb{R}^{L \times N}$，其中$\hat{X}_{i,j}$表示第$i$个子频带在第$j$个采样点被占用的概率。我们使用二元交叉熵损失函数来训练模型。
As discussed, accurate spectrum sensing below twice the Landau rate allows breaking the sampling limit in \cite{mishali2009blind}. We uniformly divide the spectrum into $L$ sub-bands, corresponding to the rows of the spectrum matrix $X$ in Eq.~\eqref{equ:cs}. \name formulates spectrum sensing as a multi-label binary classification task to predict sub-band occupancy. The true spectrum occupancy is represented by a vector of length $L$, where the $j$-th element indicates whether the $j$-th sub-band is occupied. The spectrum sensor, shown in Figure~\ref{fig:sys:ss}, consists of two Transformer encoder layers and a linear layer. To reduce complexity, a learnable [CLS] token of dimension $d_{model}$ is prepended to the embedding, independent of the input. After passing through the \rev{encoders}, the hidden state of the [CLS] token is fed to the linear layer, activated by Sigmoid to produce the output $\hat{S} \in (0, 1)^{L}$, where $\hat{S}_j$ denotes the occupancy probability of the $j$-th sub-band.

% 得到频谱感知结果$\hat{S}$后，我们将其用于信号恢复。$\hat{S}$表示了对子频带占用的预测，即频谱矩阵$X$哪些行非零。这样，我们可知Eq.~\eqref{equ:cs}中$X$的哪些行以及$A$的哪些列对结果$Y$有贡献。记$S = \{j|\hat{S}_j > \theta\}$, where $\theta \in (0, 1)$ is a threshold。我们选择与$S$对应的测量矩阵$A$的列作为$A_{S}$，将$\hat{S}$对应的频谱矩阵$X$的列记作$X_S$，则无噪声时有$Y=AX=A_S X_S$。若集合$S$的势$|S| < P$，则关于$X_S$的方程$Y=A_S X_S$为超定方程，可通过最小二乘法求解。我们的恢复算法如Algorithm~\ref{alg:recovery}所示。值得一提的是，Algorithm~\ref{alg:recovery}中的要求$|S| \le P$与Landau rate是一致的。Multi-coset采样中$P$个ADC的采样率为$B/L$，总采样率为$\frac{P}{L}B$，而我们划分的子频带频宽为$B/L$，总的占用频宽为$\frac{|S|}{L}B$。因此，$|S| \le P$与Landau rate bound相对应。
After data-driven spectrum sensing, we use $\hat{S}$ for rule-based signal recovery. $\hat{S}$ indicates the occupied sub-bands, i.e., the non-zero rows of the spectrum matrix $X$. This enables identification of the contributing rows of $X$ and columns of $A$ in Eq.~\eqref{equ:cs}. Let $S = \{j|\hat{S}_j > \theta\}$, where $\theta \in (0, 1)$ is the threshold. We select the columns of $A$ and rows of $X$ corresponding to $S$, denoted as $A_S$ and $X_S$. In the noise-free case, this gives $Y = A_S X_S$. If $|S| < P$, the equation becomes overdetermined, and $X_S$ can be solved via least squares. Our recovery algorithm is presented Algorithm~\ref{alg:recovery}. Note that the condition $|S| \le P$ aligns with the non-blind Landau rate. With $P$ ADCs sampling at $B/L$, the total sampling rate is $\frac{P}{L}B$, and the total occupied bandwidth is $\frac{|S|}{L}B$. Hence, $\lvert S \rvert \le P$ satisfies the Landau rate limit.

% \name的频谱感知和子频带信号恢复粒度为$B/L$,即multi-coset采样中每个低速ADC的采样率。然而，我们可以通过逻辑上的multi-coset ADCs实现更细粒度的频谱感知和信号恢复。具体来说，我们可以将一路ADC的采样逻辑上看作是k个ADC的交替采样，每个ADC采样率为$B/(kL)$。这样通过重新组织测量矩阵$A$和测量结果$Y$，我们可以在不改变硬件的情况下将频谱感知和信号恢复的粒度提高到$B/(kL)$。
\rev{The spectrum sensing and sub-band signal recovery granularity of \name is $B/L$, determined by the sampling rate of each low-speed ADC in multi-coset sampling. To achieve finer granularity, each single ADC can be logically treated as $k$ alternating ADCs, each sampling at $B/(kL)$. By reorganizing the measurement matrix $A$ and results $Y$, this refinement could enhance granularity to $B/(kL)$ without hardware modifications.}

\begin{algorithm}[t]
\SetKwInput{KwRequire}{Require}
\caption{Transformer-based sub-Nyquist signal recovery}
\label{alg:recovery}
\KwIn{multi-coset samples $y \in \mathbb{C}^{P \times N}$, measurement matrix $A \in \mathbb{C}^{P \times L}$, the spectrum sensing model $f \colon \mathbb{C}^{P \times N} \to (0, 1)^L$, threshold $\theta \in (0, 1)$.}
\KwOut{spectrum matrix $X_S$ for the occupied sub-bands.}
\KwRequire{$\lvert\{j | \hat{S}_j > 0\}\rvert \le P$.}

Predict occupied sub-bands $\hat{S} \gets f(y)$ \;
Predict the support set \rev{$S \gets \{j | \hat{S}_j > \theta\}$} \;
Select columns of $A$ corresponding to $S$ as $A_{S}$ \;
Transform $y$ to the measurement result $Y$ \;
Solve $X_S \gets \argmin_{\tilde{X}} \|Y - A_{S}\tilde{X}\|_2$.
\end{algorithm}

\subsection{Protocol Identification} \label{sec:sys:pkt_cls}

\begin{figure}[t]
    \centering
    \includegraphics[width=0.45\textwidth]{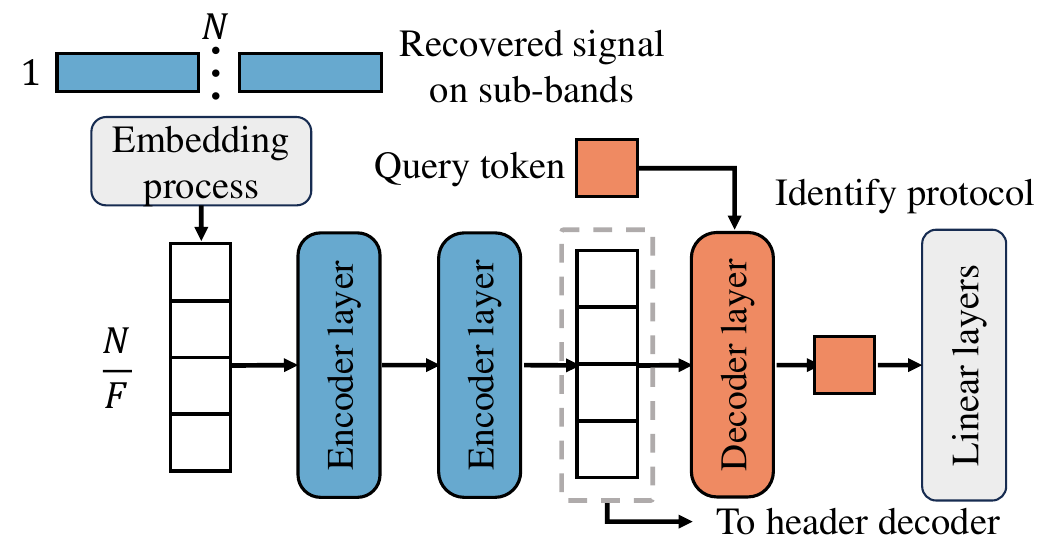}
    \caption{The architecture for protocol identification.}
    \label{fig:sys:pkt_cls}
\end{figure}

% 该模块负责识别每个子频带中的信号协议类型，将包含完整协议头的信号帧呈递给header decoding model，滤除不包含完整header的信号帧。$X_S \in \mathbb{C}^{\lvert S \rvert \times N}$表示了各个子频带的频域信号，我们将其每一行做逆傅里叶变换到时域，分别记作$x_{bb}^{(j)} \in \mathbb{C}^{1 \times N}, \quad j = 1,2,...,\lvert S \rvert$，交由protocol identification model并行处理。形式上$x_{bb}^(j)$相当于$P = 1$的采样序列，因此Section~\ref{sec:sys:pre}中的prepocessing可以直接应用。我们的protocol identification model如图~\ref{fig:sys:pkt_cls}所示。与spectrum sensor不同，此处我们不使用[CLS] token，而是使用Transformer encoder + decoder layer的组合形式，以与后续的基于Transformer decoder layer的header decoding model共享特征。
% 模型将xbb经过embedding和两层encoder layer后的feature以及一个可学习的query token输入给decoder layer。decoder layer用这个query token与encoder layer的输出做attention，萃取出全局的信号特征，最终交由两层线性层做分类。线性层的中间维度也为$d_{model}$。由于有些时候输入的信号帧中一些被占用的子频带中的信号全为负载，不包含协议头，不能够解析，因此分类时我们额外加入一个无头类别。
This model identifies the protocol in each sub-band and forwards signal frames with intact headers to the header decoding model, filtering out those without. Let $X_S \in \mathbb{C}^{\lvert S \rvert \times N}$ represent the frequency-domain signals in the selected sub-bands. Each row of $X_S$ undergoes an inverse Fourier transform to the time domain, denoted as $x_{bb}^{(j)} \in \mathbb{C}^{1 \times N}, j = 1, 2, \dots, \lvert S \rvert $, which are then processed in parallel by the protocol identification model. Structurally, each $x_{bb}^{(j)}$ can be treated as an extreme case of multi-coset sampling with only one low-speed ADC. This allows the preprocessing steps from Section~\ref{sec:sys:pre} to be applied directly, where $P = 1$ in this case.
The protocol identification model is illustrated in Figure~\ref{fig:sys:pkt_cls}. In addition to its classification role, this model also functions as a feature extractor for the subsequent header decoding model, which is primarily based on decoder layers. To enable effective feature sharing with the header decoding model, we use a combination of Transformer encoder and decoder layers rather than an encoder-only structure as in the spectrum sensor.
The model processes the embedded $x_{bb}^{(j)}$ features through two encoder layers and correlates them with a learnable query token as a substitute for the [CLS] token. \rev{This query token is unrelated to the input and serves as a part of the model's learnable parameters.} The decoder layer applies attention mechanism between the query token and encoder output to extract global signal features, which are then classified via two linear layers. The dimension of the hidden linear layer is set to $d_{model}$ with GELU activation, layer normalization and a dropout rate of 0.1. The output from the final linear layer is activated by Softmax. To account for signal frames in sub-bands lacking intact headers and therefore undecodable, a dedicated \textit{no-header} class is included in the classification. Additionally, the encoder features are forwarded to the header decoding model for feature fusion.

\subsection{Header Decoding} \label{sec:sys:decode}

\begin{figure}[t]
    \centering
    \includegraphics[width=0.45\textwidth]{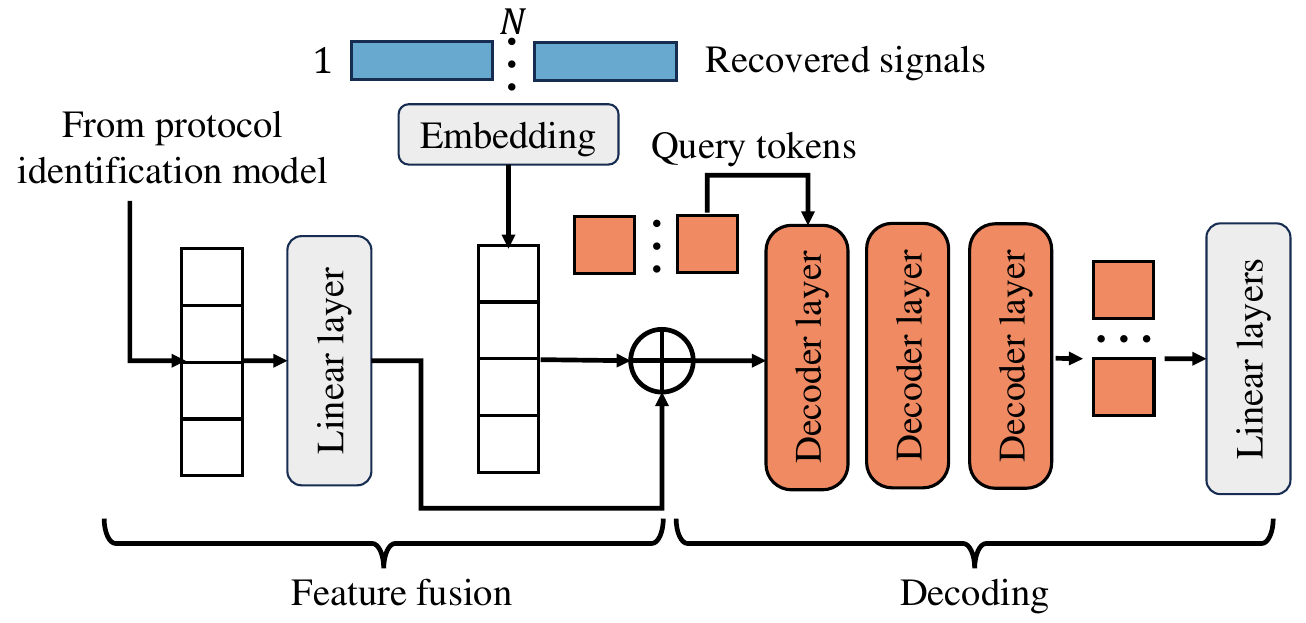}
    \caption{The architecture of header decoding models.}
    \label{fig:sys:decoder}
\end{figure}

% 该模块负责对包含完整协议头的信号帧进行解码，恢复出物理层公开包头的字段。模型包含feature fusion部分和解码部分。在feature fusion部分，首先我们同样将$x_{bb}^{(j)}$经由~\ref{sec:sys:pre}的流程处理，之后将得到的embedding与protocol identification model的encoder layers的输出特征进行融合。后者经由一层linear layer以得到与前者相同的feature维数，之后按照一定比例相加。相加的比例系数为可学习的两个参数。
This module is responsible for decoding signal frames with intact protocol headers to retrieve open fields from the physical layer packets. Frames without intact headers, as flagged by the protocol identification model, are excluded from processing here. The model consists of a feature fusion stage and a decoding stage, as illustrated in Figure~\ref{fig:sys:decoder}. In feature fusion, $x_{bb}^{(j)}$ is firstly processed following the steps in Section~\ref{sec:sys:pre}, similar to that in Section~\ref{sec:sys:pkt_cls}, although with a different embedding dimension $d_{model}$. The resulting embeddings are then merged with output features from the encoder layers of the protocol identification model. To match feature dimensions, the encoder output is passed through an adapter, which is a linear layer with GELU activation and layer normalization, before being combined with the embeddings via a weighted summation. The fusion process is expressed as
\begin{equation}
    x_{\text{fusion}} = f_{\text{embed}}(x_{bb}^{(j)}) + \alpha \cdot f_{\text{adapter}}(x_{\text{enc}}),
    \label{equ:fusion}
\end{equation}
where $f_{\text{embed}}$ and $f_{\text{adapter}}$ denote the embedding process and adapter in the header decoding model, respectively, $\alpha$ is the learnable weight parameter, and $x_{\text{enc}}$ is the output of the encoder layers in protocol identification model.

% The decoding stage只使用Transformer decoders，fusion之后的feature和一串可学习的query tokens作为decoder layers的输入。query tokens的数目与待解码的bit数相同，每个token对应一个bit的解码。例如DVB-S2协议头包含7 bit的信息，因此我们使用7个query tokens。经过decoder layers后，模型完成对信号的校准，解调和解码，最后每个token对应的hidden state经过单一的线性层，经Sigmoid激活后输出。输出的概率值表示了对应bit为1的概率。
The decoding stage employs Transformer decoders only, taking the fused features and a sequence of learnable query tokens as input. \rev{Similar to the protocol identification model, these query tokens are also irrelevant to the input samples and serve as parts of the learnable parameters}. Although each physical-layer protocol has its unique and complex calibration, demodulation, and decoding process, \rev{\name adopts a unified architecture for header decoding. For different protocols, the complex decoding procedures are abstracted into a direct signal-to-bit mapping process.} The difference lies in query tokens. The number of query tokens matches the number of unencoded bits for a given protocol, with each token dedicated to decoding a single bit. \rev{This allows flexible extension of new protocols while reserving systematic consistency, with only query tokens adjusted to align with the bit representation of the target protocol.} For example, since the DVB-S2 protocol header encodes 7 bits, its decoding model uses 7 query tokens. After passing through three decoder layers, the model performs signal calibration, demodulation, and decoding. The hidden state \rev{corresponding to each query} token is then passed through a single linear layer, followed by a Sigmoid activation, outputting the probability of each bit being 1.

\section{Implementation} \label{sec:impl}

We detail the datasets and model configurations below. To evaluate the performance of \name, we prepare both a \rev{in-band} synthetic dataset and an over-the-air collected dataset \rev{where multiple signals co-exist across 5.0-5.8~\!GHz and are sensed to get the sub-Nyquist multi-coset samples}. All models in \name are trained on the synthetic dataset.

\subsection{Synthetic Dataset} \label{sec:impl:simu}

% 我们以non-terrestrial和single-carrier的一类代表性的信号DVB-S2，以及terrestrial中OFDM的一类代表性的信号IEEE 802.11 Wi-Fi信号生成数据。其中Wi-Fi信号另外还包含high throughput的HT类型的信号。我们使用MATLAB的Communications Toolbox生成多频带信号, 信号模型如下所示：
We synthesize data using a representative single-carrier physical layer protocol, namely DVB-S2, and two representative terrestrial, OFDM-based protocols, namely IEEE 802.11g/b non-HT Wi-Fi and IEEE 802.11n HT Wi-Fi. We use MATLAB's Communications Toolbox to generate multiband signals, with the signal model described as follows,
\begin{equation}
    x(t) = \sum_{j=1}^{M} h_{j}(t) \ast [x_j(t) e^{2\pi i (f^{c}_{j} + \Delta f_{j})t}] + n(t),
    \label{equ:signal}
\end{equation}
where $x(t)$ is the synthetic multiband signal, $M$ is the number of narrowband signals existing in the spectrum and is set to 6, $h_{j}(t)$ is the channel impulse response, $x_{j}(t)$ represents each narrowband signal, $\ast$ denotes convolution, $f^{c}_{j}$ is the in-band carrier frequency, $\Delta f_{j}$ is the frequency offset, and $n(t)$ is the additive white Gaussian noise~(AWGN). For each narrowband signal, physical layer packets are generated with randomly assigned payloads, while transmission parameters such as modulation scheme, coding scheme, and packet length are randomly selected.
DVB-S2 signals have a symbol rate of 20MHz, with roll-off factors for the square-root raised cosine filter chosen randomly from $\{0.2, 0.25, 0.35\}$.
Wi-Fi signal bandwidths can be either 20~\!MHz or 40~\!MHz, as permitted by protocol specifications. The amplitude of each narrowband signal is scaled by a random factor within [0.5, 1]. Channel impulse responses are drawn from Rician and Rayleigh fading channels, featuring 3 delay paths, an average delay spread of 40~\!ns, and path gains within [-10, 0]~\!dB. We divide the spectrum into 50~\!MHz sub-bands and choose the carrier frequency $f^{c}_{j}$ from $\{ 25 + 50k \ \text{MHz} \mid k \in \mathbb{Z}, \ 0 \leq k \leq 15 \}$. The frequency offset satisfies $\lvert \Delta f_j \rvert < \text{200 KHz}$. The AWGN has a signal-to-noise ratio (SNR) of [0, 10]~\!dB for the training set.
% delay spread https://www.ieee802.org/11/Documents/DocumentArchives/1997_docs/71252.pdf
% T-Prime set the delay spread as 1.5 ns

\begin{figure}[t]
    \centering
    \includegraphics[width=0.48\textwidth]{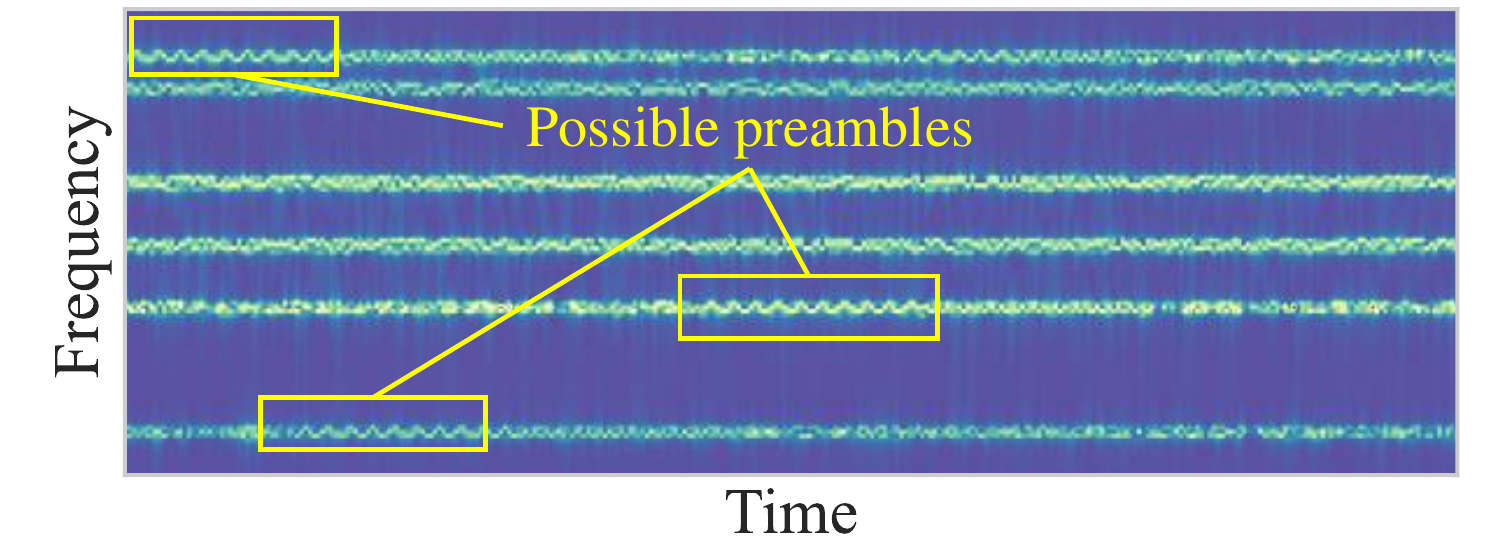}
    \caption{An example spectrogram of the multiband signals. Multi-coset sampling is then applied to get sub-Nyquist sampled dataset.}
    \label{fig:impl:spectrogram}
\end{figure}

We initially use MATLAB to generate Nyquist-rate frames of the synthetic signals at a sampling rate of 2~\!GSPS, \rev{with each frame fixed to 48~\!$\mu$s, which is generally enough to capture intact headers, enabling data-aligned parallel processing while avoiding excessive Transformer complexity due to long sequences}.
An example spectrogram of the signal is shown in Figure~\ref{fig:impl:spectrogram}. Then we apply multi-coset sampling to obtain sub-Nyquist samples, which serve as input to \name. Setting $B = 2$~\!GHz and $L = 40$ in Eq.~\eqref{equ:multicoset}, each low-speed ADC in multi-coset sampling operates at the rate of 50~\!MSPS, resulting in complex samples with a shape of ($P$, 2400), i.e., $N = 2400$. The corresponding measurement matrix $A$ and the measurement results $Y$ are constructed according to \cite{song2022approaching}. The synthetic dataset contains 300,000 samples for training, 5,000 samples for validation and 5,000 for testing.

\subsection{Over-the-Air Dataset} \label{sec:impl:real}

% \begin{figure}
%     \centering
%     \includegraphics[width=0.48\textwidth]{./figs/usrp.pdf}
%     \caption{The over-the-air dataset collection setup.}
%     \label{fig:impl:usrp}
% \end{figure}

\begin{figure}[t]
    \subfloat[DVB-S2 transmitters]{
        \centering
        \includegraphics[width=0.24\textwidth]{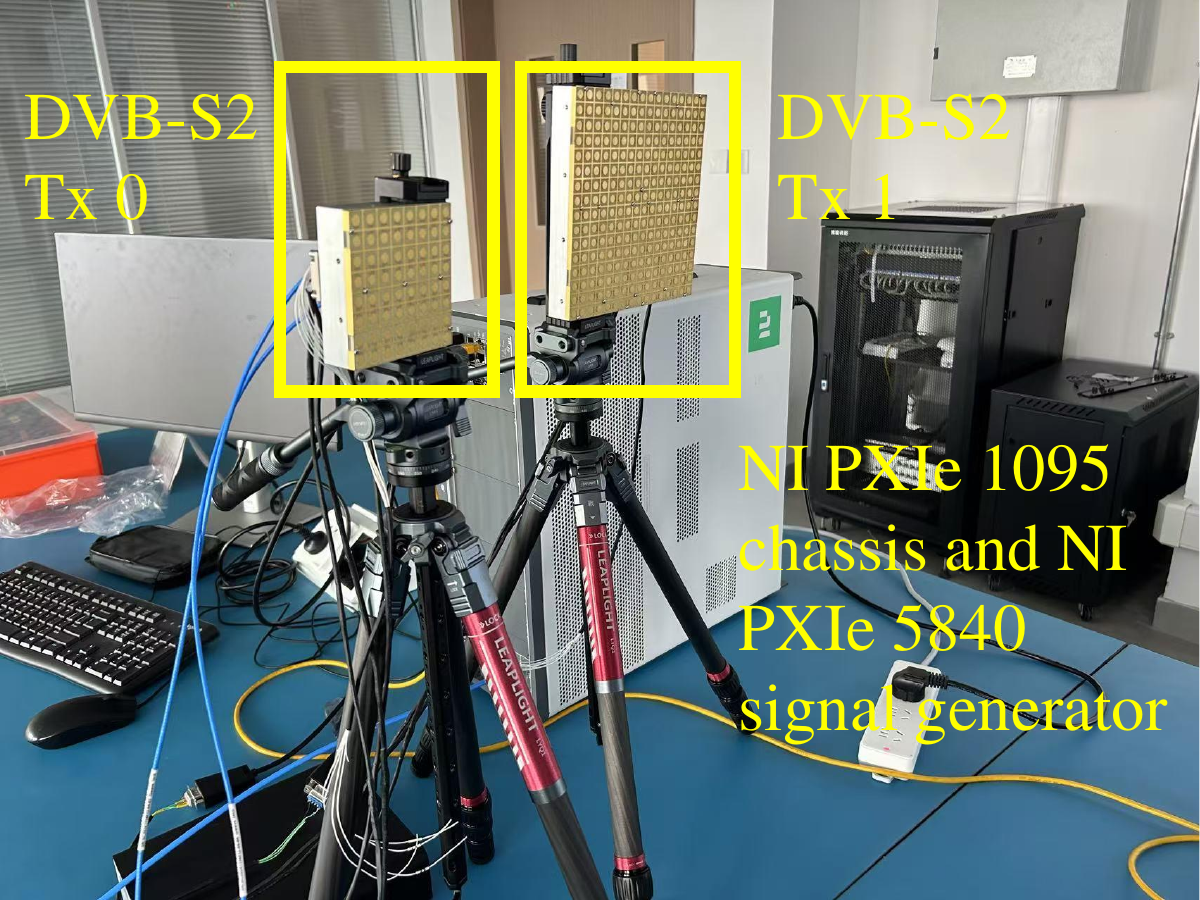}
        \label{fig:impl:usrp:dvb}
    }
    \subfloat[Wi-Fi transmitters]{
        \centering
        \includegraphics[width=0.24\textwidth]{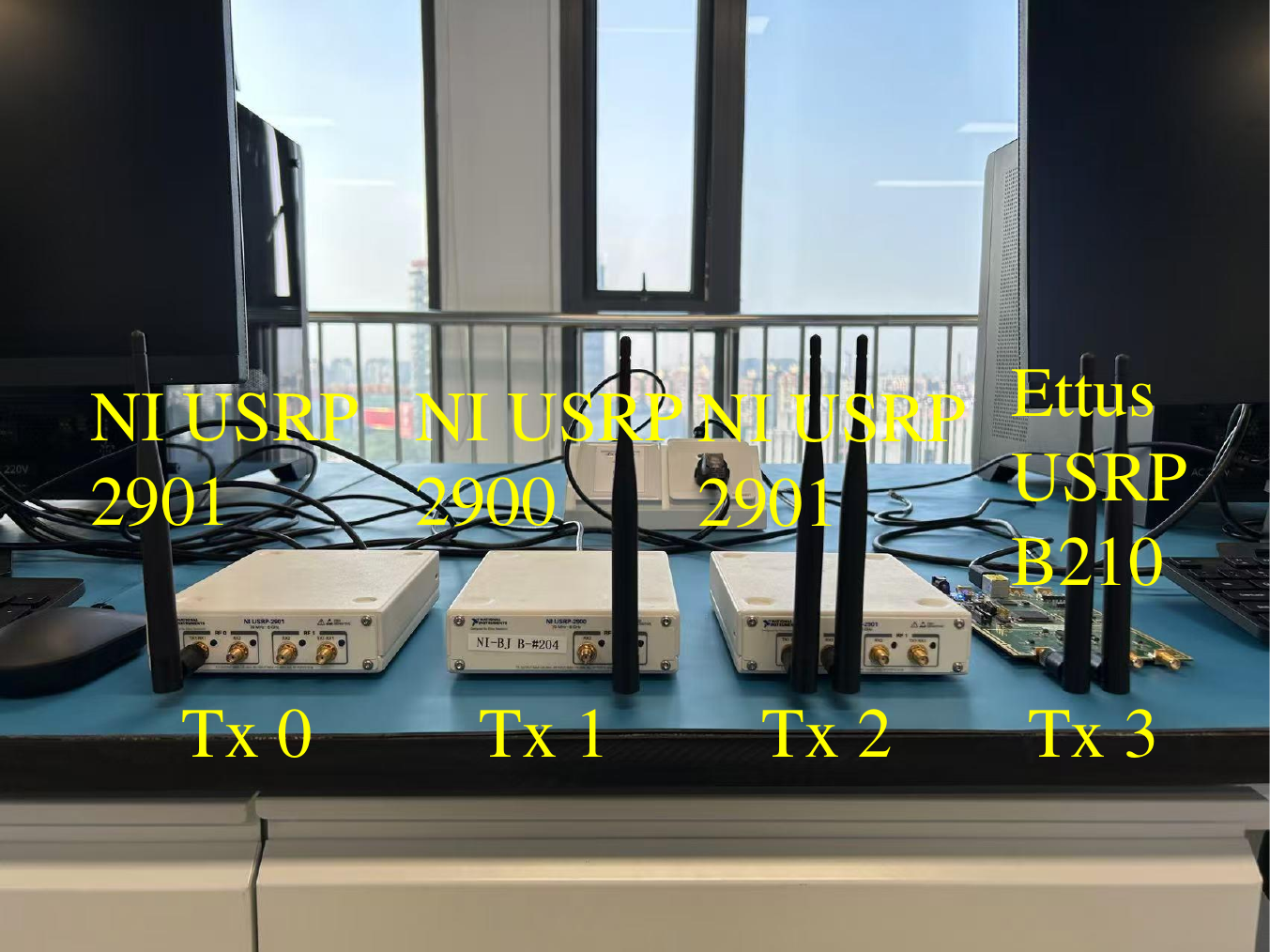}
        \label{fig:impl:usrp:wifi}
    } \\
    \subfloat[Receiver]{
        \centering
        \includegraphics[width=0.48\textwidth]{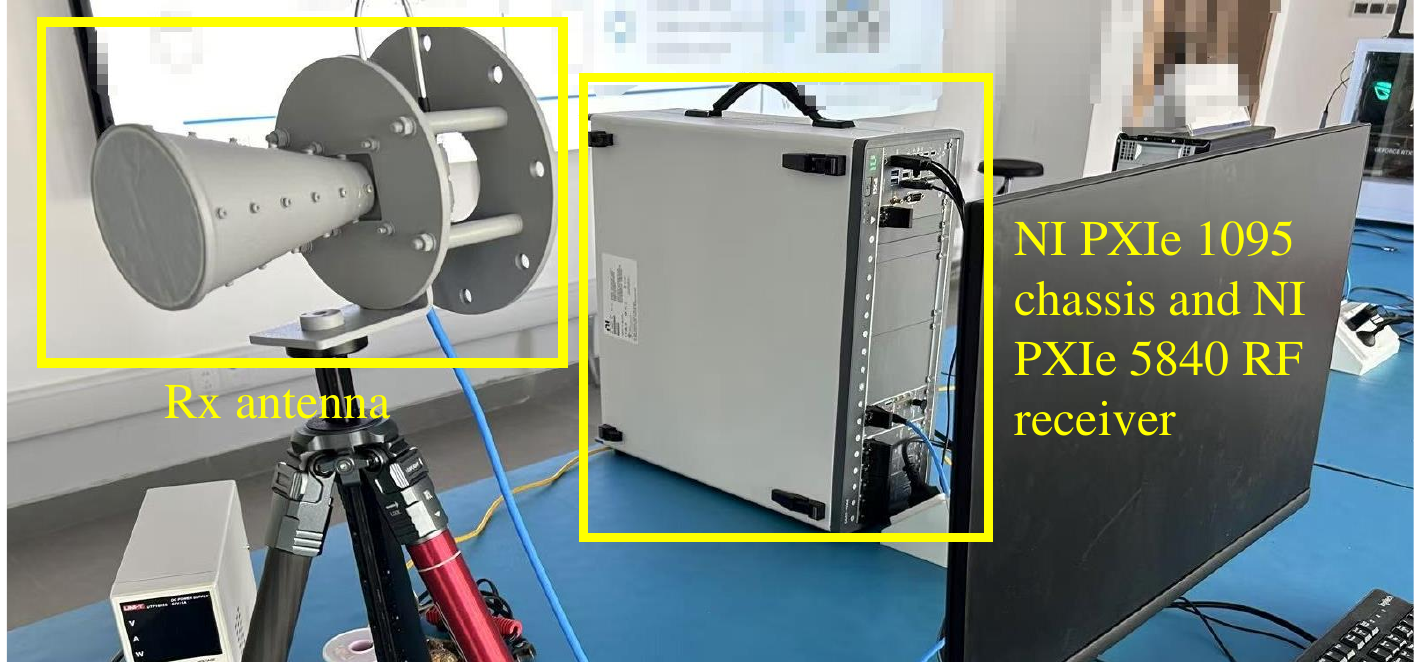}
        \label{fig:impl:usrp:receiver}
    } \\
    \caption{The over-the-air dataset collection devices.}
    \vspace{-1ex}
    \label{fig:impl:usrp}
\end{figure}

% 5 GHz ~ 5.8 GHz, receive with sampling rate of 40 MSPS. up-sampling to 2 GSPS
To validate the performance of \name beyond simulations, we collect an over-the-air dataset using Universal Software Radio Peripheral~(USRP) devices. The setup, as shown in Figure~\ref{fig:impl:usrp}, includes two National Instruments (NI) USRP 2901 devices, one NI USRP 2900, and an NI Ettus USRP B210 as Wi-Fi transmitters. For DVB-S2 signal generation, we use an NI PXIe-5840 signal generator, capable of multi-channel output and equipped with two individual transmitting antennas. The receiver is an NI PXIe-5840 configured to sample at 1~\!GSPS.

The transmitters generate 20~\!MHz-rate DVB-S2 signals, non-HT Wi-Fi and HT Wi-Fi signals, and transmits them over the air. During each transmission, each transmitter generates a batch of signals, selects a unique 50~\!MHz sub-band within 5.0-5.8~\!GHz, then transmits repeatedly to ensure successful capture at the receiver. Signals are collected over four days with varying transceiver positions to capture diverse channel conditions. In post-processing, we up-sample signals to 2~\!GSPS rate and locate header positions to label the signals. Additional AWGN of 10~\!dB SNR is added, and multi-coset sampling, consistent with that on the synthetic dataset, is then applied. The over-the-air dataset contains 1000 samples for fine-tuning and 2,000 samples for testing.

% 虽然说这样的做法并非是最真实的over the air的信号采集，但是逻辑上说，将不同时收集到的信号相加是等效的，符合多频带信号模型。

\subsection{Model Parameter Settings} \label{sec:impl:model}
We implement \name using Python 3.10.11 and PyTorch 2.0.1. The three kinds of signal analysis models are trained separately. All Transformer layers employ GELU activation. The spectrum sensor uses a folding factor of $F = 16$, $d_{model} = 128$, 4 attention heads, and a feed-forward dimension of 512. The ground truth spectrum occupancy states are represented in binary form (1 for occupied and 0 for unoccupied). We train the model with binary cross-entropy loss. The loss is summed across sub-bands and then averaged over the training batch. We use AdamW optimizer for all models, with $\beta_1 = 0.9$, $\beta_2 = 0.99$, L2 weight decay of 0.01. The initial learning rate of the spectrum sensor is set to $10^{-3}$, reduced by a factor of 0.1 when the training loss fails to decrease for 3 consecutive epochs, with a minimum learning rate of $10^{-6}$.

The protocol identification model uses $F = 32$, $d_{model} = 128$, 4 attention heads, and a feed-forward dimension of 384. The output is activated by Softmax, with cross-entropy loss incorporating a label smoothing rate of 0.1, averaged over the training batch. The initial learning rate is set at $2 \times 10^{-3}$, with linear warmup over the first 1,000 steps, followed by exponential decay at a rate of -0.5.

For header decoding, we set $F = 32$, $d_{model} = 384$, 8 attention heads, and a feed-forward dimension of 1,536, using 3 decoder layers for all DVB-S2, non-HT Wi-Fi, and HT Wi-Fi decoding models. The feature fusion weight $\alpha$ is initialized to 1.0. We decode all header fields in DVB-S2, all L-SIG fields for non-HT Wi-Fi, and both L-SIG and HT-SIG1 fields for HT Wi-Fi, corresponding to 7, 24, and 48 query tokens, respectively. The final output is activated by Softmax, with cross-entropy loss and a label smoothing rate of 0.1, averaged over all bits in a training batch. The initial learning rate is $6.25 \times 10^{-4}$, with linear warmup for the first 10,000 steps, followed by exponential decay at -0.5. All the learnable tokens, including the [CLS] token and the query tokens, are initialized with Gaussian distribution $\mathcal{N}(0, 1)$. The loss is not back-propagated to the protocol identification model, although features from encoders are fused here.

\section{Experiment Results} \label{sec:exp}

In this section, we evaluate the performance of \name in spectrum sensing, signal recovery, protocol identification, and header decoding tasks at sampling rates below the sub-Nyquist sampling limit. All the models are trained on the synthetic dataset where 6 of the 50~\!MHz sub-bands are occupied: the spectrum sensor is trained with a batch size of 512 for 100 epochs, the protocol identification model with a batch size of 128 for 300 epochs, and the header decoding models with a batch size of 128 for 600 epochs. \rev{We only fine-tune the 3-layer decoding models on 1,000 over-the-air training signals with a batch size of 128 for 5 epochs. To evaluate generalization, all the spectrum sensing and protocol identification models, as well as a 4-layer decoding model, are tested on the over-the-air dataset without any fine-tuning.} All training and evaluation are conducted on Ubuntu 22.04 using an NVIDIA RTX 4090 GPU with CUDA 11.8, paired with an Intel Platinum 8352V CPU. Parameters of models in \name and the baseline models used in the experiments are listed in Table~\ref{tab:n_params}.

% \begin{table}[b]
%     \vspace{-2ex}
%   \caption{Number of parameters for each model in \name and benchmarks. $5 \le P \le 11$ is the number of low-speed ADCs. The parameter difference between header decoders lies in query tokens.}
%   \label{tab:n_params}
%   \begin{center}
%   \footnotesize
%   \begin{tabular}{cc}
%     \toprule
%     Model&No. of Params\\
%     \midrule
%     Spectrum sensor &630 + 8$P$~\!K\\
%     WrT~\cite{zhang2024spectrum} &812 + 4$P$~\!K\\
%     DCS~\cite{wu2019deep} & 15.9~\!K\\
%     Protocol identifier&706~\!K\\
%     T-Prime~\cite{belgiovine2024tprime}&755~\!K\\
%     DVB-S2 header decoder&8.67~\!M\\
%     Non-HT header decoder&8.68~\!M\\
%     HT header decoder&8.69~\!M\\
%   \bottomrule
% \end{tabular}
% \end{center}
% \end{table}

\begin{table}[b]
    \vspace{-2ex}
    \caption{\rev{Model parameters, memory usage, and inference time for one batch (size = 1,024, representing data with duration 49~\!ms). The batch contains 1,340 DVB-S2, 1,416 Non-HT Wi-Fi, and 1,345 HT Wi-Fi headers. $5 \le P \le 11$ denotes the number of low-speed ADCs.
    }}
    \label{tab:n_params}
    \begin{center}
    \footnotesize
    \begin{tabular}{cccc}
        \toprule
        Model & No. of Params & GPU Mem (MB) & Inference Time \\
        \midrule
        Spectrum sensor & 630 + 8$P$~\!K & 1677$^\star$ & 17.0~\!ms$^\star$ \\
        WrT~\cite{zhang2024spectrum} & 812 + 4$P$~\!K & 2619$^\star$ & 27.9~\!ms$^\star$ \\
        \name signal recovery & N/A & 1679$^{\star}$ & 29.1~\!ms$^{\star}$ \\
        DTMP~\cite{song2022approaching} & N/A & 539$^\star$ & 4.36~\!s$^\star$ \\
        DCS~\cite{wu2019deep} & 15.9~\!K & 391$^{\star\dagger}$ & 204~\!s$^{\star\dagger}$ \\
        Protocol identifier & 706~\!K & 3631 & 45.3~\!ms \\
        T-Prime~\cite{belgiovine2024tprime} & 755~\!K & 5135 & 81.5~\!ms \\
        DVB-S2 header decoder & 8.67~\!M & 2607 & 26.8~\!ms \\
        Non-HT header decoder & 8.68~\!M & 3347 & 39.5~\!ms \\
        HT header decoder & 8.69~\!M & 4615 & 52.7~\!ms \\
        \bottomrule
    \end{tabular}
    \end{center}
    \begin{flushleft}
    \footnotesize $^\star$Measured with $P = 8$. \\
    \footnotesize $^\dagger$Measured with batch size = 1.
    \end{flushleft}
    % \vspace{-6ex}
\end{table}

% 基本实验，通常压缩感知场景下的嗅探
% 频谱感知：传统压缩感知算法（PCSBL-GAMP），深度学习算法（WrT），恢复的MSE
% 协议分类：先压缩感知恢复，再对比T-PRIME，MCformer神经网络
% 协议解析：（和Swirls场景不一样。。。）
% 消融实验: embedding，feature fusion
\subsection{Spectrum Sensing and Signal Recovery} \label{sec:exp:ss}

\begin{figure}[t]
    \subfloat[False alarm]{
        \centering
        \includegraphics[width=0.24\textwidth]{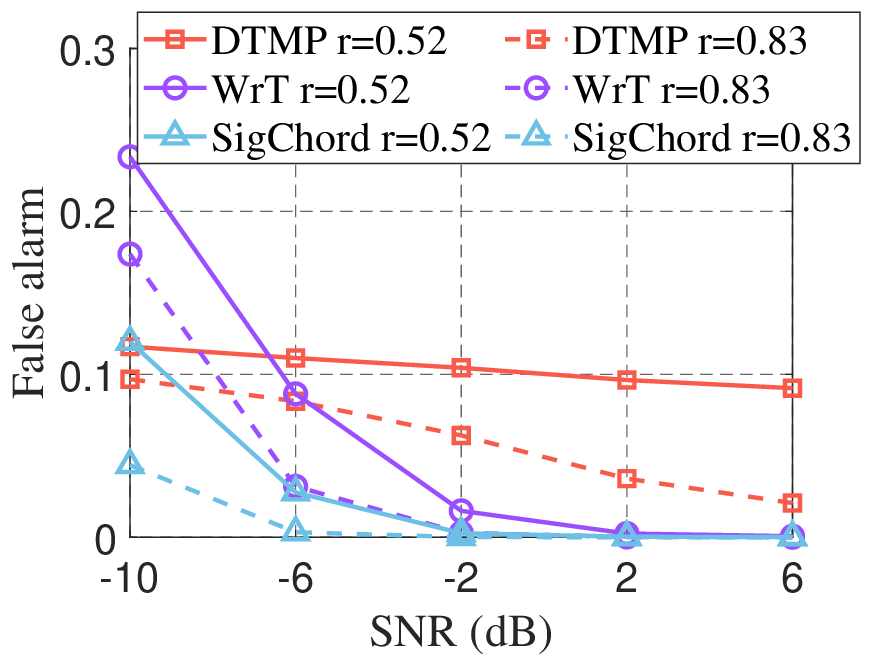}
        \label{fig:exp:ss:fa}
    }
    \subfloat[Miss detection]{
        \centering
        \includegraphics[width=0.24\textwidth]{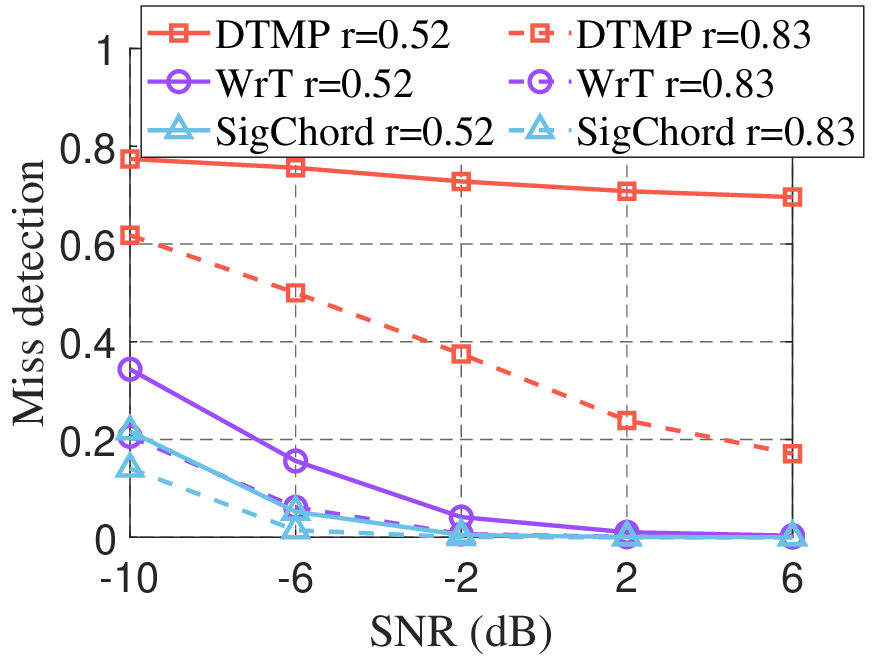}
        \label{fig:exp:ss:md}
    }
    \caption{Spectrum sensing performance on the synthetic dataset under different SNRs. The threshold $\theta$ for \name and WrT~\cite{zhang2024spectrum} is set to 0.5.}
    \label{fig:exp:ss}
\end{figure}

In this section, we evaluate the performance of \name in spectrum sensing and signal recovery. We compare \name with a Transformer-based SOTA spectrum sensing model WrT~\cite{zhang2024spectrum}, the CS recovery algorithm DTMP~\cite{song2022approaching}, and a deep compressed sensing~(DCS) recovery algorithm~\cite{wu2019deep}. Firstly, we show the spectrum sensing performance of \name, WrT and DTMP. Since WrT is designed for Nyquist-rate spectrum sensing only, we adjust its spectra size and patch size to accommodate sub-Nyquist samples. Specifically, we reshape and unfold a ($P$, 2400) complex sample to (2400, $2P$) real sample, and set the spectra size and patch size to (2400, $2P$) and (16, $2P$), respectively, corresponding to the folding factor $F = 16$ in \name. We implement WrT with 3 Transformer encoder layers, $d_{model} = 128$, 4 attention heads and feeding forward dimension of 512. The number of parameters of WrT is similar to that of \name. WrT is trained on the synthetic dataset with batch size of 512 for 100 epochs as well. For the CS recovery algorithm DTMP, we recover $X$ with the number of narrowband signals as the prior information. Apart from the synthetic dataset and over-the-air dataset, we generate a Gaussian random signal dataset to validate the generalization of \name. Unless otherwise specified, each random signal contains 6 narrowband signals, each with a 50~\!MHz bandwidth and frequency components generated randomly following a Gaussian distribution.

% $r=0.52$ 和 $r=0.83$ 对应了低速采样的ADC数目 $P = 5$ and $P = 8$。可以看到，在SNR大于-2 dB的情况下，传统的压缩感知算法DTMP不能准确地在$r<1$时，即低于blind sub-Nyquist sampling limit的采样率下检测到信号，特别是在$r=0.52$接近Landau rate时，DTMP的性能急剧恶化。而基于Transformer深度学习网络的\name和WrT均能在$r < 1$时准确做到频谱感知。\name相较于WrT在低SNR下有明显更好的性能，得益于用于处理欠采样输入的embedding process。
Results on the synthetic dataset are shown in Figure~\ref{fig:exp:ss}. Let $r$ represent the ratio of the total sampling rate to twice the Landau rate, with $r = 0.5$ corresponding to the Landau rate, and $r = 1$ marking the sub-Nyquist sampling limit.
The cases $r = 0.52$ and $r = 0.83$ correspond to $P = 5$ and $P = 8$ low-rate ADCs, respectively.
The conventional compressed sensing algorithm, DTMP, struggles to reliably detect signals when $r < 1$, particularly as $r = 0.52$ nears the Landau rate, where its performance deteriorates sharply.
In contrast, both Transformer-based deep learning models, \name and WrT, accurately perform spectrum sensing at $r < 1$. Importantly, despite WrT having more parameters in our experiments, \name outperforms WrT at low SNRs, leveraging its embedding process, which is specifically tailored for sub-Nyquist sampling inputs.

\begin{figure}[t]
    \subfloat[Over-the-air signals]{
        \centering
        \includegraphics[width=0.24\textwidth]{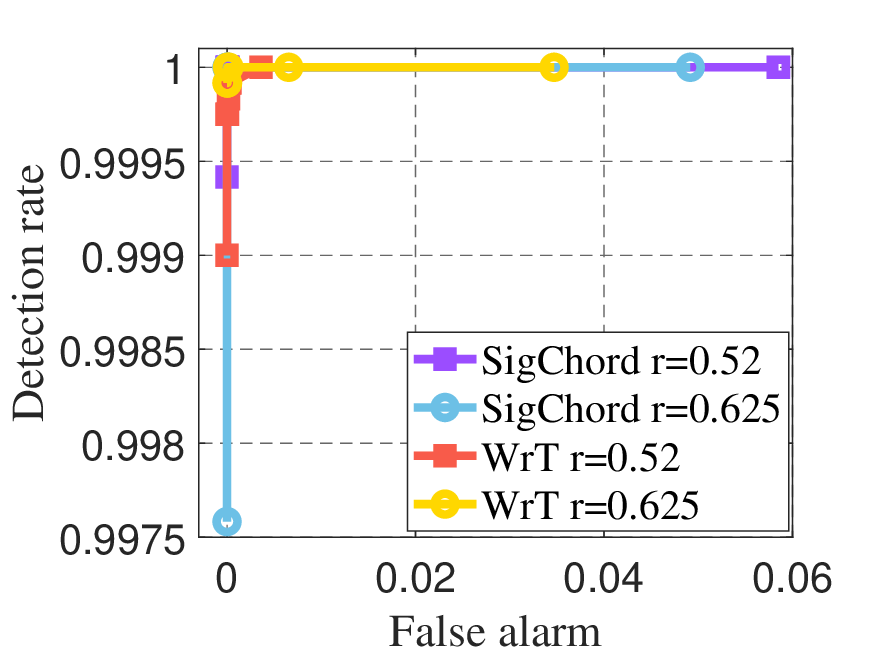}
        \label{fig:exp:roc:ota}
    }
    \subfloat[Random signals]{
        \centering
        \includegraphics[width=0.24\textwidth]{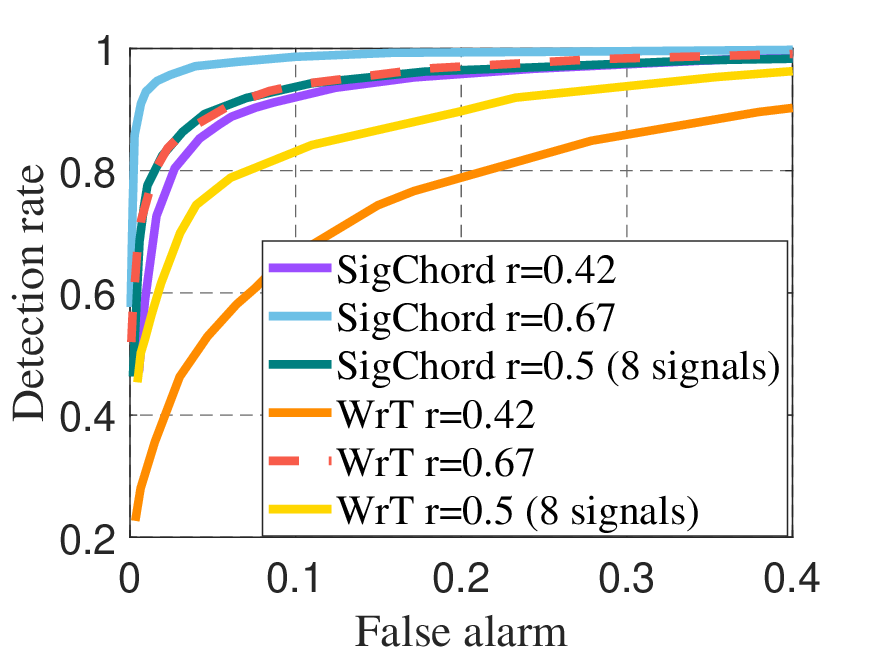}
        \label{fig:exp:roc:random}
    } \\
    \caption{ROC curves of spectrum sensing from unseen environments. The over-the-air signals have SNR of 10~\!dB, and the random signals have SNR of -6~\!dB.}
    \label{fig:exp:roc}
\end{figure}

% 调整阈值$\theta$，图~\ref{}展示了\name和WrT在未见过的数据集下的ROC曲线。其中Random signals数据集是按照高斯分布生成的完全随机的多频带信号，其包含6个随机分布于各子频带的窄带信号，每个窄带信号占宽50MHz，频域按照高斯分布完全随机生成。
% \name和WrT均能准确迁移到over-the-air数据集，虽然$r < 1$，在良好的SNR条件下几乎没有错误。在数据分布完全不同的random signals中，\name取得了明显由于WrT的更好的性能。即使在包含8个信号的拥挤频谱中，\name也能准确地检测到信号。甚至当采样率低于Landau rate时($r=0.42$)，\name也有可观的性能。
By adjusting the threshold $\theta$, Figure~\ref{fig:exp:roc} shows the ROC curves of \name and WrT at sampling rates below the sub-Nyquist limit, evaluated on unseen over-the-air signals and random Gaussian signals. \name and WrT generalize well to over-the-air datasets, showing minimal errors under favorable SNR that could achieve perfect detection and zero false alarms. For the random signal dataset, where data distributions differ significantly from the training data, \name consistently outperforms WrT. Its ROC curves are closer to the top-left corner, reflecting superior accuracy.
Notably, \name achieves comparable performance at $r = 0.5$ with dense spectra containing 8 signals, while WrT requires $r = 0.67$ to handle only 6 signals.
Furthermore, even at an extremely low sampling rate below the Landau rate ($r = 0.42$), \name demonstrates robust performance, maintaining both false alarm and miss detection rates below 0.1.

\begin{figure}[t]
    \subfloat[Over-the-air signals]{
        \centering
        \includegraphics[width=0.24\textwidth]{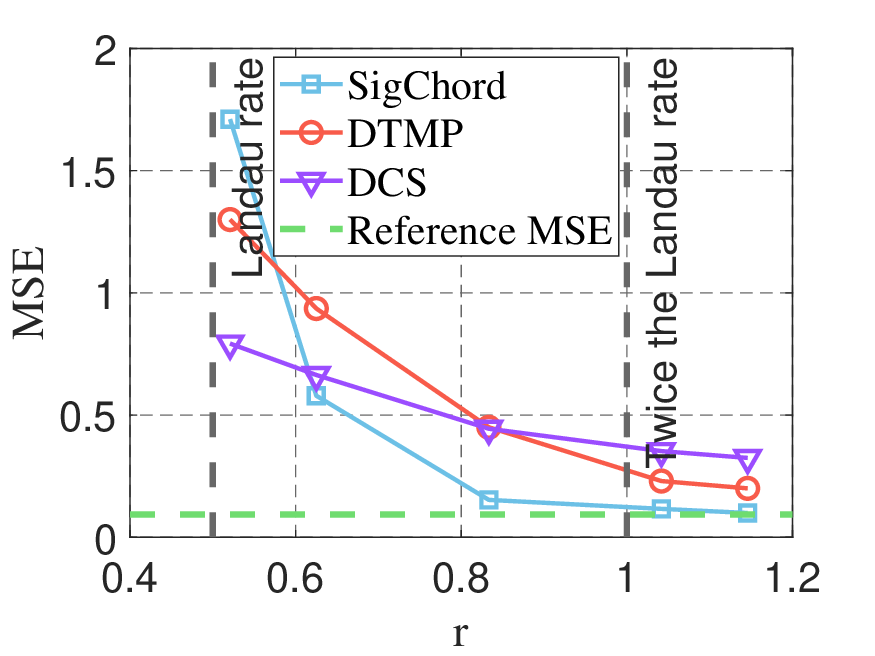}
        \label{fig:exp:rec:phy}
    }
    \subfloat[Random signals]{
        \centering
        \includegraphics[width=0.24\textwidth]{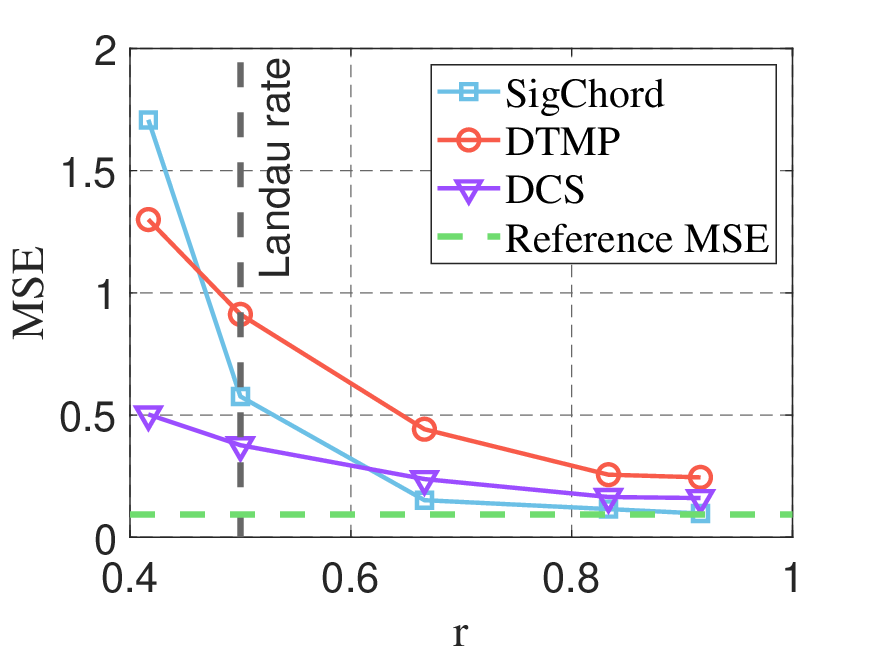}
        \label{fig:exp:rec:rand}
    }
    \caption{Recovery performance on signals from unseen environments at different sampling rates. The signals have SNR of 10~\!dB.}
    \label{fig:exp:rec}
\end{figure}

\begin{figure}[t]
    \centering
    \subfloat{
        \includegraphics[width=0.48\textwidth]{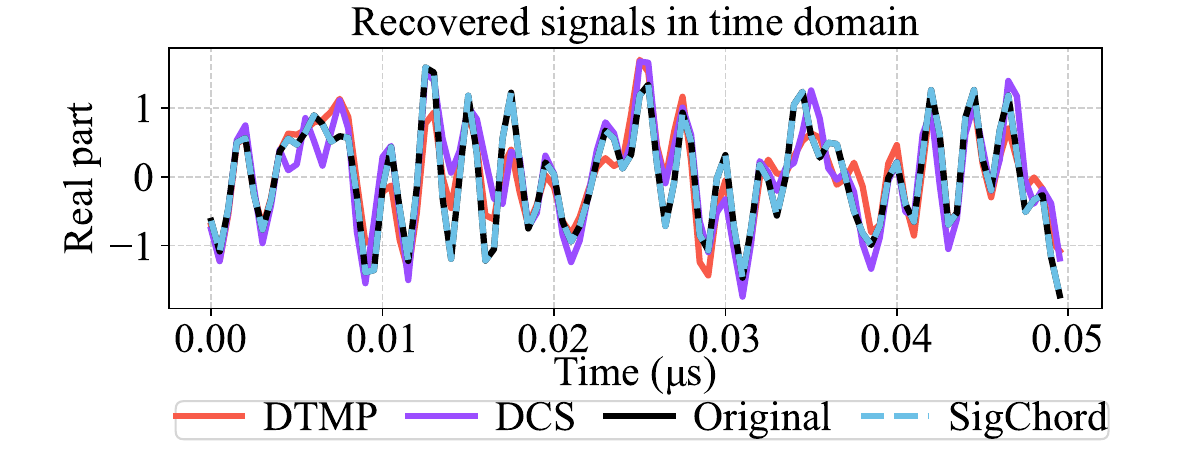}
    } \\
    \vspace{-2ex}
    \subfloat{
        \centering
        \includegraphics[width=0.11\textwidth]{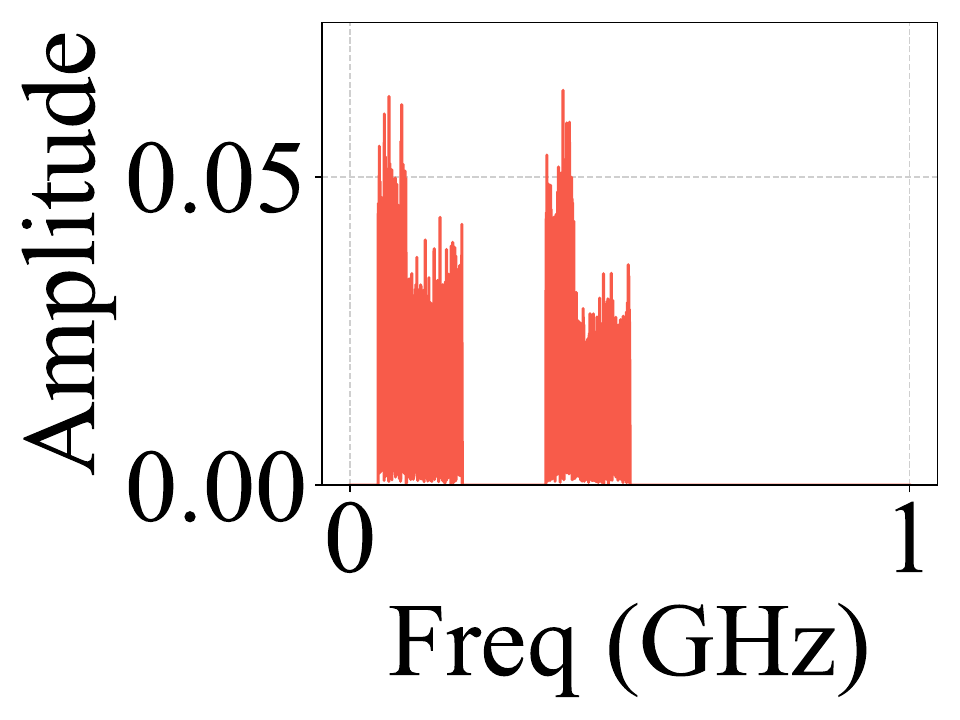}
    }
    \subfloat{
        \centering
        \includegraphics[width=0.11\textwidth]{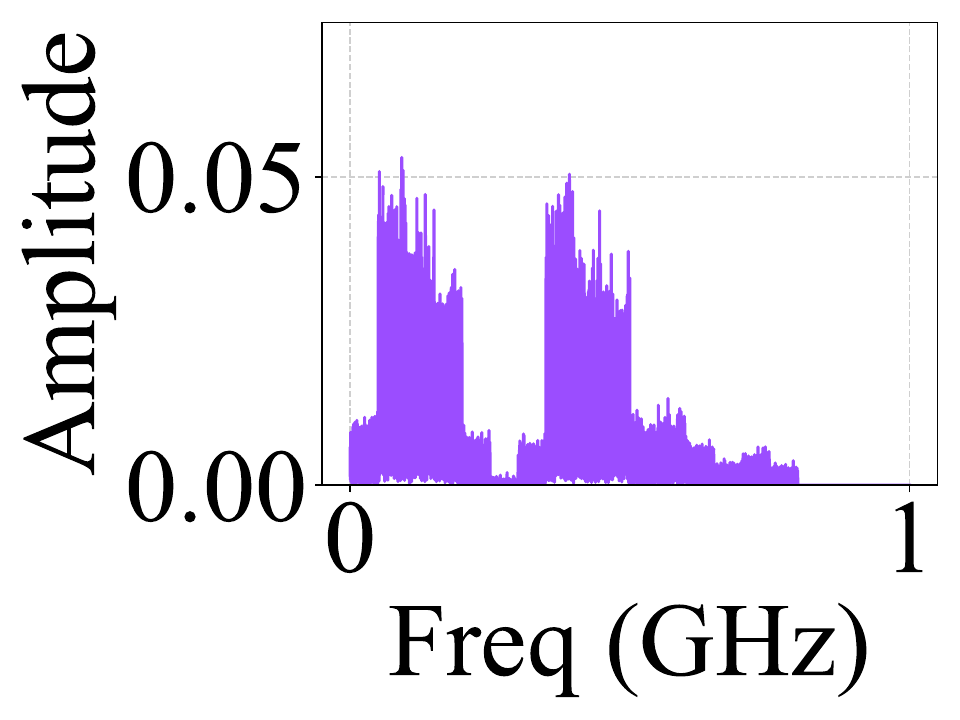}
    }
    \subfloat{
        \centering
        \includegraphics[width=0.11\textwidth]{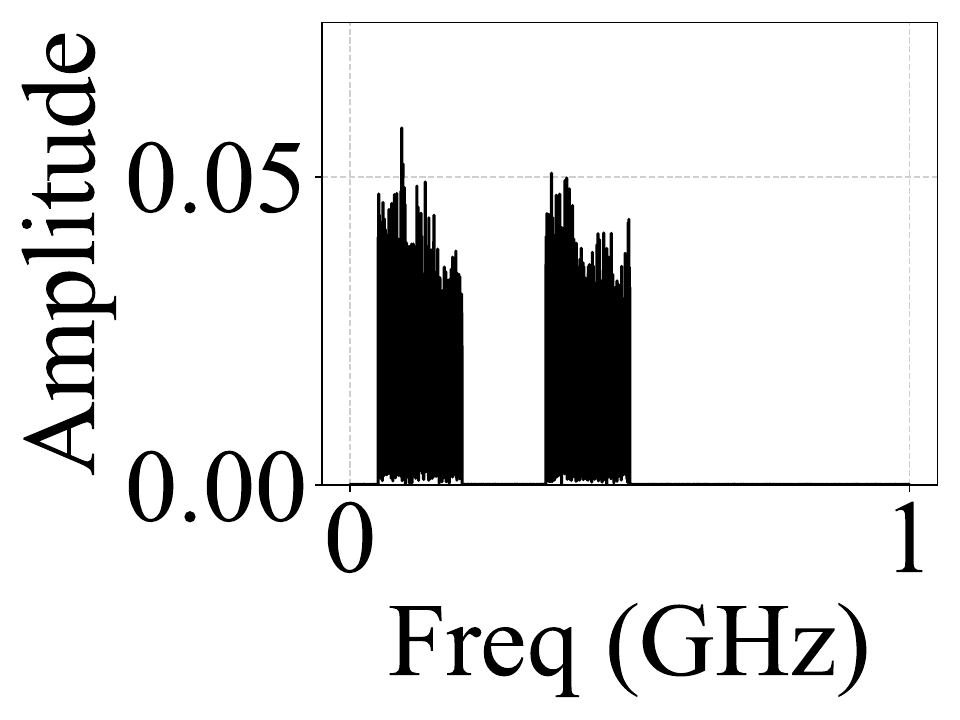}
    }
    \subfloat{
        \centering
        \includegraphics[width=0.11\textwidth]{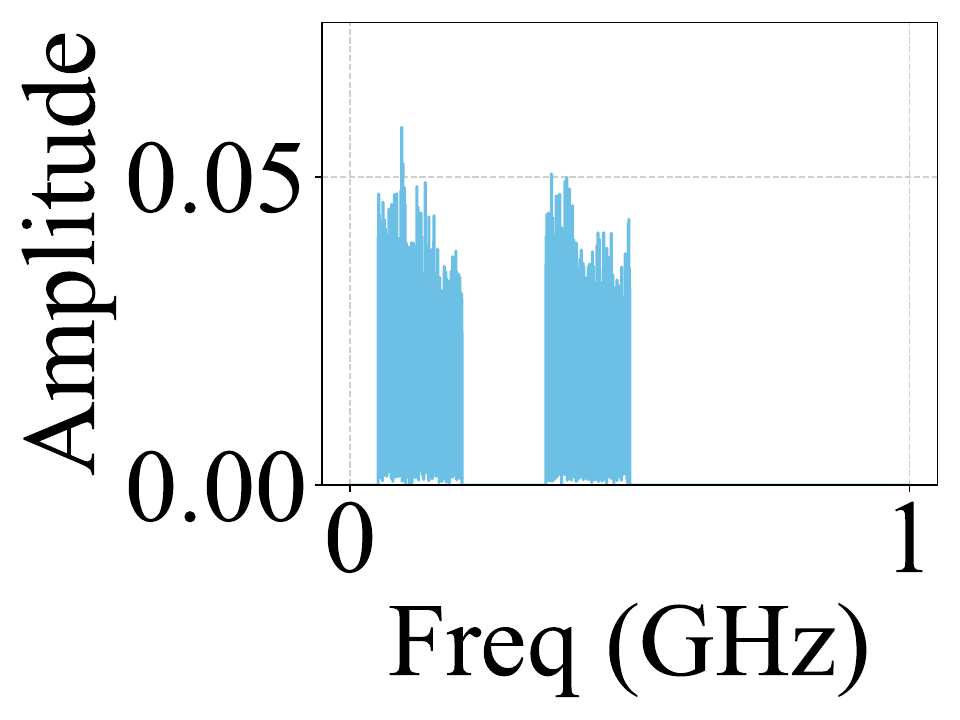}
    }

    \vspace{-2ex}
    \caption{Examples of noise-free random Gaussian signal recovery in time and frequency domains by \name, DTMP and DCS at $r=0.67$.}
    \label{fig:exp:rec:example}
    \vspace{-3ex}
\end{figure}

% 之后我们展示算法1的信号恢复性能。可以看到，基于传统CS的DTMP在$r < 1$时性能明显下降，并且即使是在$r > 1$时也与reference MSE存在差距。\name中频谱感知的泛化性能使得算法1也能在未曾见过的数据上展现出优秀的泛化性。在over-the-air信号上在$r > 0.83$时，在random Gaussian signals上在$r > 0.67$时恢复的MSE就已经贴近了reference MSE，突破了blind sub-Nyquist sampling limit实现了非稀疏信号的高质量恢复。当$r$接近0.5，\name的性能也出现了恶化，这源于算法1中$A_{S}$的列满秩性质的不再成立，也符合Landau rate的理论。
Next, We evaluate the signal recovery mean square error (MSE) of Algorithm~\ref{alg:recovery}, DTMP, and DCS across various sampling rates. DCS, which requires seconds to converge \rev{even for a single input}, is iterated for 20,000 steps. The reference MSE is calculated by comparing the noisy Nyquist-rate signal against its noise-free counterpart.
% The $r$ values for random Gaussian signals are lower due to each narrowband signal occupying more bandwidth, yet the number of low-speed ADCs is identical for corresponding points in Figure~\ref{fig:exp:rec:phy} and Figure~\ref{fig:exp:rec:rand}.
\name's robust spectrum sensing ensures Algorithm~\ref{alg:recovery} generalizes effectively to unseen signals. For over-the-air signals with $r > 0.83$ and random Gaussian signals with $r > 0.67$, the recovery MSE aligns closely with the optimal reference, enabling high-quality non-sparse signal recovery below the sub-Nyquist limit. In contrast, DTMP suffers significant degradation when $r < 1$ and retains a notable gap from the reference even at $r > 1$. While DCS performs adequately on random Gaussian signals, it struggles with over-the-air signals. For noise-free random Gaussian signals shown in Figure~\ref{fig:exp:rec:example}, DCS fails in faithful spectrum recovery, and only \name achieves perfect recovery due to its rule-based Transformer design, as shown in Figure~\ref{fig:exp:rec:example}. At $r \approx 0.5$, \name’s performance declines due to the loss of the full column rank of $A_{S}$ in Algorithm~\ref{alg:recovery}, consistent with the Landau rate. \rev{The drop occurs earlier for over-the-air signals since they do not fully occupy each sub-band, causing the full-rank transition ($P = \lvert S \rvert$) to occur at an $r$ slightly above 0.5.}

% 在模型推理阶段，我们设置batch size为1,024，$P = 8$时, \name的频谱感知模块处理一个batch所需要的时间为14.62 +- 0.12 ms，平均每条数据的处理时间为14.35 us. 而算法1恢复这些信号所需要的时间为26.60 +- 0.24 ms，平均恢复时间为25.98 us. 对于窄带信号数目为8的random Gaussian signals，算法1恢复一个batch所需要的时间为29.48 +- 0.34 ms, 平均恢复时间为28.79 us。注意到每条信号的长度为48 us，因此\name的频谱感知和信号恢复模块均能在实时性要求下用消费级显卡完成任务。
During the inference phase, we set the batch size to 1,024. For $P = 8$, the spectrum sensing module processes one batch in \rev{17.0 $\pm$ 0.33~\!ms}, corresponding to an average per-sample processing time of \rev{16.6 $\mu$s}. Meanwhile, the remaining part of Algorithm~\ref{alg:recovery} takes \rev{29.1 $\pm$ 4.0 ms} to recover these signals, resulting in an average recovery time of \rev{28.42 $\mu$s}. For random Gaussian signals containing eight narrowband signals, the recovery phase requires \rev{31.7 $\pm$ 4.0 ms} per batch, averaging \rev{30.96 $\mu$s} per sample. \rev{The recovery speed is significantly faster than DTMP and DCS, which require seconds to recover signals.} Given that the duration of each multiband signal is 48 $\mu$s, the spectrum sensing and recovery modules in \name can achieve real-time performance using a consumer-level GPU.

\subsection{Protocol Identification} \label{sec:exp:pkt_cls}

% T-Prime为encoder-only的信号分类模型，与\name不同，其不对输入信号做特别的embedding process，仅做reshape。T-Prime最后将encoder输出的所有token展平后输入到全连接层做分类。在实验中，我们实现的T-Prime包含3层encoder layer，$d_{model}$为64，每层包含8个attention heads，feeding forward dimension设置为1,024，sequence length设置为75与folding factor $F=32$的\name相同。
This section evaluates the protocol identification performance of \name, comparing it with the Transformer-based SOTA model T-Prime~\cite{belgiovine2024tprime}. For both \name and T-Prime, signals are recovered by Algorithm~\ref{alg:recovery} assuming accurate spectrum sensing. T-Prime is an encoder-only model designed for signal classification. Unlike \name, it does not apply a specialized embedding process to the input signal, using only a simple reshaping operation instead. T-Prime flattens all output tokens from the encoders and feeds them into a fully connected layer for classification. We implement T-Prime with 3 encoder layers, $d_{model} = 64$, 8 attention heads, and a feedforward dimension of 1,024. The sequence length is thus 75, same as \name with folding factor $F=32$. We train T-Prime on the synthetic dataset with a batch size of 128 for 300 epochs.

\begin{figure}[t]
    \centering
    \includegraphics[width=0.4\textwidth]{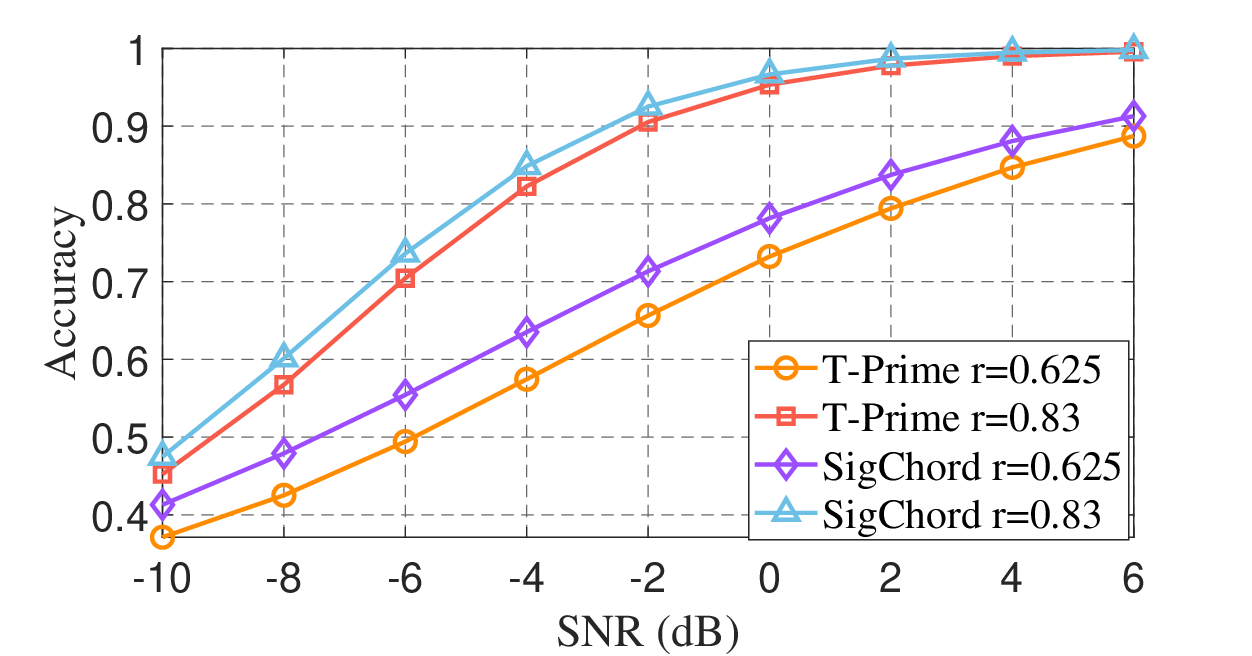}
    \caption{The protocol identification accuracy of \name compared with T-Prime~\cite{belgiovine2024tprime} on synthetic dataset at varying SNRs.}
    \label{fig:exp:pkt_cls}
\end{figure}

% 我们首先在synthetic dataset上对比T-Prime和\name在不同SNR下的分类性能，如图~\ref{}所示。可以看到，虽然T-Prime的参数量更多，但\name在所有SNR下，不论是在$r = 0.625$还是$r = 0.83$的低采样率下都有更好的性能。在极低采样率$r = 0.625$下，\name相对于T-Prime有大约2 dB的增益，SNR为6 dB时\name的准确率能够突破90\%。
First, we compare the classification performance on the synthetic dataset under varying SNR conditions, as shown in Figure~\ref{fig:exp:pkt_cls}. \rev{Here and in Section.~\ref{sec:exp:decode}, we choose $r = 0.625$ and $r = 0.83$, representing situations with mild recovery errors and near-Nyquist optimal recovery as shown in Figure~\ref{fig:exp:rec}, respectively.}
Although T-Prime has more parameters, \name consistently achieves superior performance across all SNR levels. At an extremely low sampling rate of $r = 0.625$, both models show performance degradation due to increased recovery errors as shown in Figure~\ref{fig:exp:rec:phy}, but \name demonstrates approximately a 2~\!dB performance gain over T-Prime, achieving over 90\% accuracy at an SNR of 6~\!dB.

\begin{figure}[t]
    \subfloat[T-Prime~\cite{belgiovine2024tprime} $r=0.625$]{
        \centering
        \includegraphics[width=0.24\textwidth]{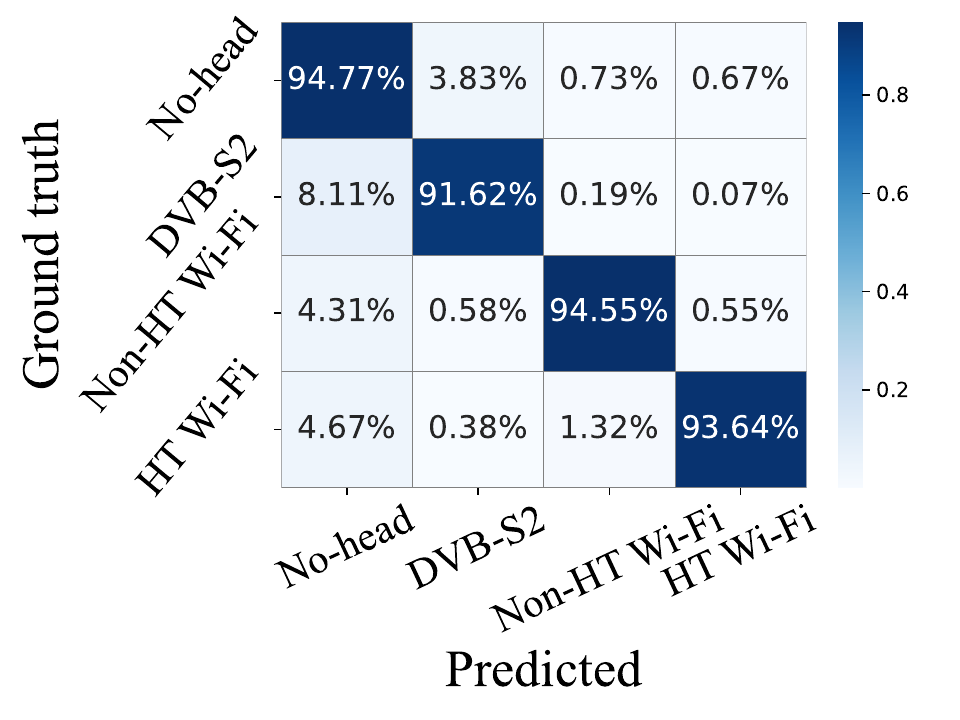}
    }
    \subfloat[T-Prime~\cite{belgiovine2024tprime} $r=0.83$]{
        \centering
        \includegraphics[width=0.24\textwidth]{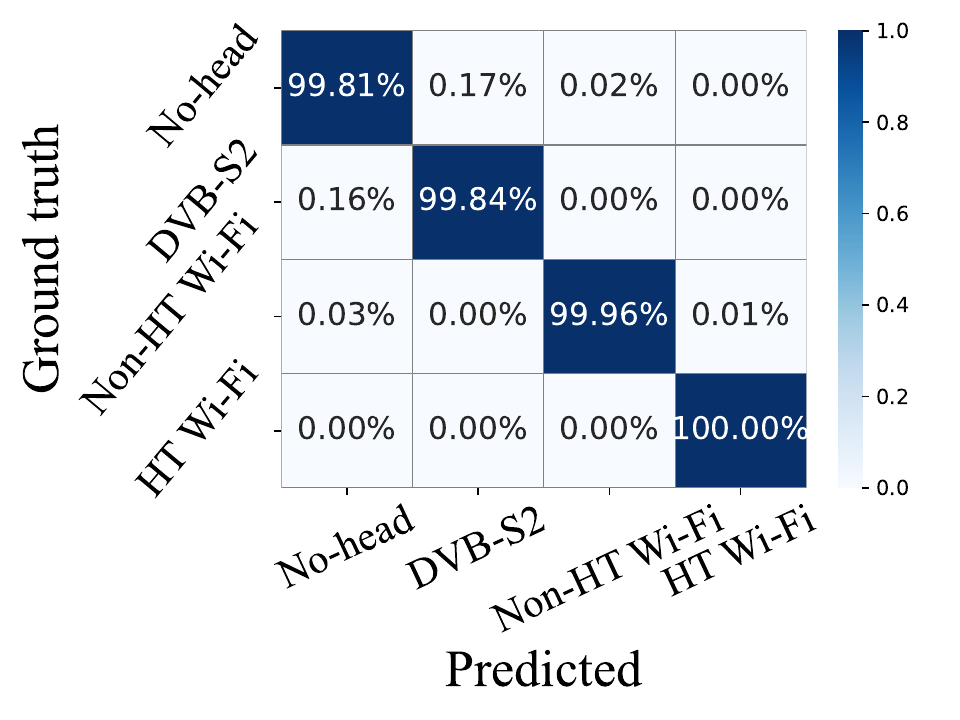}
    } \\
    \subfloat[\name $r=0.625$]{
        \centering
        \includegraphics[width=0.24\textwidth]{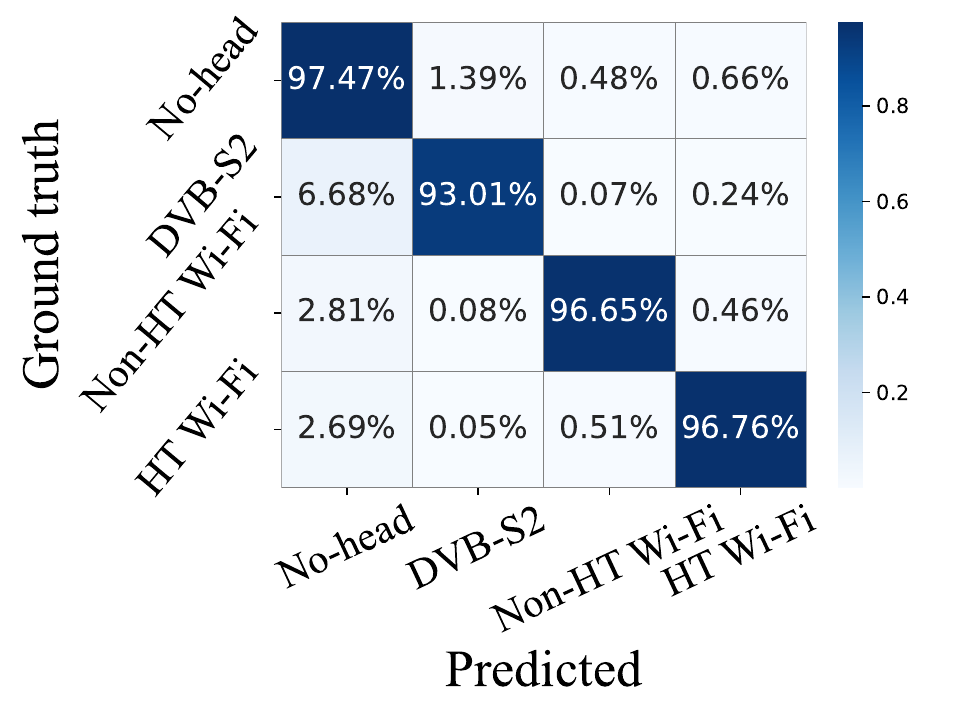}
    }
    \subfloat[\name $r=0.83$]{
        \centering
        \includegraphics[width=0.24\textwidth]{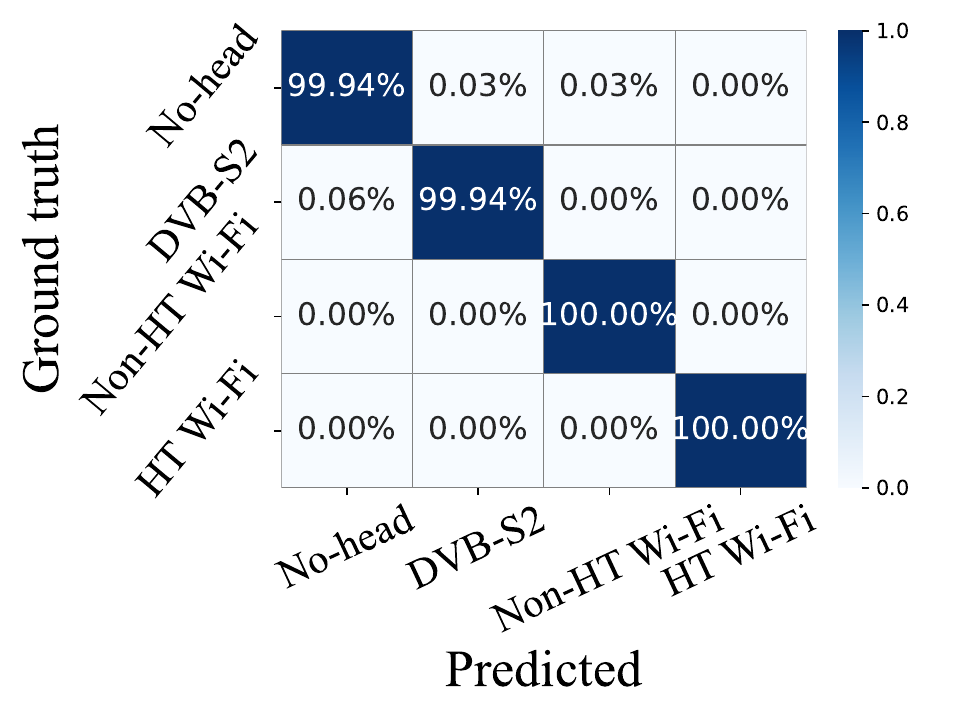}
    }
    \caption{The confusion matrix of protocol identification on synthetic dataset at SNR 10~\!dB.}
    \label{fig:exp:pkt_cls:conf_m}
\end{figure}

\begin{figure}[t]
    \subfloat[T-Prime~\cite{belgiovine2024tprime} $r=0.625$]{
        \centering
        \includegraphics[width=0.24\textwidth]{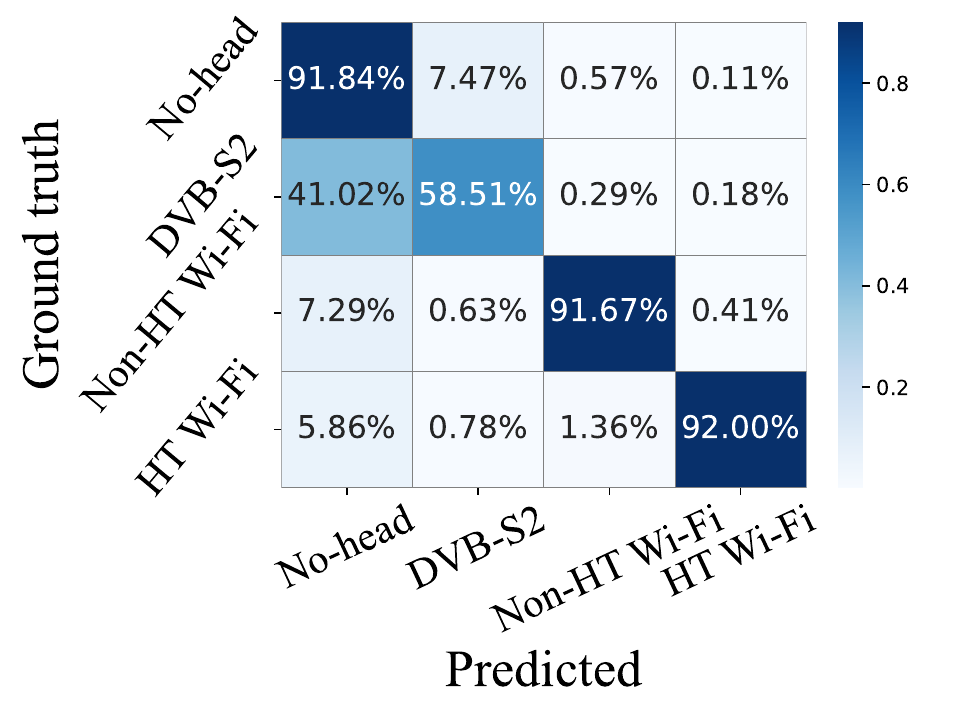}
    }
    \subfloat[T-Prime~\cite{belgiovine2024tprime} $r=0.83$]{
        \centering
        \includegraphics[width=0.24\textwidth]{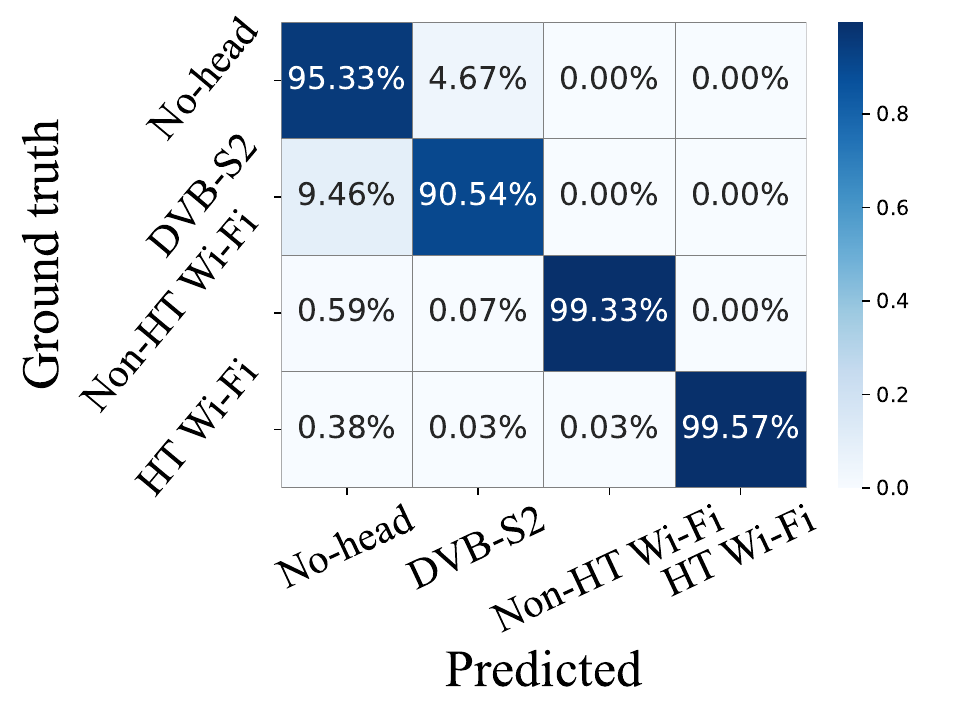}
    } \\
    \subfloat[\name $r=0.625$]{
        \centering
        \includegraphics[width=0.24\textwidth]{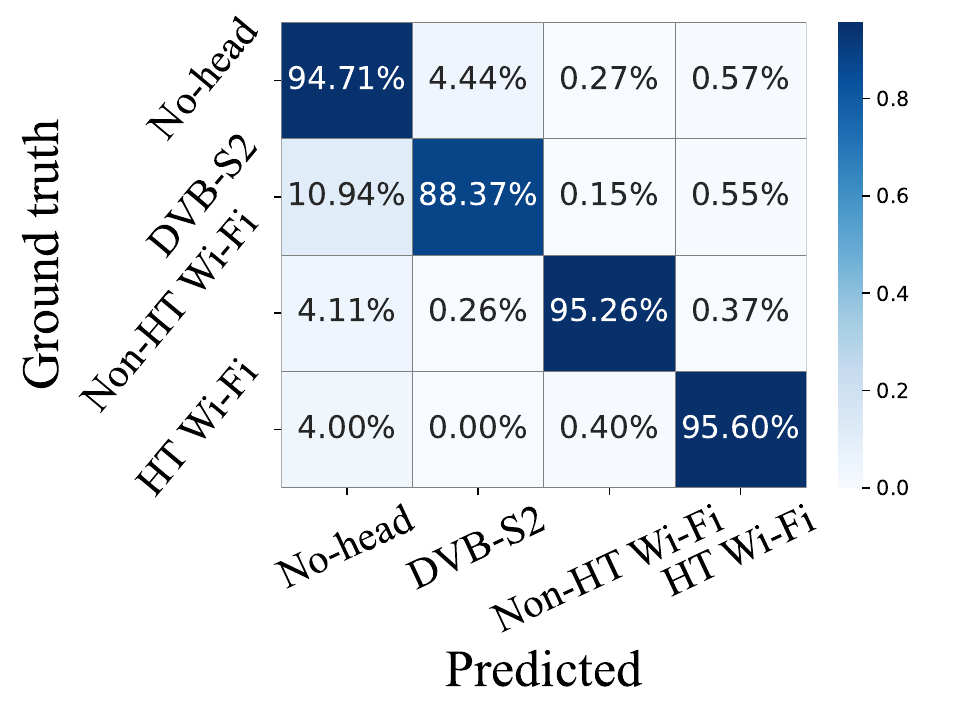}
    }
    \subfloat[\name $r=0.83$]{
        \centering
        \includegraphics[width=0.24\textwidth]{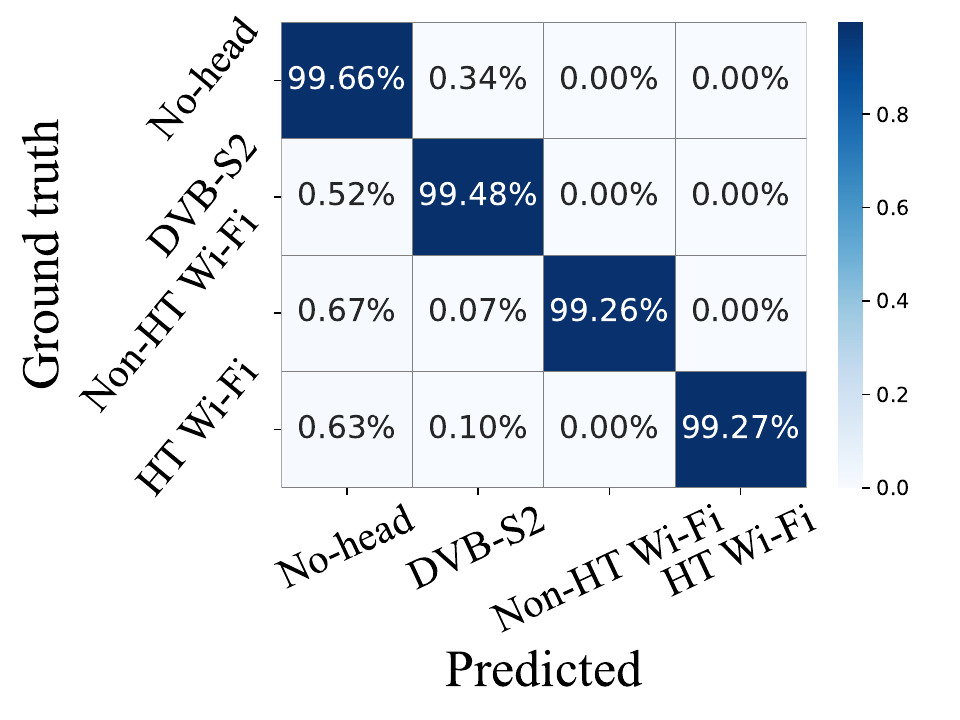}
    }
    \caption{The confusion matrix of protocol identification on the unseen over-the-air signals.}
    \label{fig:exp:pkt_cls:conf_m_ota}
\end{figure}

Figure~\ref{fig:exp:pkt_cls:conf_m} shows the classification confusion matrices on the synthetic dataset at an SNR of 10~\!dB. At $r = 0.625$, both models effectively distinguish Wi-Fi headers from DVB-S2. They show slightly lower accuracy in differentiating more similar non-HT and HT Wi-Fi formats. \name outperforms T-Prime by 1.4\% for DVB-S2, 2.1\% for non-HT Wi-Fi, 3.1\% for HT Wi-Fi, and 2.7\% in recognizing \textit{no-head} frames.
\rev{Both models struggle to distinguish DVB-S2 from payloads (no-head) due to its short preamble in the header, which provides limited distinctive features.} The entire header spans only 4.5~\!$\mu$s, with just 1.3~\!$\mu$s for the SOF preamble. In contrast, Wi-Fi headers (L-STF, L-LTF, HT-STF, HT-LTF) are at least 18 $\mu$s long \rev{thus have higher accuracies.} At $r = 0.83$, both models perform accurately, with \name exceeding 99.9\% accuracy in all categories.

% 图~\ref{}展示了T-Prime和\name在over-the-air数据集上的分类混淆矩阵。两个模型并没有在训练时见过over-the-air数据。可以看到T-Prime出现了泛化性不足的问题，其对DVB-S2头的识别准确率下降明显，在$r=0.83$信号恢复误差相对较小时识别准确率下降到90\%左右，特别是在$r = 0.625$信号恢复误差较大的极低采样率情况下，其识别准确率仅有58\%左右。而\name则在各采样率下展现出良好的泛化性，在$r = 0.83$时各类别的识别准确率都在99\%以上。

\begin{table*}[t]
  \caption{Header fields decoded by \name}
  \label{tab:fields}
  \centering
  \footnotesize
  \begin{tabular}{lcccccc}
    \toprule
    Protocol & Sampling Rate & Avg. Bit Acc & \makecell{MCS Acc} & \makecell{Frame Size Acc / \\Packet Length Error} & Pilot Acc / Duration Error & \makecell{CBW Acc} \\
    \midrule
    \multirow{2}{*}{DVB-S2} 
    & $r=0.625$ & 0.937 & 0.846 & Frame Size Acc: 0.915 & Pilot: 0.934 & - \\
    & $r=0.83$ & 0.998 & 0.990 & Frame Size Acc: 0.999 & Pilot: 0.999 & - \\
    \midrule
    \multirow{2}{*}{Non-HT Wi-Fi} 
    & $r=0.625$ & 0.965 & 0.943 & 95\%-tile of Err: 1.06~\!KB & - & - \\
    & $r=0.83$ & 0.993 & 0.998 & 95\%-tile of Err: 0.00~\!KB & - & - \\
    \midrule
    \multirow{2}{*}{HT Wi-Fi} 
    & $r=0.625$ & 0.931 & 0.893 & 95\%-tile of Err: 12.5~\!KB& 95\%-tile of Err: 1.18~\!ms & 0.985 \\
    & $r=0.83$ & 0.987 & 0.995 & 95\%-tile of Err: 0.224~\!KB& 95\%-tile of Err: 0.0~\!ms  & 0.997 \\
    \bottomrule
  \end{tabular}
\end{table*}

\begin{figure*}[t]
    \subfloat[DVB-S2, $r=0.625$]{
        \centering
        \includegraphics[width=0.32\textwidth]{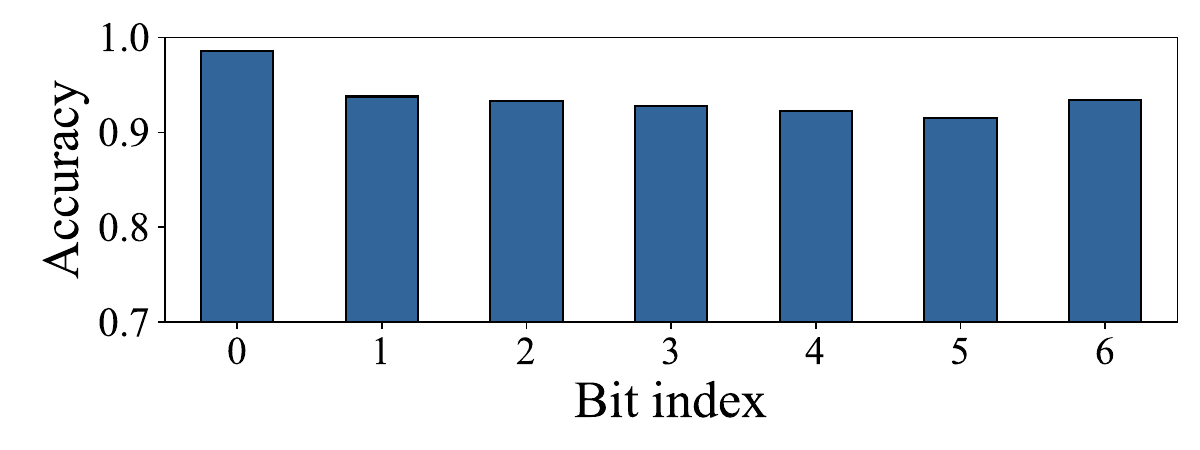}
        \label{fig:exp:decode:dvbs2:C6}
        % \vspace{-2ex}
    }
    \subfloat[Non-HT Wi-Fi, $r=0.625$]{
        \centering
        \includegraphics[width=0.32\textwidth]{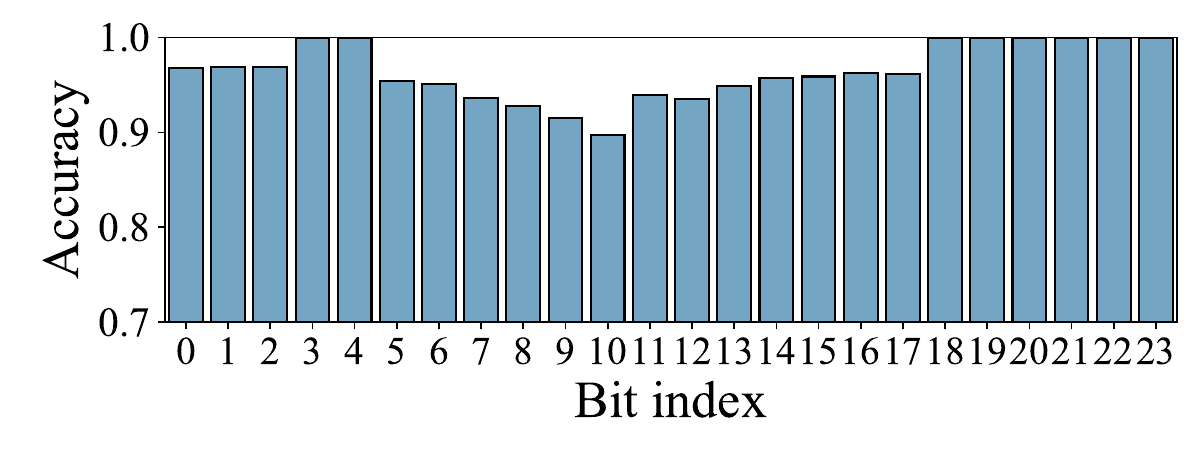}
        \label{fig:exp:decode:nonHT:C6}
        % \vspace{-2ex}
    }
    \subfloat[HT Wi-Fi, $r=0.625$]{
        \centering
        \includegraphics[width=0.32\textwidth]{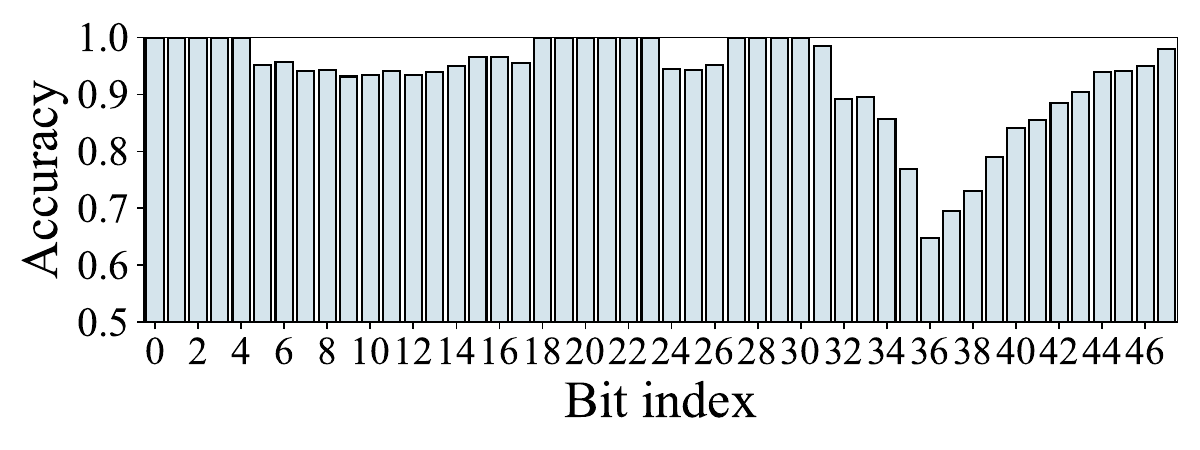}
        \label{fig:exp:decode:HT:C6}
    } \\
    % \vspace{-2ex}
    \subfloat[DVB-S2, $r=0.83$]{
        \centering
        \includegraphics[width=0.32\textwidth]{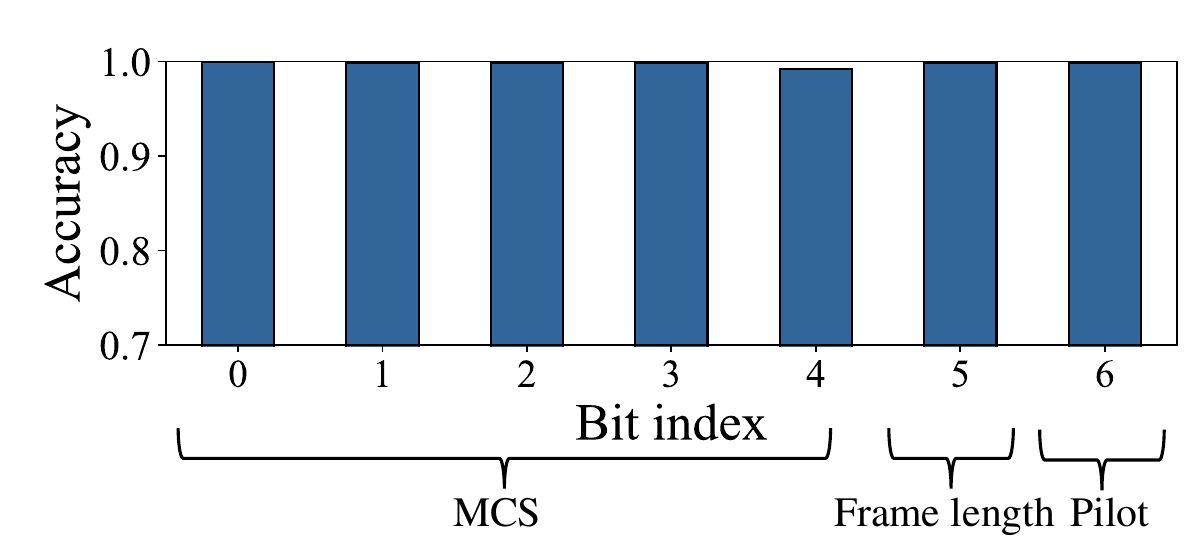}
        \label{fig:exp:decode:dvbs2:C8}
    }
    \subfloat[Non-HT Wi-Fi, $r=0.83$]{
        \centering
        \includegraphics[width=0.32\textwidth]{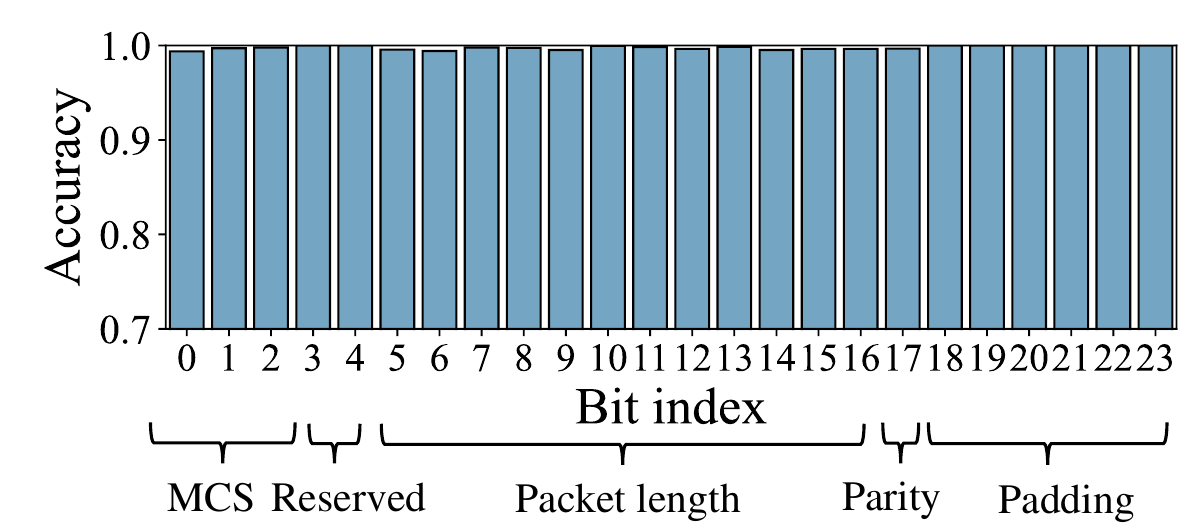}
        \label{fig:exp:decode:nonHT:C8}
    }
    \subfloat[HT Wi-Fi, $r=0.83$]{
        \centering
        \includegraphics[width=0.32\textwidth]{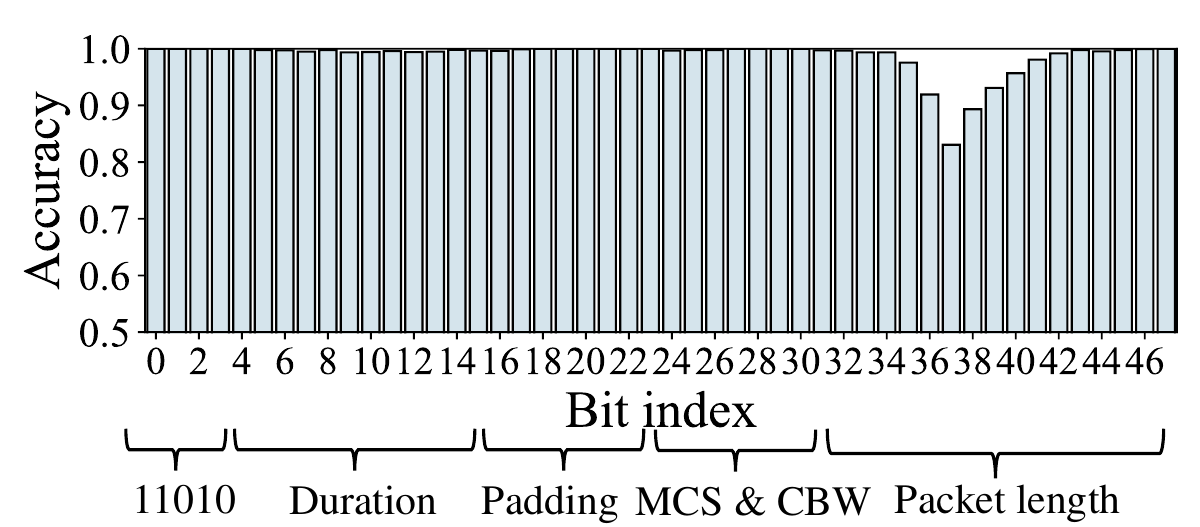}
        \label{fig:exp:decode:HT:C8}
    }
    % \vspace{-3ex}
    \caption{Decoding accuracy on each bit.}
    \label{fig:exp:decode}
    % \vspace{-3ex}
\end{figure*}

Figure~\ref{fig:exp:pkt_cls:conf_m_ota} presents the classification confusion matrices on over-the-air signals, despite neither model being trained on such data. T-Prime demonstrates limited generalization, particularly with DVB-S2 headers. At $r = 0.83$, where signal recovery error is relatively low, its classification accuracy for DVB-S2 drops to around 90\%. This issue is pronounced at $r = 0.625$, where increased recovery errors reduce its accuracy to approximately 58\%. In contrast, \name exhibits robust generalization across all sampling rates. It achieves over 99\% accuracy in identifying each category at $r = 0.83$, and over 88\% accuracy in recognizing DVB-S2 headers at $r = 0.625$.

The protocol identification model can achieve high performance and real-time processing at the same time. Setting the inference batch size to 1,024, \name processes one batch in \rev{45.3 $\pm$ 0.4~\!ms}, averaging \rev{44.23}~\!$\mu$s per sample. By organizing the spectrum sensing, signal recovery and protocol identification model as a pipeline, those wide multiband signal analysis tasks can be completed in real-time on a consumer-level GPU. These analysis results along with recovered signals can be further forwarded to downstream modules.

\subsection{Header Decoding} \label{sec:exp:decode}
\begin{figure}[t]
    \centering
    \subfloat[Non-HT packet length error]{
        \centering
        \includegraphics[width=0.23\textwidth]{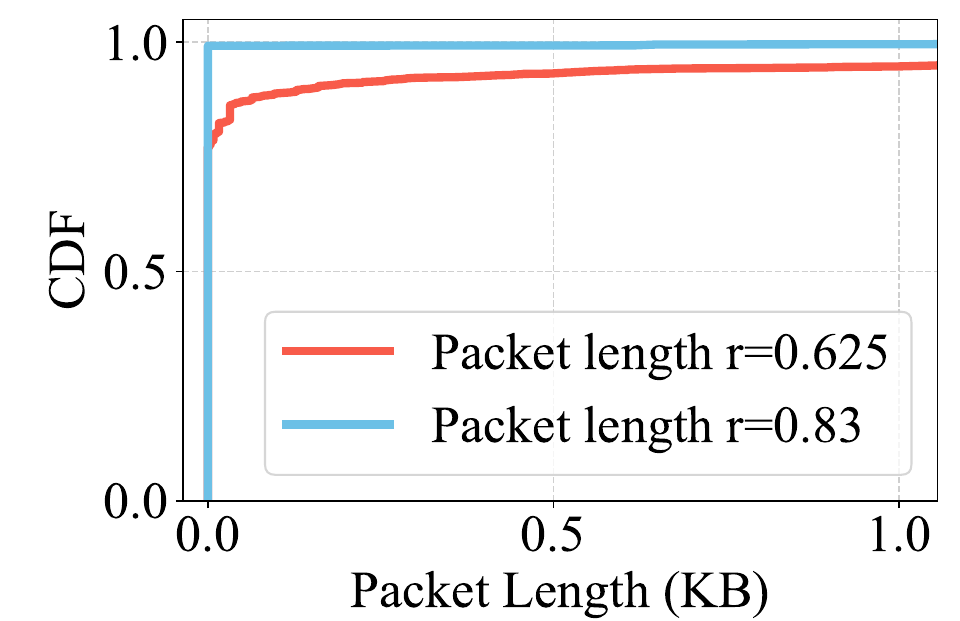}
        \label{fig:exp:decode:cdf:nonHT}
    }
    \subfloat[HT packet length and duration errors]{
        \centering
        \includegraphics[width=0.23\textwidth]{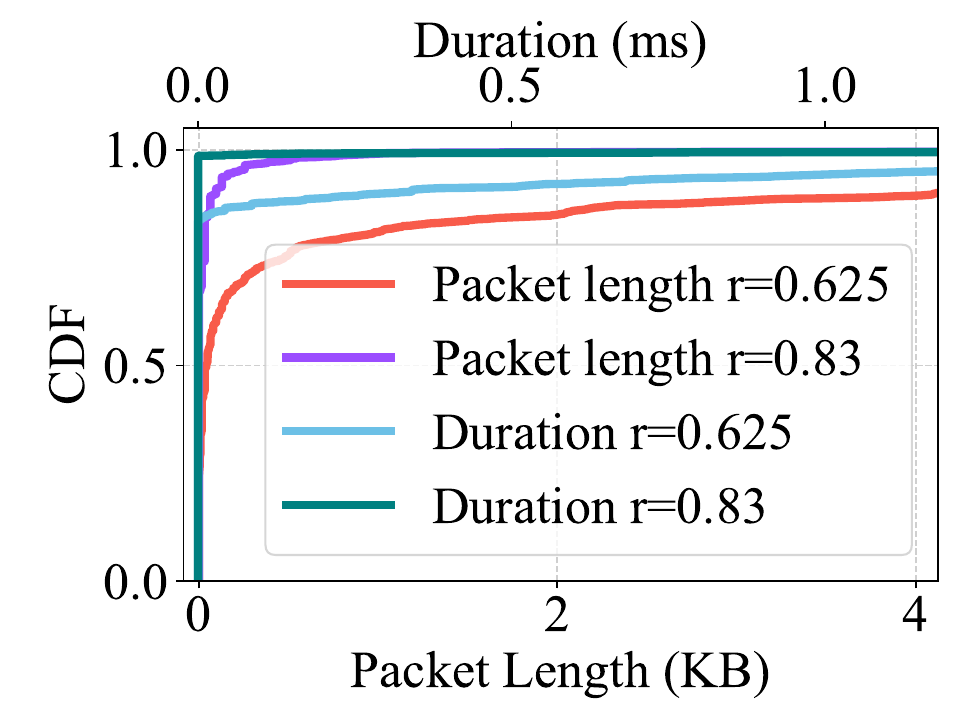}
        \label{fig:exp:decode:cdf:HT}
    }
    \caption{The cumulative distribution function of the predicted length and duration errors of non-HT and HT signals.}
    \label{fig:exp:decode:cdf}
\end{figure}

% \vspace{-1ex}
% 可以看到在$r=0.83$时，\name在DVB-S2和Non-HT Wi-Fi头的解码准确率非常高，平均每个bit的准确率分别达到了99.8\%和99.3\%。在最复杂的HT Wi-Fi头的解码上，packet length字段的部分低权重bits的解析准确率偏低，但还是能准确解析其它位置的bits。HT Wi-Fi这些解析准确率较低的bits对于无线网络sniffing的影响较小，因为这些bits表示的是packet length的低位，从Table~\ref{}中可以看出，\name对packet length预测的95th percentile of error仅有0.224 KB，平均误差仅有0.07 KB. 此外，我们发现以500, 750, 1000个数据点为微调的HT解析最低比特准确率分别为0.7835, 0.814, 0.8301，将decoder层数增加到4层后，最低比特准确率提升到0.9059，
% 0.7835, 0.814, 0.8301（最低bit准确率）, (500, 750, 1000, 微调数据集大小)，4层0.9059
% \name对其它字段的解码准确率非常准确，在$r=0.83$的采样率下，三类协议的MCS预测准确率均高于99\%。\name对DVB-S2的帧长格式，payload的pilots格式参数的解码准确率均有99.9\%, 对于HT Wi-Fi channel bandwidth的解码准确率高达99.7\%. \name对Non-HT Wi-Fi包长和HT Wi-Fi的传输duration解码也非常准确，如Table.~\ref{}和Figure~\ref{}所示。对Non-HT Wi-Fi的解析，\name对至少95\%的包都能做到无误差的包长预测，平均的预测误差仅有0.014 KB。对至少95\%的包的传输时间预测无误差，平均的传输时间预测误差仅为13.57$\mu$s
This section evaluates the header decoding performance of \name on DVB-S2, IEEE 802.11g/b non-HT Wi-Fi and IEEE 802.11n HT Wi-Fi signals, assuming the protocol headers are identified correctly. With seconds of fine-tuning on 1,000 data points for 5 epochs, we show the decoding accuracy on over-the-air signals. The decoding accuracies of each bit in the headers are shown in Figure~\ref{fig:exp:decode}. \name demonstrates exceptionally high decoding accuracy across various protocol fields at $r=0.83$, with MCS prediction accuracy exceeding 99\% for all three protocol types. For DVB-S2, frame length and payload pilot format parameters are decoded with 99.9\% accuracy, while the channel bandwidth for HT Wi-Fi reaches 99.7\% accuracy.
Additionally, \name excels in decoding Non-HT Wi-Fi packet length and HT Wi-Fi transmission duration fields, as shown in Table~\ref{tab:fields} and Figure~\ref{fig:exp:decode:cdf}. Specifically, for Non-HT Wi-Fi, \name decodes packet length with zero error for at least 95\% of packets, achieving an average packet length prediction error of only 0.014~\!KB. Similarly, it predicts transmission time with zero error for at least 95\% of packets, with an average prediction error of just 13.57~\!$\mu$s.

For HT Wi-Fi headers, the bit-level decoding accuracy is slightly lower for certain low-weight bits in the packet length field, but this has minimal impact on wireless network sniffing since these bits are less significant. As shown in Table~\ref{tab:fields}, the 95th percentile error in packet length prediction is only 0.224~\!KB, with an average error of 0.07~\!KB, which remain well within acceptable thresholds. Additionally, we observe that the accuracy in HT decoding easily improves with increased data and model scale. Fine-tuning with 500, 750, and 1,000 data points raises the lowest accuracy from 0.7835 to 0.8140 and 0.8301. Further, increasing decoder layers to four boosts the accuracy to 0.9059, with the 95th percentile packet length error of 0.064~\!KB and the average error of 0.042~\!KB. Remarkably, even without any fine-tuning, the four-layer model achieves the lowest bit accuracy of 0.8973. These results suggest that with an appropriately scaled dataset and model, even minor errors can be further minimized, supporting the broader potential of \name for effective wireless sniffing applications.

% 在极低采样率$r=0.625$时，\name的字段解析准确率有所下降，不过对DVB-S2的帧长格式，payload pilot格式，Non-HT Wi-Fi的MCS，HT Wi-Fi的channel bandwidth这些字段的解析依然有90\%以上的准确率。对于Non-HT Wi-Fi的包长，95th percentile误差为1.06 KB, 平均误差为0.136 KB. 对于HT Wi-Fi的传输duration，95th percentile误差为1.18 ms，平均误差为0.145 ms。在接近Landau rate的情况下，\name依然能相对准确地解析这些字段，这对于无线网络sniffing的应用仍然具有一定的实用性。
At $r=0.625$, \name shows a reduction in decoding accuracy across some fields; however, it still maintains over 90\% accuracy for key fields such as DVB-S2 frame length, payload pilot format, Non-HT Wi-Fi MCS, and HT Wi-Fi channel bandwidth. For Non-HT Wi-Fi packet length, the 95th percentile error is 1.06~\!KB, with an average error of 0.136~\!KB. For HT Wi-Fi transmission duration, the 95th percentile error is 1.18~\!ms, with an average error of 0.145~\!ms. Despite operating near the Landau rate, \name achieves relatively reliable header decoding, indicating practical applicability for wireless network sniffing tasks for non-sparse signals with low cost.

% 之后讨论实时性
% DVB-S2 27.63959382660687 +- 0.0949321356437758 ms, 1374条数据
% nonHT 42.79770053457469 +- 0.09138195594989777 ms, 1509条数据
% HT 54.8213 +- 0.106 ms, 1378条数据
% 推理速度方面，在batch size为1,024时，\name解析batch中1374条DVB-S2信号头所需的时间为27.64 +- 0.09 ms；解析batch中1509条Non-HT Wi-Fi信号头所需的时间为42.80 +- 0.09 ms；解析batch中1378条HT Wi-Fi信号头所需的时间为54.82 +- 0.11 ms。以1024条欠采样的多频带信号做平均，\name解析三类头所需要的时间分别为26.99 us, 41.80 us, 53.54 us，处理时间与输入信号的时长48 us相当。考虑到实际的信号中信号头只占整个信号的一小部分（例如DVB-S2头最多只占信号长度的1/37），经过protocol identifier的分类和滤除掉payload frames后，\name实际需要解码的frame数量会远小于batch size。因此，\name的实时性能能够满足实际应用的需求。
In terms of inference speed, for a batch size of 1,024 sub-sampled multiband signals, which include \rev{1,340} DVB-S2 headers, \rev{1,416} Non-HT Wi-Fi headers, and \rev{1,345} HT Wi-Fi headers, \name processes them in \rev{26.8 $\pm$ 0.25~\!ms}, \rev{39.5 $\pm$ 0.22~\!ms}, and \rev{52.7 $\pm$ 0.22~\!ms}, respectively. Averaged over 1,024 sub-samples multiband signals, the processing times are \rev{26.17~\!$\mu$s}, \rev{41.80~\!$\mu$s}, and \rev{51.46~\!$\mu$s}, comparable to the 48~\!$\mu$s duration of the input. Given that signal headers represent only a small fraction of the whole signal (e.g., DVB-S2 headers account for at most 1/37 of the total length), and that payload frames are filtered out by the protocol identifier of \name, the actual frames requiring processing are much fewer than those in the dataset.

\rev{GPU memory usage and inference time are summarized in Table~\ref{tab:n_params}, where \name demonstrates moderate GPU memory consumption and real-time processing capability.} \rev{The computational cost of \name scales linearly as the number of signals increases. And due to the modular and software-based design, \name can employ software engineering techniques such as load balancing and dynamic scaling to handle increased signal loads. In exceptionally high-density scenarios, the challenge shifts from individual module efficiency to specific scheduling algorithm design, akin to upper-layer network telemetry under heavy traffic. We leave such algorithm design for future exploration.}

\section{Related Work} \label{sec:related}

% 序列扫描的一些工作，例如infocom, mobicom，nsdi的一些扫描工作
% 朗道，non-uniform采样的理论，mishali所提出的二倍采样率工作，以及后续的MWC工作。song的approaching blind limit的工作。
% 一些突破尝试突破稀疏性的工作，power spectrum，nsdi谐振腔（谐振腔在模拟域工作，丢失信息），深度学习在图像CS上的应用，zhang han的工作
% 单独的信号分析工作，调制识别，信号恢复，信号解调，信号解码等

Sniffing physical-layer signals is far more challenging than upper-layer analysis due to the complexities of electromagnetic characteristics. Key difficulties include the constraints imposed by Nyquist sampling theory and the need to parse raw electromagnetic waves instead of clean bit streams.

\textbf{Sequential scanning:} There are existing sequential scanning studies to address the first challenge. SpecInsight~\cite{shi2015beyond} achieves fine-grained spectrum sensing via carefully scheduled scanning. SweepSense~\cite{guddeti2019sweepsense} improves scanning speed and enables protocol identification using cyclostationary analysis. Crescendo~\cite{subbaraman2023crescendo} further improves SweepSense with power control and locked VCOs for high-fidelity. However, sequential scanning inherently fails to capture entire headers or monitor the full spectrum simultaneously.

\textbf{Universal sub-Nyquist sampling:} Sub-Nyquist sampling reduces the cost of wideband monitoring. Landau~\cite{landau1967necessary} proved that the average sampling rate must be greater than the sum of occupied bandwidths if the spectrum occupancy is known. Under the CS framework, Tropp et al.\cite{tropp2009beyond} showed multi-tone recovery via random sub-Nyquist sampling. Mishali et al.~\cite{mishali2009blind} demonstrated stable sampling above twice the Landau rate without occupancy knowledge and empirically validated practical schemes such as multi-coset sampling and modulated wideband converters~\cite{mishali2010from}, achieving blind sampling at approximately 5.98 times the Landau rate. More recently, Song et al.~\cite{song2022approaching} optimize sub-sampling and recovery algorithms to achieve sampling rates as low as 5/2 of the Landau rate for spectrum sensing tasks.

\textbf{Beyond sparsity:} The twice-Landau-rate bound limits wideband sniffing to sparse signals. Some studies bypass this sparsity restriction by focusing on less detailed properties~\cite{lexa2011compressive,leus2011power,ariananda2012compressive,cohen2014sub,cohen2017sub}. Cohen and Eldar~\cite{cohen2014sub,cohen2017sub} proved that twice-Landau-rate sampling is not necessary of power spectrum and cyclic spectrum recovery. Guan et al.~\cite{guan2022efficient} use MEMS acoustic resonators to sparsify analog signals, enabling non-sparse spectrum sensing. WISE~\cite{luo2022wise} leverage UWB signals for spectrum sensing, bypassing sparse recovery altogether. Zhang et al.~\cite{zhang2022machine} incorporate deep learning for sub-Nyquist spectrum sensing, outperforming traditional methods. While these approaches mitigate sparsity constraints, they lose detailed information, precluding deep physical-layer analysis.

\textbf{Passive signal analysis:} Non-cooperative signal analysis is challenging due to the complexity of electromagnetic waves and limited prior knowledge. Early work~\cite{swami2000hierarchical} used higher-order cumulants for modulation classification. CNNs~\cite{o2018over} demonstrate superior performance over handcrafted features, and recent advancements shift to Transformer models~\cite{belgiovine2024tprime,cai2022signal,hamidi-rad2021mcformer} for modulation and protocol classification. T-Prime~\cite{belgiovine2024tprime} classifies fine-grained Wi-Fi signals even with spectrum overlapping. Beyond classification, deep learning has been applied to channel state estimation~\cite{kang2024cross,hanna2021signal}, signal demodulation~\cite{hanna2021signal,wu2019cnn,zheng2022demodnet} and decoding~\cite{wang1996artificial,henarejos2020decoding}. Li et al.~\cite{li2021nelora} address ultra-low SNR demodulation of LoRa signals with deep learning. Cammerer et al.~\cite{cammerer2023neural} implement a deep learning receiver for 5G NR signals, though it lacks adaptability to varying parameters. These methods focus on single-signal analysis and are unsuitable for wide multiband scenarios.

\section{Conclusion} \label{sec:conclusion}
% \name的limiations包括如下方面：（1）\name频谱感知粒度和信号恢复以sub-band为单位，取决于低速ADC的速率$B/L$。细粒度的感知和恢复可以以降低ADC采样率，增加ADC数目的方式实现。Alternatively, 在不改变硬件的情况下，可以将单个ADC采样的结果视作多个ADC的交错采样，逻辑上实现细粒度的感知和恢复。（2）\name的信号恢复算法只恢复被占据的子频带信号，未占据的子频带的噪声会聚集到恢复的信号中，降低信号的恢复质量，如图~\ref{}所示。我们可以通过统计已有信号的模式优化前端滤波器减轻其它频带噪声的干扰。(3) header decoding models虽然结构相同但参数不同。可以设计更大的统一模型，参考大语言模型的思路，通过prompt的方式完成不同任务的解码，然而这方面需要更多的研究，因为这需要更大的模型容量降低推理效率，实时性可能受到影响。

We present \name, a Transformer-based system designed to sniff physical layer signals in wide and non-sparse spectra. Existing methods face significant trade-offs, either failing to capture the entire spectrum concurrently, losing critical details, or relying heavily on sparsity constraints. \name addresses these challenges by integrating sub-Nyquist sampling with the powerful representation capabilities of Transformer networks. It achieves signal recovery exceeding the sub-Nyquist sampling limit through a rule-based Transformer network, reducing the SOTA sampling rate from 2.5 times the Landau rate to 1.66 and lower. And signal analysis process is simplified with cascaded Transformer networks. Experimental results show that \name excels in key tasks such as spectrum sensing, signal recovery, protocol identification, and header decoding, all while achieving real-time on a consumer-level GPU. As a potential future direction, we are looking forward to extending our method to improve the performance of various applications, such as large language models~\cite{lin2024splitlora,wang2025contemporary,hu2024agentscodriver,qiu2024ifvit,hu2024agentscomerge,zhou2025survey}, multimodal training~\cite{tang2024merit,fang2024ic3m}, distributed learning system~\cite{lin2024hierarchical,zhang2024fedac,lin2024adaptsfl,lyu2023optimal,lin2025leo,fang2024automated,lin2024fedsn}, and secure system~\cite{zhang2025state,duan2025rethinking,zhang2025robust}.

% fine-grained sub-bands

% \begin{acks}
% To Gale, Wyll, Shadowheart, Astarion, Lae'zel, and the rest of the party, for the many hours of entertainment and inspiration.
% \end{acks}

%%
%% The next two lines define the bibliography style to be used, and
%% the bibliography file.
% \bibliographystyle{ACM-Reference-Format}
\bibliographystyle{unsrt}
\bibliography{ref}

\begin{thebibliography}{100}

\bibitem{shi2015beyond}
Lixin Shi, Paramvir Bahl, and Dina Katabi.
\newblock {Beyond Sensing: Multi-GHz Realtime Spectrum Analytics}.
\newblock In {\em NSDI '15}, pages 159--172, Oakland, CA, May 2015. USENIX Association.

\bibitem{guddeti2019sweepsense}
Yeswanth Guddeti, Raghav Subbaraman, Moein Khazraee, Aaron Schulman, and Dinesh Bharadia.
\newblock {SweepSense: Sensing 5 GHz in 5 Milliseconds with Low-cost Radios}.
\newblock In {\em NSDI '19}, pages 317--330, Boston, MA, February 2019. USENIX Association.

\bibitem{subbaraman2023crescendo}
Raghav Subbaraman, Kevin Mills, Aaron Schulman, and Dinesh Bharadia.
\newblock {Crescendo: Towards Wideband, Real-time, High-Fidelity Spectrum Sensing Systems}.
\newblock In {\em ACM MobiCom '23}, New York, NY, USA, 2023. Association for Computing Machinery.

\bibitem{wireshark}
Wireshark.
\newblock {Wireshark}, 2024.

\bibitem{mishali2009blind}
Moshe Mishali and Yonina~C. Eldar.
\newblock {Blind Multiband Signal Reconstruction: Compressed Sensing for Analog Signals}.
\newblock {\em IEEE Transactions on Signal Processing}, 57(3):993--1009, 2009.

\bibitem{hassanieh2014ghz}
Haitham Hassanieh, Lixin Shi, Omid Abari, Ezzeldin Hamed, and Dina Katabi.
\newblock {GHz-wide Sensing and Decoding using the Sparse Fourier Transform}.
\newblock In {\em IEEE INFOCOM '14}, pages 2256--2264. IEEE, 2014.

\bibitem{qin2018sparse}
Zhijin Qin, Jiancun Fan, Yuanwei Liu, Yue Gao, and Geoffrey~Ye Li.
\newblock {Sparse Representation for Wireless Communications: A Compressive Sensing Approach}.
\newblock {\em IEEE Signal Processing Magazine}, 35(3):40--58, 2018.

\bibitem{song2022approaching}
Zihang Song, Jian Yang, Han Zhang, and Yue Gao.
\newblock {Approaching Sub-nyquist Boundary: Optimized Compressed Spectrum Sensing based on Multicoset Sampler for Multiband Signal}.
\newblock {\em IEEE Transactions on Signal Processing}, 70:4225--4238, 2022.

\bibitem{dvbs2}
ETSI Standard.
\newblock {Digital Video Broadcasting (DVB), Second Generation Framing Structure, Channel Coding and Modulation Systems for Broadcasting, Interactive Services, News Gathering and other Broadband Satellite Applications (DVB-S2)}.
\newblock {\em European Telecommunications Standards Institute (ETSI) EN}, 302(307):V1, 2014.

\bibitem{ieee802.11}
{IEEE Standard for Information Technology--Telecommunications and Information Exchange between Systems - Local and Metropolitan Area Networks--Specific Requirements - Part 11: Wireless LAN Medium Access Control (MAC) and Physical Layer (PHY) Specifications}.
\newblock {\em IEEE Std 802.11-2020 (Revision of IEEE Std 802.11-2016)}, pages 1--4379, 2020.

\bibitem{lin2021spatial}
Zheng Lin, Lifeng Wang, Bo~Tan, and Xiang Li.
\newblock Spatial-spectral terahertz networks.
\newblock {\em IEEE Transactions on Wireless Communications}, 21(6):3881--3892, 2021.

\bibitem{ding2023hidden}
Fangqiang Ding, Andras Palffy, Dariu~M Gavrila, and Chris~Xiaoxuan Lu.
\newblock {Hidden Gems: 4d Radar Scene Flow Learning using Cross-modal Supervision}.
\newblock In {\em Proc of IEEE/CVF CVPR}, pages 9340--9349, 2023.

\bibitem{yang2020magprint}
Lanqing Yang, Yi-Chao Chen, Hao Pan, Dian Ding, Guangtao Xue, Linghe Kong, Jiadi Yu, and Minglu Li.
\newblock {Magprint: Deep Learning based User Fingerprinting using Electromagnetic Signals}.
\newblock In {\em IEEE INFOCOM'20}, pages 696--705. IEEE, 2020.

\bibitem{yang2023slnet}
Zheng Yang, Yi~Zhang, Kun Qian, and Chenshu Wu.
\newblock {SLNet: A Spectrogram Learning Neural Network for Deep Wireless Sensing}.
\newblock In {\em Proc. of USENIX NSDI'23}, pages 1221--1236, 2023.

\bibitem{lin2022tracking}
Zheng Lin, Lifeng Wang, Jie Ding, Yuedong Xu, and Bo~Tan.
\newblock Tracking and transmission design in terahertz v2i networks.
\newblock {\em IEEE Transactions on Wireless Communications}, 22(6):3586--3598, 2022.

\bibitem{huang2013depth}
Junxian Huang, Feng Qian, Yihua Guo, Yuanyuan Zhou, Qiang Xu, Z~Morley Mao, Subhabrata Sen, and Oliver Spatscheck.
\newblock {An In-depth Study of LTE: Effect of Network Protocol and Application Behavior on Performance}.
\newblock {\em ACM SIGCOMM Computer Communication Review}, 43(4):363--374, 2013.

\bibitem{li2016mobileinsight}
Yuanjie Li, Chunyi Peng, Zengwen Yuan, Jiayao Li, Haotian Deng, and Tao Wang.
\newblock {Mobileinsight: Extracting and Analyzing Cellular Network Information on Smartphones}.
\newblock In {\em Proc. of MobiCom'16}, pages 202--215, 2016.

\bibitem{li2021experience}
Yuanjie Li, Chunyi Peng, Zhehui Zhang, Zhaowei Tan, Haotian Deng, Jinghao Zhao, Qianru Li, Yunqi Guo, Kai Ling, Boyan Ding, et~al.
\newblock {Experience: a Five-year Retrospective of MobileInsight}.
\newblock In {\em Proc. of Mobicom'21}, pages 28--41, 2021.

\bibitem{yaseen2021towards}
Nofel Yaseen, Behnaz Arzani, Krishna Chintalapudi, Vaishnavi Ranganathan, Felipe Frujeri, Kevin Hsieh, Daniel~S Berger, Vincent Liu, and Srikanth Kandula.
\newblock {Towards a Cost vs. Quality Sweet Spot for Monitoring Networks}.
\newblock In {\em Proc. of ACM HotNets'21}, pages 38--44, 2021.

\bibitem{huang2017sketchvisor}
Qun Huang, Xin Jin, Patrick~PC Lee, Runhui Li, Lu~Tang, Yi-Chao Chen, and Gong Zhang.
\newblock Sketchvisor: Robust network measurement for software packet processing.
\newblock In {\em Proc. of SIGCOMM'17}, pages 113--126, 2017.

\bibitem{gadre2020frequency}
Akshay Gadre, Revathy Narayanan, Anh Luong, Anthony Rowe, Bob Iannucci, and Swarun Kumar.
\newblock {Frequency Configuration for Low-Power Wide-Area Networks in a Heartbeat}.
\newblock In {\em 17th USENIX Symposium on Networked Systems Design and Implementation (NSDI 20)}, pages 339--352, 2020.

\bibitem{zhang2023dependent}
Zhehui Zhang, Yanbing Liu, Qianru Li, Zizheng Liu, Chunyi Peng, and Songwu Lu.
\newblock {Dependent Misconfigurations in 5G/4.5 G Radio Resource Control}.
\newblock {\em Proceedings of the ACM on Networking}, 1(CoNEXT1):1--20, 2023.

\bibitem{guirguis2018primary}
Arsany Guirguis, Fadel Digham, Karim~G Seddik, Mohamed Ibrahim, Khaled~A Harras, and Moustafa Youssef.
\newblock {Primary User-aware Optimal Discovery Routing for Cognitive Radio Networks}.
\newblock {\em IEEE Transactions on Mobile Computing}, 18(1):193--206, 2018.

\bibitem{cano2017fair}
Cristina Cano, Douglas~J Leith, Andres Garcia-Saavedra, and Pablo Serrano.
\newblock {Fair Coexistence of Scheduled and Random Access Wireless Networks: Unlicensed LTE/WiFi}.
\newblock {\em IEEE/ACM Transactions on Networking}, 25(6):3267--3281, 2017.

\bibitem{zheng2005collaboration}
Haitao Zheng and Chunyi Peng.
\newblock {Collaboration and Fairness in Opportunistic Spectrum Access}.
\newblock In {\em IEEE ICC 2005}, volume~5, pages 3132--3136. IEEE, 2005.

\bibitem{pawelczak2005cognitive}
Przemyslaw Pawelczak, R~Venkatesha Prasad, Liang Xia, and Ignas~GMM Niemegeers.
\newblock {Cognitive Radio Emergency Networks-Requirements and Design}.
\newblock In {\em Proc. of IEEE DySPAN}, pages 601--606. IEEE, 2005.

\bibitem{brik2005dsap}
Vladimir Brik, Eric Rozner, Suman Banerjee, and Paramvir Bahl.
\newblock {DSAP: a Protocol for Coordinated Spectrum Access}.
\newblock In {\em Proc. of IEEE DySPAN}, pages 611--614. IEEE, 2005.

\bibitem{guan2017smart}
Chaowen Guan, Aziz Mohaisen, Zhi Sun, Lu~Su, Kui Ren, and Yaling Yang.
\newblock {When Smart TV Meets CRN: Privacy-preserving Fine-grained Spectrum Access}.
\newblock In {\em 2017 IEEE 37th International Conference on Distributed Computing Systems (ICDCS)}, pages 1105--1115. IEEE, 2017.

\bibitem{visser2008multinode}
Frank Visser, Gerard~JM Janssen, and Przemyslaw Pawelczak.
\newblock {Multinode Spectrum Sensing based on Energy Detection for Dynamic Spectrum Access}.
\newblock In {\em IEEE Vehicular Technology Conference}, pages 1394--1398. IEEE, 2008.

\bibitem{valls2016maximizing}
V{\'\i}ctor Valls, Andr{\'e}s Garcia-Saavedra, Xavier Costa, and Douglas~J Leith.
\newblock {Maximizing LTE Capacity in Unlicensed Bands (LTE-U/LAA) while Fairly Coexisting with 802.11 WLANs}.
\newblock {\em IEEE Communications Letters}, 20(6):1219--1222, 2016.

\bibitem{peng2024sums}
Jinbo Peng, Zhe Chen, Zheng Lin, Haoxuan Yuan, Zihan Fang, Lingzhong Bao, Zihang Song, Ying Li, Jing Ren, and Yue Gao.
\newblock Sums: Sniffing unknown multiband signals under low sampling rates.
\newblock {\em IEEE Transactions on Mobile Computing}, 2024.

\bibitem{yuan2023graph}
Haoxuan Yuan, Zhe Chen, Zheng Lin, Jinbo Peng, Zihan Fang, Yuhang Zhong, Zihang Song, Xiong Wang, and Yue Gao.
\newblock Graph learning for multi-satellite based spectrum sensing.
\newblock In {\em Proc. ICCT}, pages 1112--1116, 2023.

\bibitem{al-jumaily2022evaluation}
Abdulmajeed Al-Jumaily, Aduwati Sali, Vćtor P.~Gil Jiménez, Fernando~Pérez Fontán, Mandeep~Jit Singh, Alyani Ismail, Qusay Al-Maatouk, Ali~M. Al-Saegh, and Dhiya Al-Jumeily.
\newblock {Evaluation of 5G Coexistence and Interference Signals in the C-Band Satellite Earth Station}.
\newblock {\em IEEE Transactions on Vehicular Technology}, 71(6):6189--6200, 2022.

\bibitem{yuan2024satsense}
Haoxuan Yuan, Zhe Chen, Zheng Lin, Jinbo Peng, Zihan Fang, Yuhang Zhong, Zihang Song, and Yue Gao.
\newblock Satsense: Multi-satellite collaborative framework for spectrum sensing.
\newblock {\em IEEE Transactions on Cognitive Communications and Networking}, 2024.

\bibitem{yuan2025constructing}
Haoxuan Yuan, Zhe Chen, Zheng Lin, Jinbo Peng, Yuhang Zhong, Xuanjie Hu, Songyan Xue, Wei Li, and Yue Gao.
\newblock Constructing 4d radio map in leo satellite networks with limited samples.
\newblock {\em arXiv preprint arXiv:2501.02775}, 2025.

\bibitem{hauri2020internet}
Yannick Hauri, Debopam Bhattacherjee, Manuel Grossmann, and Ankit Singla.
\newblock " internet from space" without inter-satellite links.
\newblock In {\em Proc. of ACM HotNets'20}, pages 205--211, 2020.

\bibitem{gao2023swirls}
Zhihui Gao, Yiran Chen, and Tingjun Chen.
\newblock {Swirls: Sniffing Wi-Fi Using Radios with Low Sampling Rates}.
\newblock In {\em Mobihoc '23}, pages 260--269, 2023.

\bibitem{kochut2004sniffing}
Andrzej Kochut, Arunchandar Vasan, A~Udaya Shankar, and Ashok Agrawala.
\newblock {Sniffing out the Correct Physical Layer Capture Model in 802.11 b}.
\newblock In {\em Proc. of ICNP 2004.}, pages 252--261. IEEE, 2004.

\bibitem{albazrqaoe2016practical}
Wahhab Albazrqaoe, Jun Huang, and Guoliang Xing.
\newblock Practical bluetooth traffic sniffing: Systems and privacy implications.
\newblock In {\em Proc. of MobiSys'16}, pages 333--345, 2016.

\bibitem{xie2022ng}
Yaxiong Xie and Kyle Jamieson.
\newblock {Ng-scope: Fine-grained Telemetry for NextG Cellular Networks}.
\newblock {\em Proc. ACM Meas. Anal. Comput. Syst.}, 6(1):1--26, 2022.

\bibitem{hu2019nb}
Zhenxian Hu, Guangtao Xue, Yi-Chao Chen, and Minglu Li.
\newblock {NB-IoT network monitoring and diagnosing}.
\newblock In {\em Proc. of IEEE SECON}, pages 1--9. IEEE, 2019.

\bibitem{nika2014towards}
Ana Nika, Zengbin Zhang, Xia Zhou, Ben~Y Zhao, and Haitao Zheng.
\newblock {Towards Commoditized Real-time Spectrum Monitoring}.
\newblock In {\em Proc. of ACM HotNets'14}, pages 25--30, 2014.

\bibitem{zhang2022machine}
Han Zhang, Jian Yang, and Yue Gao.
\newblock {Machine Learning Empowered Spectrum Sensing under a Sub-sampling Framework}.
\newblock {\em IEEE transactions on wireless communications}, 21(10):8205--8215, 2022.

\bibitem{guan2022efficient}
Junfeng Guan, Jitian Zhang, Ruochen Lu, Hyungjoo Seo, Jin Zhou, Songbin Gong, and Haitham Hassanieh.
\newblock {Efficient Wideband Spectrum Sensing using MEMS Acoustic Resonators}.
\newblock {\em GetMobile: Mobile Computing and Communications}, 25(3):23--27, 2022.

\bibitem{zhang2014distributed}
Wenlin Zhang, Yi~Guo, Hongbo Liu, Yingying Chen, Zheng Wang, and Joseph Mitola~III.
\newblock {Distributed Consensus-based Weight Design for Cooperative Spectrum Sensing}.
\newblock {\em IEEE Transactions on Parallel and Distributed Systems}, 26(1):54--64, 2014.

\bibitem{zhang2014vehicle}
Tan Zhang, Ning Leng, and Suman Banerjee.
\newblock {A vehicle-based measurement framework for enhancing whitespace spectrum databases}.
\newblock In {\em Proc. of MobiCom'14}, pages 17--28, 2014.

\bibitem{pu2021spectrum}
Yiping Pu, Fengyuan Zhu, Mingqi Xie, Meng Jin, and Xiaohua Tian.
\newblock {Spectrum Analysis of 2.4 GHz Band Based on Successive Signal Detection}.
\newblock In {\em Proc. of WCSP}, pages 1--6. IEEE, 2021.

\bibitem{mishali2010from}
Moshe Mishali and Yonina~C. Eldar.
\newblock {From Theory to Practice: Sub-Nyquist Sampling of Sparse Wideband Analog Signals}.
\newblock {\em IEEE Journal of Selected Topics in Signal Processing}, 4(2):375--391, 2010.

\bibitem{hassanieh2012faster}
Haitham Hassanieh, Fadel Adib, Dina Katabi, and Piotr Indyk.
\newblock {Faster GPS via the Sparse Fourier Transform}.
\newblock In {\em Proc. of MobiCom'12}, pages 353--364, 2012.

\bibitem{candes2006robust}
E.J. Candes, J.~Romberg, and T.~Tao.
\newblock {Robust Uncertainty Principles: Exact Signal Reconstruction from Highly Incomplete Frequency Information}.
\newblock {\em IEEE Transactions on Information Theory}, 52(2):489--509, 2006.

\bibitem{landau1967necessary}
HJ~Landau.
\newblock {Necessary Density Conditions for Sampling and Interpolation of Certain Entire Functions}.
\newblock 1967.

\bibitem{wu2019deep}
Yan Wu, Mihaela Rosca, and Timothy Lillicrap.
\newblock {Deep compressed sensing}.
\newblock In {\em International Conference on Machine Learning}, pages 6850--6860. PMLR, 2019.

\bibitem{bora2017compressed}
Ashish Bora, Ajil Jalal, Eric Price, and Alexandros~G Dimakis.
\newblock {Compressed Sensing using Generative Models}.
\newblock In {\em International Conference on Machine Learning}, pages 537--546. PMLR, 2017.

\bibitem{zhou2024larger}
Lexin Zhou, Wout Schellaert, Fernando Mart{\'\i}nez-Plumed, Yael Moros-Daval, C{\`e}sar Ferri, and Jos{\'e} Hern{\'a}ndez-Orallo.
\newblock {Larger and More Instructable Language Models Become Less Reliable}.
\newblock {\em Nature}, pages 1--8, 2024.

\bibitem{lewkowycz2022solving}
Aitor Lewkowycz, Anders Andreassen, David Dohan, Ethan Dyer, Henryk Michalewski, Vinay Ramasesh, Ambrose Slone, Cem Anil, Imanol Schlag, Theo Gutman-Solo, et~al.
\newblock {Solving Quantitative Reasoning Problems with Language Models}.
\newblock {\em Advances in Neural Information Processing Systems}, 35:3843--3857, 2022.

\bibitem{satpute2024can}
Ankit Satpute, Noah Gie{\ss}ing, Andr{\'e} Greiner-Petter, Moritz Schubotz, Olaf Teschke, Akiko Aizawa, and Bela Gipp.
\newblock {Can Llms Master Math? Investigating Large Language Models on Math Stack Exchange}.
\newblock In {\em Proc. of ACM SIGIR '24}, pages 2316--2320, 2024.

\bibitem{chi2024rf}
Guoxuan Chi, Zheng Yang, Chenshu Wu, Jingao Xu, Yuchong Gao, Yunhao Liu, and Tony~Xiao Han.
\newblock {RF-Diffusion: Radio Signal Generation via Time-Frequency Diffusion}.
\newblock In {\em MobiCom'24}, pages 77--92, 2024.

\bibitem{tropp2005simultaneous}
J.A. Tropp, A.C. Gilbert, and M.J. Strauss.
\newblock {Simultaneous Sparse Approximation via Greedy Pursuit}.
\newblock In {\em ICASSP '05}, volume~5, pages v/721--v/724 Vol. 5, 2005.

\bibitem{lin2024efficient}
Zheng Lin, Guangyu Zhu, Yiqin Deng, Xianhao Chen, Yue Gao, Kaibin Huang, and Yuguang Fang.
\newblock Efficient parallel split learning over resource-constrained wireless edge networks.
\newblock {\em IEEE Transactions on Mobile Computing}, 2024.

\bibitem{hu2024accelerating}
Mingda Hu, Jingjing Zhang, Xiong Wang, Shengyun Liu, and Zheng Lin.
\newblock Accelerating federated learning with model segmentation for edge networks.
\newblock {\em IEEE Transactions on Green Communications and Networking}, 2024.

\bibitem{lin2022channel}
Zheng Lin, Lifeng Wang, Jie Ding, Bo~Tan, and Shi Jin.
\newblock Channel power gain estimation for terahertz vehicle-to-infrastructure networks.
\newblock {\em IEEE Communications Letters}, 27(1):155--159, 2022.

\bibitem{zhang2024satfed}
Yuxin Zhang, Zheng Lin, Zhe Chen, Zihan Fang, Wenjun Zhu, Xianhao Chen, Jin Zhao, and Yue Gao.
\newblock Satfed: A resource-efficient leo satellite-assisted heterogeneous federated learning framework.
\newblock {\em arXiv preprint arXiv:2409.13503}, 2024.

\bibitem{lin2024split}
Zheng Lin, Guanqiao Qu, Xianhao Chen, and Kaibin Huang.
\newblock Split learning in 6g edge networks.
\newblock {\em IEEE Wireless Communications}, 2024.

\bibitem{zhang2024spectrum}
Weishan Zhang, Yue Wang, Xiang Chen, Zhipeng Cai, and Zhi Tian.
\newblock {Spectrum Transformer: An Attention-based Wideband Spectrum Detector}.
\newblock {\em IEEE Transactions on Wireless Communications}, 2024.

\bibitem{gao2021fedswap}
Zhihui Gao, Ang Li, Yunfan Gao, Bing Li, Yu~Wang, and Yiran Chen.
\newblock {FedSwap: A Federated Learning based 5G Decentralized Dynamic Spectrum Access System}.
\newblock In {\em Proc. of IEEE/ACM ICCAD}, pages 1--6. IEEE, 2021.

\bibitem{belgiovine2024tprime}
Mauro Belgiovine, Joshua Groen, Miquel Sirera, Chinenye Tassie, Ayberk~Yarkın Yıldız, Sage Trudeau, Stratis Ioannidis, and Kaushik Chowdhury.
\newblock {T-PRIME: Transformer-based Protocol Identification for Machine-learning at the Edge}.
\newblock In {\em IEEE INFOCOM '24}. IEEE, 2024.

\bibitem{vaswani2017attention}
A~Vaswani.
\newblock {Attention is All You Need}.
\newblock {\em Advances in Neural Information Processing Systems}, 2017.

\bibitem{ling2022multi}
Fenghua Ling, Jing-Jia Luo, Yue Li, Tao Tang, Lei Bai, Wanli Ouyang, and Toshio Yamagata.
\newblock {Multi-task Machine Learning Improves Multi-seasonal Prediction of the Indian Ocean Dipole}.
\newblock {\em Nature Communications}, 13(1):7681, 2022.

\bibitem{zaheer2020big}
Manzil Zaheer, Guru Guruganesh, Kumar~Avinava Dubey, Joshua Ainslie, Chris Alberti, Santiago Ontanon, Philip Pham, Anirudh Ravula, Qifan Wang, Li~Yang, et~al.
\newblock {Big Bird: Transformers for Longer Sequences}.
\newblock {\em Advances in neural information processing systems}, 33:17283--17297, 2020.

\bibitem{song2024nonuniform}
Zihang Song, Yiyuan She, Jian Yang, Jinbo Peng, Yue Gao, and Rahim Tafazolli.
\newblock {Nonuniform Sampling Pattern Design for Compressed Spectrum Sensing in Mobile Cognitive Radio Networks}.
\newblock {\em IEEE Transactions on Mobile Computing}, pages 1--14, 2024.

\bibitem{cai2022signal}
Jingjing Cai, Fengming Gan, Xianghai Cao, and Wei Liu.
\newblock {Signal Modulation Classification Based on the Transformer Network}.
\newblock {\em IEEE Transactions on Cognitive Communications and Networking}, 8(3):1348--1357, 2022.

\bibitem{hamidi-rad2021mcformer}
Shahab Hamidi-Rad and Swayambhoo Jain.
\newblock {MCformer: A Transformer Based Deep Neural Network for Automatic Modulation Classification}.
\newblock In {\em IEEE GLOBECOM' 21}, pages 1--6, 2021.

\bibitem{tropp2009beyond}
Joel~A Tropp, Jason~N Laska, Marco~F Duarte, Justin~K Romberg, and Richard~G Baraniuk.
\newblock {Beyond Nyquist: Efficient Sampling of Sparse Bandlimited Signals}.
\newblock {\em IEEE transactions on information theory}, 56(1):520--544, 2009.

\bibitem{lexa2011compressive}
Michael~A Lexa, Mike~E Davies, John~S Thompson, and Janosch Nikolic.
\newblock {Compressive Power Spectral Density Estimation}.
\newblock In {\em 2011 IEEE ICASSP}, pages 3884--3887. IEEE, 2011.

\bibitem{leus2011power}
Geert Leus and Dyonisius~Dony Ariananda.
\newblock {Power Spectrum Blind Sampling}.
\newblock {\em IEEE Signal Processing Letters}, 18(8):443--446, 2011.

\bibitem{ariananda2012compressive}
Dyonisius~Dony Ariananda and Geert Leus.
\newblock {Compressive Wideband Power Spectrum Estimation}.
\newblock {\em IEEE Transactions on signal processing}, 60(9):4775--4789, 2012.

\bibitem{cohen2014sub}
Deborah Cohen and Yonina~C Eldar.
\newblock {Sub-Nyquist Sampling for Power Spectrum Sensing in Cognitive Radios: A Unified Approach}.
\newblock {\em IEEE Transactions on Signal Processing}, 62(15):3897--3910, 2014.

\bibitem{cohen2017sub}
Deborah Cohen and Yonina~C Eldar.
\newblock {Sub-Nyquist Cyclostationary Detection for Cognitive Radio}.
\newblock {\em IEEE Transactions on Signal Processing}, 65(11):3004--3019, 2017.

\bibitem{luo2022wise}
Zhicheng Luo, Qianyi Huang, Rui Wang, Hao Chen, Xiaofeng Tao, Guihai Chen, and Qian Zhang.
\newblock {WISE: Low-cost Wide Band Spectrum Sensing using UWB}.
\newblock In {\em Proc. of SenSys '22}, pages 651--666, 2022.

\bibitem{swami2000hierarchical}
Ananthram Swami and Brian~M Sadler.
\newblock {Hierarchical Digital Modulation Classification using Cumulants}.
\newblock {\em IEEE Transactions on communications}, 48(3):416--429, 2000.

\bibitem{o2018over}
Timothy~James O'Shea, Tamoghna Roy, and T~Charles Clancy.
\newblock {Over-the-air Deep Learning based Radio Signal Classification}.
\newblock {\em IEEE Journal of Selected Topics in Signal Processing}, 12(1):168--179, 2018.

\bibitem{kang2024cross}
Hua Kang, Qingyong Hu, Huangxun Chen, Qianyi Huang, Qian Zhang, and Min Cheng.
\newblock {Cross-shaped Separated Spatial-Temporal UNet Transformer For Accurate Channel Prediction}.
\newblock In {\em IEEE INFOCOM'24}, pages 2079--2088. IEEE, 2024.

\bibitem{hanna2021signal}
Samer Hanna, Chris Dick, and Danijela Cabric.
\newblock {Signal Processing-based Deep Learning for Blind Symbol Decoding and Modulation Classification}.
\newblock {\em IEEE Journal on Selected Areas in Communications}, 40(1):82--96, 2021.

\bibitem{wu2019cnn}
Tian Wu.
\newblock {CNN and RNN-based Deep Learning Methods for Digital Signal Demodulation}.
\newblock In {\em Proc. of IVSP'19}, pages 122--127, 2019.

\bibitem{zheng2022demodnet}
Shilian Zheng, Xiaoyu Zhou, Shichuan Chen, Peihan Qi, Caiyi Lou, and Xiaoniu Yang.
\newblock {DemodNet: Learning Soft Demodulation from Hard Information using Convolutional Neural Network}.
\newblock In {\em IEEE ICC'22}, pages 1--6. IEEE, IEEE, 2022.

\bibitem{wang1996artificial}
Xiao-An Wang and Stephen~B Wicker.
\newblock {An Artificial Neural Net Viterbi Decoder}.
\newblock {\em IEEE Transactions on communications}, 44(2):165--171, 1996.

\bibitem{henarejos2020decoding}
Pol Henarejos and Miguel~{\'A}ngel V{\'a}zquez.
\newblock Decoding 5g-nr communications via deep learning.
\newblock In {\em 2020 IEEE ICASSP}, pages 3782--3786. IEEE, 2020.

\bibitem{li2021nelora}
Chenning Li, Hanqing Guo, Shuai Tong, Xiao Zeng, Zhichao Cao, Mi~Zhang, Qiben Yan, Li~Xiao, Jiliang Wang, and Yunhao Liu.
\newblock {NELoRa: Towards ultra-low SNR LoRa communication with neural-enhanced demodulation}.
\newblock In {\em Proc. of ACM SenSys'21}, pages 56--68, 2021.

\bibitem{cammerer2023neural}
Sebastian Cammerer, Fay{\c{c}}al~A{\"\i}t Aoudia, Jakob Hoydis, Andreas Oeldemann, Andreas Roessler, Timo Mayer, and Alexander Keller.
\newblock {A Neural Receiver for 5G NR Multi-user MIMO}.
\newblock In {\em 2023 IEEE Globecom Workshops (GC Wkshps)}, pages 329--334. IEEE, 2023.

\bibitem{lin2024splitlora}
Zheng Lin, Xuanjie Hu, Yuxin Zhang, Zhe Chen, Zihan Fang, Xianhao Chen, Ang Li, Praneeth Vepakomma, and Yue Gao.
\newblock Splitlora: A split parameter-efficient fine-tuning framework for large language models.
\newblock {\em arXiv preprint arXiv:2407.00952}, 2024.

\bibitem{wang2025contemporary}
Jiayimei Wang, Tao Ni, Wei-Bin Lee, and Qingchuan Zhao.
\newblock A contemporary survey of large language model assisted program analysis.
\newblock {\em arXiv preprint arXiv:2502.18474}, 2025.

\bibitem{hu2024agentscodriver}
Senkang Hu, Zhengru Fang, Zihan Fang, Yiqin Deng, Xianhao Chen, and Yuguang Fang.
\newblock Agentscodriver: Large language model empowered collaborative driving with lifelong learning.
\newblock {\em arXiv preprint arXiv:2404.06345}, 2024.

\bibitem{qiu2024ifvit}
Yuhang Qiu, Honghui Chen, Xingbo Dong, Zheng Lin, Iman~Yi Liao, Massimo Tistarelli, and Zhe Jin.
\newblock Ifvit: Interpretable fixed-length representation for fingerprint matching via vision transformer.
\newblock {\em IEEE Transactions on Information Forensics and Security}, 2024.

\bibitem{hu2024agentscomerge}
Senkang Hu, Zhengru Fang, Zihan Fang, Yiqin Deng, Xianhao Chen, Yuguang Fang, and Sam Kwong.
\newblock Agentscomerge: Large language model empowered collaborative decision making for ramp merging.
\newblock {\em arXiv preprint arXiv:2408.03624}, 2024.

\bibitem{zhou2025survey}
Yihe Zhou, Tao Ni, Wei-Bin Lee, and Qingchuan Zhao.
\newblock A survey on backdoor threats in large language models (llms): Attacks, defenses, and evaluations.
\newblock {\em arXiv preprint arXiv:2502.05224}, 2025.

\bibitem{tang2024merit}
Yongyang Tang, Zhe Chen, Ang Li, Tianyue Zheng, Zheng Lin, Jia Xu, Pin Lv, Zhe Sun, and Yue Gao.
\newblock Merit: Multimodal wearable vital sign waveform monitoring.
\newblock {\em arXiv preprint arXiv:2410.00392}, 2024.

\bibitem{fang2024ic3m}
Zihan Fang, Zheng Lin, Senkang Hu, Hangcheng Cao, Yiqin Deng, Xianhao Chen, and Yuguang Fang.
\newblock Ic3m: In-car multimodal multi-object monitoring for abnormal status of both driver and passengers.
\newblock {\em arXiv preprint arXiv:2410.02592}, 2024.

\bibitem{lin2024hierarchical}
Zheng Lin, Wei Wei, Zhe Chen, Chan-Tong Lam, Xianhao Chen, Yue Gao, and Jun Luo.
\newblock Hierarchical split federated learning: Convergence analysis and system optimization.
\newblock {\em arXiv preprint arXiv:2412.07197}, 2024.

\bibitem{zhang2024fedac}
Yuxin Zhang, Haoyu Chen, Zheng Lin, Zhe Chen, and Jin Zhao.
\newblock Fedac: A adaptive clustered federated learning framework for heterogeneous data.
\newblock {\em arXiv preprint arXiv:2403.16460}, 2024.

\bibitem{lin2024adaptsfl}
Zheng Lin, Guanqiao Qu, Wei Wei, Xianhao Chen, and Kin~K Leung.
\newblock Adaptsfl: Adaptive split federated learning in resource-constrained edge networks.
\newblock {\em arXiv preprint arXiv:2403.13101}, 2024.

\bibitem{lyu2023optimal}
Song Lyu, Zheng Lin, Guanqiao Qu, Xianhao Chen, Xiaoxia Huang, and Pan Li.
\newblock Optimal resource allocation for u-shaped parallel split learning.
\newblock In {\em 2023 IEEE Globecom Workshops (GC Wkshps)}, pages 197--202, 2023.

\bibitem{lin2025leo}
Zheng Lin, Yuxin Zhang, Zhe Chen, Zihan Fang, Cong Wu, Xianhao Chen, Yue Gao, and Jun Luo.
\newblock Leo-split: A semi-supervised split learning framework over leo satellite networks.
\newblock {\em arXiv preprint arXiv:2501.01293}, 2025.

\bibitem{fang2024automated}
Zihan Fang, Zheng Lin, Zhe Chen, Xianhao Chen, Yue Gao, and Yuguang Fang.
\newblock Automated federated pipeline for parameter-efficient fine-tuning of large language models.
\newblock {\em arXiv preprint arXiv:2404.06448}, 2024.

\bibitem{lin2024fedsn}
Zheng Lin, Zhe Chen, Zihan Fang, Xianhao Chen, Xiong Wang, and Yue Gao.
\newblock Fedsn: A federated learning framework over heterogeneous leo satellite networks.
\newblock {\em IEEE Transactions on Mobile Computing}, 2024.

\bibitem{zhang2025state}
Zongyuan Zhang, Tianyang Duan, Zheng Lin, Dong Huang, Zihan Fang, Zekai Sun, Ling Xiong, Hongbin Liang, Heming Cui, and Yong Cui.
\newblock State-aware perturbation optimization for robust deep reinforcement learning.
\newblock {\em arXiv preprint arXiv:2503.20613}, 2025.

\bibitem{duan2025rethinking}
Tianyang Duan, Zongyuan Zhang, Zheng Lin, Yue Gao, Ling Xiong, Yong Cui, Hongbin Liang, Xianhao Chen, Heming Cui, and Dong Huang.
\newblock Rethinking adversarial attacks in reinforcement learning from policy distribution perspective.
\newblock {\em arXiv preprint arXiv:2501.03562}, 2025.

\bibitem{zhang2025robust}
Zongyuan Zhang, Tianyang Duan, Zheng Lin, Dong Huang, Zihan Fang, Zekai Sun, Ling Xiong, Hongbin Liang, Heming Cui, Yong Cui, et~al.
\newblock Robust deep reinforcement learning in robotics via adaptive gradient-masked adversarial attacks.
\newblock {\em arXiv preprint arXiv:2503.20844}, 2025.

\end{thebibliography}

%%
%% If your work has an appendix, this is the place to put it.
% \appendix

\end{document}